\documentclass[preprint,12pt]{elsarticle}

\usepackage{fullpage}

\usepackage{subcaption}
\usepackage{amssymb}
\usepackage{bm}
\usepackage{amsmath}
\usepackage{algorithm}
\usepackage{algpseudocode}
\usepackage{caption}
\usepackage{etoolbox} 

\journal{Journal of Computational Physics}

\begin{document}

\begin{frontmatter}



\title{Asymptotic-preserving semi-implicit finite volume scheme for Extended Magnetohydrodynamics}


\author{Yi Han Toh, Joshua Dolence, Karthik Duraisamy} %% Author name

\affiliation{organization={University of Michigan \& Los Alamos National Laboratory}}

%% Abstract
\begin{abstract}
A Finite Volume (FV) scheme is developed for solving the extended magnetohydrodynamic (XMHD) equations, yielding accurate results in the ideal, resistive, and Hall MHD limits.
%A new Finite Volume (FV) scheme is developed for solving the extended magnetohydrodynamic (XMHD) equations while providing good results in the ideal, resistive and Hall MHD limits. 
This is accomplished by first re-writing the XMHD equations such that it allows the algorithm to retain the use of ideal MHD Riemann solvers and the constrained transport method to preserve  divergence-free magnetic fields. Incorporation of electron inertia and displacement current introduces additional numerical stiffness which motivates a semi-implicit FV scheme that re-formulates the XMHD model as a relaxation system. The equations are then advanced in time using an explicit 2nd-order Runge–Kutta scheme with operator splitting applied to the implicit source term updates at each sub-stage. For additional numerical
stability, a density-dependent slope limiter is implemented to increase flux diffusivity at low density regions where non-ideal effects become significant. The algorithm is subsequently implemented in a scalable adaptive mesh refinement (AMR) framework. As the new algorithm retains many aspects of the ideal MHD formulations, it  asymptotes naturally to  the ideal MHD limit. Moreover, it shows promising results at the resistive and Hall MHD limits. This is verified against reference test problems for ideal, resistive and Hall MHD. 
\end{abstract}

%% Keywords

\end{frontmatter}



\section{Introduction}
\label{sec1}
%% Labels are used to cross-reference an item using \ref command.
In high-energy-density (HED) plasma systems such as magnetized liner inertial fusion (MagLIF) implosions, inertial confinement fusion (ICF) experiments \cite{walsh2020}, Z-pinch implosions \cite{Woolstrum2020}, X-pinches \cite{young2021}, tokamak disruptions \cite{Kleiner2025} and pulsed-power systems \cite{Hamlin2019}, one must model plasmas that span a wide dynamic range of current-carrying densities. At sufficiently low densities, where the characteristic scale lengths become comparable to the ion inertial length, the single-fluid description provided by ideal magnetohydrodynamics (MHD) breaks down. This occurs because ideal MHD, despite its broad applicability, is valid only in certain regimes that satisfy certain simplifying assumptions \cite{Hazeltine2004}. Ideal MHD is typically derived as a limiting case of the two-fluid model which is obtained from the moments of the Boltzmann equation \cite{Hazeltine2004,krall1973principles}. When the two-fluid equations are reformulated in terms of one-fluid variables (the center-of-mass velocity and the current density) together with some other concomitant simplifications, the resulting framework is known as Extended MHD (XMHD) \cite{Hazeltine2004}.

XMHD incorporates two key two-fluid effects necessary for HED plasma systems: (i) the Hall drift arising from the difference between ion and electron fluid velocities and (ii) electron inertia which reflects the finite but small electron mass. Neglecting these two effects recovers the ideal MHD limit, whereas assuming massless electrons leads to the Hall MHD model. Although the concept of XMHD has been recognized at least since the 1950s \cite{1956pfig.book.....S}, it has largely been explored from a theoretical standpoint until recent years. As one approaches smaller spatial scales, extended MHD effects become increasingly significant. Specifically, the ideal MHD regime applies for $L>\lambda_i$, the Hall MHD regime applies for $\lambda_i>L>\lambda_e$ while the electron-inertia-dominated regime for $L<\lambda_e$, where L is a characteristic scale length and $\lambda_s=c/\omega_{ps}$ denotes the skin depth of species s. While extended MHD provides a more general description than ideal MHD, it remains a fluid model and thus does not capture kinetic phenomena such as Landau damping or dissipative effects such as viscosity and resistivity unless these are explicitly included.

Having established that two-fluid MHD model is preferable to the single-fluid MHD model for high-energy-density (HED) plasma systems, especially in low-density regimes where characteristic scale lengths approach the ion inertial length, the applicability of existing numerical methods \cite{CHACON2003573,HAKIM2006418,Hakim2008,Huba2005,LOVERICH2005251,SHUMLAK2003620} in this regime however remains limited. Chacon and Knoll \cite{CHACON2003573} develop a fully implicit Hall MHD scheme through the use of a Jacobian-Free Newton–Krylov (JFNK) method to overcome stiffness associated with dispersive waves, but the solver is incompressible. For compressible flow, Hakim et al. \cite{Hakim2008} solved the ten-moment two-fluid plasma model which evolves the full pressure tensor to capture Hall and finite Larmor radius (FLR) effects, enabling simulation of small-scale instabilities and magnetic reconnection beyond ideal MHD. However, the model ignores collisional relaxation, causing the solutions to deviate significantly from ideal MHD in models of magnetic shock tubes due to the production of dispersive waves arising from the source terms.

Another type of two-fluid plasma model involves coupling the electron and ion fluids with electromagnetic fields through an approximate Riemann solver and a locally implicit treatment of source terms to account for stiffness \cite{SHUMLAK2003620}. This allows for the representation of phenomena beyond standard MHD models like Hall effects and diamagnetic currents but it is unclear if the solver is able to preserve $\nabla\cdot \mathbf{B}=0$ since all variables, including the magnetic field, are co-located. An alternative approach that explicitly enforces a divergence-free magnetic field is the Naval Research Laboratory (NRL)'s 3D Hall MHD code VooDoo \cite{Huba2005}. It employs a finite volume method on a staggered Yee grid, with the magnetic field $\mathbf{B}$ defined on cell faces, the electric field $\mathbf{E}$ along cell edges, and all other variables at the cell centers. The code uses a distribution function approach to compute mass, momentum, and energy fluxes as well as the convective electric field \cite{Huba1999}, combined with a partial donor-cell method for flux limiting. However, electron inertia effects were ignored and the Hall term is sub-cycled for efficiency.  To address these issues, Zhao et al. \cite{ZHAO2014400} developed a positivity-preserving discontinuous Galerkin (DG) scheme that ensures both the preservation of positive physical quantities and a locally divergence-free magnetic field in XMHD systems, thereby maintaining numerical stability.

In their approach, Zhao et al. employ a relaxation algorithm to solve the coupled generalized Ohm’s law (GOL) and Maxwell–Ampere’s law for the electric field. This relaxation method is a semi-implicit time discretization technique that converges to the solution of the algebraic equations obtained by setting the stiff source terms to zero. For the XMHD system, this implies convergence to the GOL when electron inertial terms are neglected, and to Ampere’s law when the displacement current term is neglected.
By using this relaxation approach, the algorithm overcomes constraints associated with under-resolved stiff source terms, allowing time steps to exceed electron plasma and cyclotron frequency scales, therefore enabling the XMHD equations to be integrated on MHD time scales. Moreover, the Hall term is treated locally and implicitly which avoids the computational cost of solving large linear systems.

An additional advantage of the XMHD formulation is that in low-density regions, the current is naturally suppressed by both Hall and electron inertia effects. This enables the use of the unmodified Spitzer resistivity, ensuring a smooth and physically consistent plasma-vacuum transition. This relaxation algorithm has also been implemented in an XMHD code using 2nd-order finite volume (FV) scheme named PERSEUS (Plasma as an Extended-MHD Relaxation System using an Efficient Upwind Scheme) \cite{10.1063/1.3543799}. Using this framework, Zhao et al. successfully simulated a low-density magnetic shock tube problem, demonstrating the robustness and applicability of their method. However, the results remain oscillatory both at low and high density (e.g., Brio Wu shocktube). This motivates the current work which is to develop an XMHD solver that is not only stable at low density but asymptotic-preserving (AP) at the high density limit where it will produce similar results to the already well-established ideal MHD results.

The remainder of this paper  is organized as follows. Section 2 describes the governing equations of the XMHD model and a re-formulation  into a new form that allows one to retain most of the ideal MHD framework as well as its non-dimensional form and associated relaxation parameters. Section 3 focuses on the numerical setup for solving the new form of XMHD equations, including discretization, interpolation and order of accuracy. Section 4 presents  a range of test cases  to test the accuracy of the XMHD solver at ideal vs resistive vs Hall MHD limits. Classical ideal MHD problems are also tested but at different densities to show the deviation from ideal MHD results due to the increment of non-ideal effects. Concluding remarks are then provided in the final section.

%% Use \subsection commands to start a subsection.
\section{Governing equations}
\label{sec1}
The quasi-neutral ($n_e=Z n_i = nZ$) extended magnetohydrodynamics (XMHD) model is described as \cite{ZHAO2014400}
\begin{align}
\frac{\partial \rho}{\partial t} + \nabla \cdot (\rho \mathbf{u}) &= 0 \label{DG_eq:mass} \\
\frac{\partial}{\partial t}(\rho \mathbf{u}) + \nabla \cdot (\rho \mathbf{u} \mathbf{u} + P\mathbb{I}) &= \mathbf{J} \times \mathbf{B} \label{DG_eq:momentum} \\
\frac{\partial E_n}{\partial t} + \nabla \cdot \left[ \mathbf{u}(E_n + P) \right] &= \mathbf{u} \cdot (\mathbf{J} \times \mathbf{B}) + \eta J^2 \label{DG_eq:energy} \\
\frac{\partial \mathbf{B}}{\partial t} + \nabla \times \mathbf{E} &= 0 \label{DG_eq:faraday} \\
\frac{\partial \mathbf{E}}{\partial t} - c^2 \nabla \times \mathbf{B} &= -\frac{1}{\epsilon_0} \mathbf{J} \label{DG_eq:ampere} \\
\frac{\partial \mathbf{J}}{\partial t} + \nabla \cdot \left[ \mathbf{u} \mathbf{J} + \mathbf{J} \mathbf{u} - \frac{1}{n_e e} \mathbf{J} \mathbf{J} - \frac{e}{m_e} P_e\mathbb{I} \right] &= \frac{n_e e^2}{m_e} \left[ \mathbf{E} + \mathbf{u} \times \mathbf{B} - \eta \mathbf{J} - \frac{1}{n_e e} \mathbf{J} \times \mathbf{B} \right] \label{DG_eq:current} \\
\frac{\partial S_e}{\partial t} + \nabla \cdot (\mathbf{u}_e S_e) &= (\gamma - 1) n_e^{1 - \gamma} \eta J^2, \label{DG_eq:entropy}
\end{align}
where $\eta$ is resistivity, $\rho=m_i n_i + m_e n_e$, $\mathbf{J}=e(Zn_i \mathbf{u}_i-n_e \mathbf{u}_e)$, $\mathbf{u}_e=\mathbf{u}-\frac{\mathbf{J}}{n_e e}$, Z is net ion charge, $P_e=S_e n_e^{\gamma-1}$ is electron pressure, $E_n=\rho e + \frac{1}{2}\rho u^2$ is hydrodynamic  energy density and assumed ideal equation of state thus, $\rho e=\frac{P}{\gamma-1}$. Including electron inertia allows for a consistent treatment of low-density or vacuum regions, eliminating the need to introduce an artificial vacuum resistivity. The complete electromagnetic formulation of Maxwell’s equations, together with the Hall term, ensures a self-consistent calculation of all electric field components and allows one to properly apply general boundary conditions to the problem \cite{10.1063/1.3543799}. Moreover, including both electron inertia and displacement current ensures Whistler waves to resonate at the cyclotron frequency which prevents the wave velocity to grow unbounded with frequency, hence helping to stabilize the Hall term \cite{SHUMLAK2003620}. However, they introduced additional numerical stiffness which motivates a semi-implicit scheme that will be detailed below. As for the electron equation of state, it is modeled using the electron entropy density $S_e=\frac{P_e}{n_e^{\gamma-1}}$ as it gives rise to a positive definite source term which guarantees that the electron pressure is positive for numerical stability \cite{ZHAO2014400}. The limitation is that one cannot capture weak solutions in the form of electron shocks like Langmuir shocks. But since the focus here is on low frequency phenomena, well below the electron plasma frequency, this use of electron entropy is reasonable in this context.

%% Use \subsubsection, \paragraph, \subparagraph commands to 
%% start 3rd, 4th and 5th level sections.
%% Refer following link for more details.
%% https://en.wikibooks.org/wiki/LaTeX/Document_Structure#Sectioning_commands

\subsection{Reformulation of governing equations}
\begin{figure}[htbp]
    \centering
    \begin{subfigure}[b]{\linewidth}
        \centering
        \includegraphics[width=\linewidth]{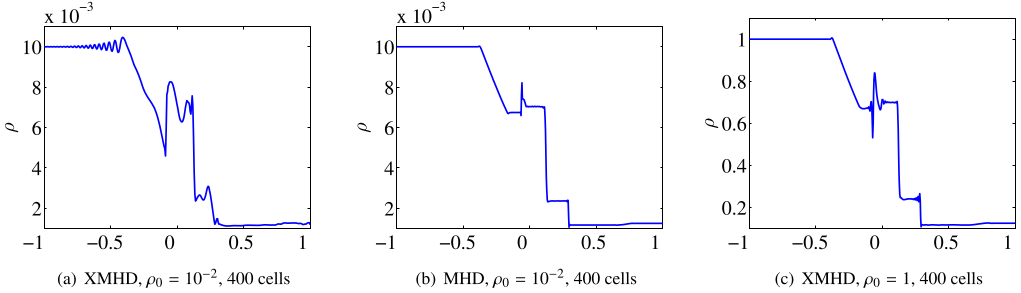}
    \end{subfigure}
    \vspace{1em} 
    \begin{subfigure}[b]{0.9\linewidth}
        \centering
        \includegraphics[width=\linewidth]{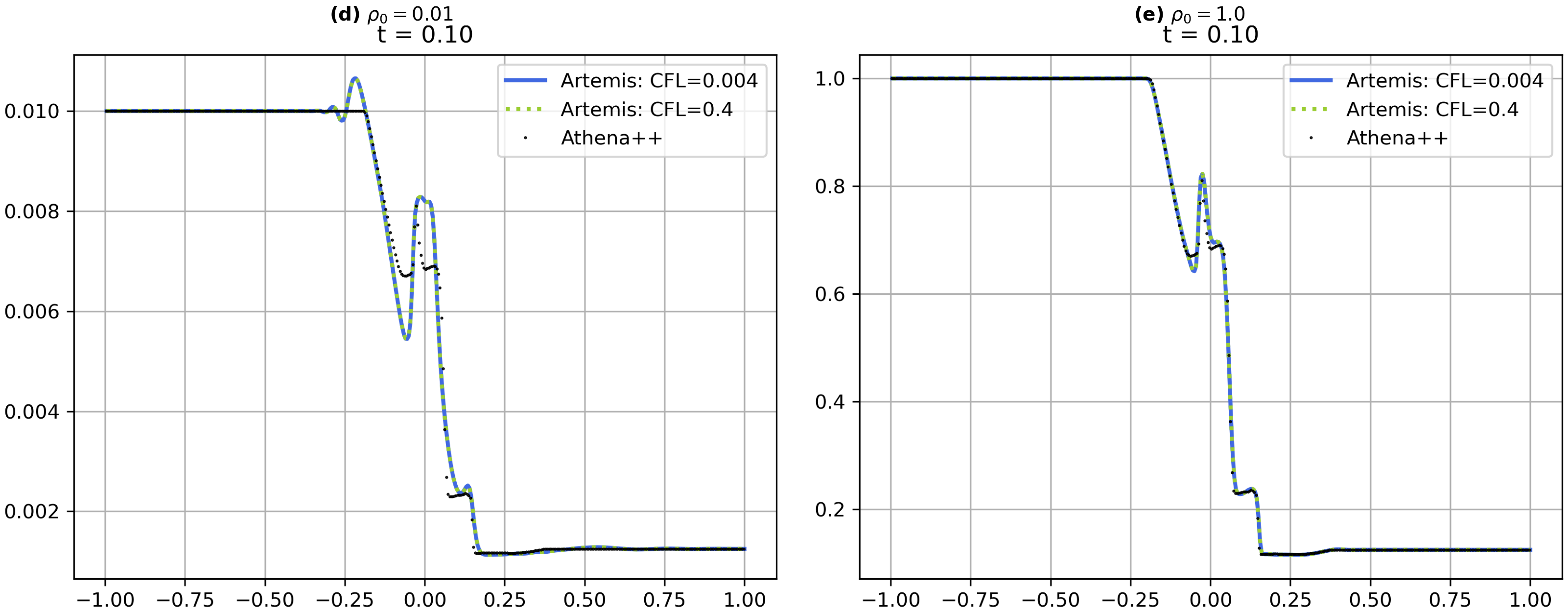}
    \end{subfigure}    
    \caption{Brio-Wu shock tube solution from Zhao et. al. \cite{ZHAO2014400} using LLF Riemann solver at cell interface whereby the left figure (a) contains oscillation due to whistler wave introduced by Hall effect which is important at low density. However, the figure (c) on the right is unable to replicate the same solution as figure (b) at the ideal MHD limit. Figures (d) and (e) are produced using XMHD (LLF) on Artemis and compared against Athena++ ideal MHD solver (LLF) to show the increment of non-ideal effects at lower density. Different CFL numbers are also used to ensure that the solution is consistent at different time resolution.}
    \label{fig:DG_XMD_bad_BWtube}
\end{figure}

In both the momentum (\ref{DG_eq:momentum}) and energy (\ref{DG_eq:energy}) equations, the magnetic field contributions are treated as source terms. As a result, the Rankine–Hugoniot conditions are not fully captured by the flux formulation and must be corrected by the source term, which may still lead to inaccuracies near discontinuities like in Fig. \ref{fig:DG_XMD_bad_BWtube} from Zhao et. al. \cite{ZHAO2014400}. Therefore, Eqs. \ref{DG_eq:mass} to \ref{DG_eq:faraday} must be re-written to preserve ideal MHD equations on the left-hand side while the additional physics associated with extended MHD will be placed at the right-hand side as source term. The new form of extended MHD based on above Eqs. \ref{DG_eq:mass} to \ref{DG_eq:entropy} is then written as follows

\begin{align}
\frac{\partial \rho}{\partial t} + \nabla \cdot (\rho \mathbf{u}) = 0 \label{XMHDv1:mass} 
\\
\frac{\partial}{\partial t}(\rho \mathbf{u}) + \nabla \cdot \left[\rho \mathbf{u} \mathbf{u} - \frac{1}{\mu_0}\mathbf{B} \mathbf{B} + \left(P + \frac{1}{2\mu_0}B^2 \right)\mathbb{I}\right] = \left(\mathbf{J} - \frac{1}{\mu_0}\nabla \times \mathbf{B}\right) \times \mathbf{B} \label{XMHDv1:momentum} 
\\
\frac{\partial}{\partial t} \left(E_n + \frac{1}{2\mu_0}B^2 \right) + \nabla \cdot \left[ \mathbf{u} \left(E_n + P + \frac{1}{\mu_0}B^2 \right) - \frac{1}{\mu_0} (\mathbf{B} \cdot \mathbf{u})\mathbf{B} \right] \label{XMHDv1:energy} \\ = -\frac{1}{\mu_0} \mathbf{B} \cdot [\nabla \times(\mathbf{E}+\mathbf{u}\times \mathbf{B})]   \notag
+ \mathbf{u} \cdot \left(\left(\mathbf{J} - \frac{1}{\mu_0} \nabla \times \mathbf{B} 
\right) \times \mathbf{B} \right) + \eta J^2  
\\
\frac{\partial \mathbf{B}}{\partial t} - \nabla \cdot (\mathbf{B}\mathbf{u}-\mathbf{u}\mathbf{B}) = -\nabla \times (\mathbf{E} + \mathbf{u} \times \mathbf{B}) \label{XMHDv1:faraday} 
\\
\frac{\partial \mathbf{E}}{\partial t} - c^2 \nabla \times \mathbf{B} = -\frac{1}{\epsilon_0} \mathbf{J} \label{XMHDv1:ampere} \\
\frac{\partial \mathbf{J}}{\partial t} + \nabla \cdot \left[ \mathbf{u} \mathbf{J} + \mathbf{J} \mathbf{u} - \frac{1}{n_e e} \mathbf{J} \mathbf{J} - \frac{e}{m_e} P_e\mathbb{I} \right] = \frac{n_e e^2}{m_e} \left[ \mathbf{E} + \mathbf{u} \times \mathbf{B} - \eta \mathbf{J} - \frac{1}{n_e e} \mathbf{J} \times \mathbf{B} \right] \label{XMHDv1:current} \\
\frac{\partial S_e}{\partial t} + \nabla \cdot (\mathbf{u}_e S_e) = (\gamma - 1) n_e^{1 - \gamma} \eta J^2. \label{XMHDv1:entropy}
\end{align}
The above derivation can be found in \ref{App1}. Extended comments on treatment of the electron entropy equation are provided in section~\ref{Subsect:Strongly_collision}.

For ideal MHD, plasma density is high which means that a large $n_e$ will render the left-hand side (electron inertia and pressure term) and the Hall-term on the right-hand side of Eq. \ref{XMHDv1:current} to negligible as compared to the specific form of resistive Ohm's law. Additionally, ideal MHD assumed infinite conductivity and thus, Eq. \ref{XMHDv1:current} reduced to $\mathbf{E}=-\mathbf{u}\times \mathbf{B}$. This will subsequently reduce Eq. \ref{XMHDv1:faraday} to that of ideal MHD's induction equation. Additionally, ideal MHD assumes a non-relativistic regime, which means that the Ampere's law in Eq. \ref{XMHDv1:ampere} (or $\frac{1}{c^2}\frac{\partial\mathbf{E}}{\partial t}-\nabla\times\mathbf{B}=-\mu_0 \mathbf{J}$) will simplify to $\mathbf{J}=\frac{1}{\mu_0}\nabla \times \mathbf{B}$. This will then eliminate the source term of Eq. \ref{XMHDv1:momentum}, recovering the ideal MHD's momentum equation. The above reductions make the right-hand side of Eq. \ref{XMHDv1:energy} go to zero, simplifying it to the ideal MHD energy equation. Most importantly, the jump conditions across the shocks will be preserved at the ideal MHD limit and this also allows the user to use existing MHD Riemann solvers like LLF or HLLD to solve for the left-hand side of Eqs. \ref{XMHDv1:mass} to \ref{XMHDv1:faraday}.

Eqs. \ref{XMHDv1:ampere} and \ref{XMHDv1:current} can then be non-dimensionalized (details in \ref{App2}) to obtain the non-dimensional parameters which characterize relaxation \cite{ZHAO2014400} as follows,
\begin{align}
\frac{\partial \mathbf{E}}{\partial t}
&= \frac{c^{2}}{v^{2}}\bigl(\nabla\times\mathbf{B}-\mathbf{J}\bigr)
\label{XMHDv1:ampere_ND}\\[1ex]
\frac{\partial \mathbf{J}}{\partial t}
&+\nabla\cdot\left[
   \mathbf{u}\mathbf{J}
 + \mathbf{J}\mathbf{u}
 - \frac{\lambda_{i}}{L_{0}n}\mathbf{J}\mathbf{J}
 - \frac{m_{i}L_{0}}{m_{e}\lambda_{i}} P_e  \mathbb{I}
\right]
= \frac{L_{0}^{2} n_e}{\lambda_{e}^{2}}
   \left[
     \mathbf{E}
   + \mathbf{u}\times\mathbf{B}
   - \frac{\lambda_{i}}{L_{0}n_e}\mathbf{J}\times\mathbf{B}
   - \eta\mathbf{J}
   \right],
\label{XMHDv1:current_ND}
\end{align}
where $v=\frac{L_0}{t_0}$ is the characteristic speed while $L_0$ and $t_0$ are the representative length and time, respectively. The electron and ion inertial lengths are $\lambda_j^2=\frac{m_j}{n_0 e^2 \mu_0}$. These give the relaxation parameters $\frac{c^2}{v^2}$ and $\frac{L_0^2 n}{\lambda_e^2}$ in above Eqs. \ref{XMHDv1:ampere_ND} and \ref{XMHDv1:current_ND}, respectively. The remaining non-dimensionalized equations are
\begin{align}
\frac{\partial \rho}{\partial t} + \nabla \cdot (\rho \mathbf{u}) = 0 \label{XMHDv1:mass_ND} 
\\
\frac{\partial}{\partial t}(\rho \mathbf{u}) + \nabla \cdot \left[\rho \mathbf{u} \mathbf{u} - \frac{B_0^2}{\mu_0 n_0 m_i v^2}\mathbf{B} \mathbf{B} + \left(P + \frac{1}{2}\frac{B_0^2}{\mu_0 n_0 m_i v^2}B^2 \right)\mathbb{I}\right] \label{XMHDv1:momentum_ND}
\\
= \frac{B_0^2}{\mu_0 n_0 m_i v^2}\left(\mathbf{J} - \nabla \times \mathbf{B}\right) \times \mathbf{B}  \notag
\\
\frac{\partial}{\partial t} \left(E_n + \frac{1}{2}\frac{B_0^2}{\mu_0 n_0 m_i v^2}B^2 \right) + \nabla \cdot \left[ \mathbf{u} \left(E_n + P + \frac{B_0^2}{\mu_0 n_0 m_i v^2}B^2 \right) - \frac{B_0^2}{\mu_0 n_0 m_i v^2}(\mathbf{B} \cdot \mathbf{u})\mathbf{B} \right]
\label{XMHDv1:energy_ND} 
\\
= \frac{B_0^2}{\mu_0 n_0 m_i v^2}\left( - \mathbf{B} \cdot [\nabla \times(\mathbf{E}+\mathbf{u}\times \mathbf{B})] \right.   \notag
\left. + \mathbf{u} \cdot \left(\left(\mathbf{J} -  \nabla \times \mathbf{B} 
\right) \times \mathbf{B} \right) + \eta J^2 \right) 
\\
\frac{\partial \mathbf{B}}{\partial t} - \nabla \cdot (\mathbf{B}\mathbf{u}-\mathbf{u}\mathbf{B}) = -\nabla \times (\mathbf{E} + \mathbf{u} \times \mathbf{B}) \label{XMHDv1:faraday_ND} 
\\
\frac{\partial S_e}{\partial t} + \nabla \cdot (\mathbf{u}_e S_e) = \frac{B_0^2}{\mu_0 n_0 m_i v^2} (\gamma - 1) n_e^{1 - \gamma} \eta J^2, \label{XMHDv1:entropy_ND}
\end{align}
where $\frac{B_0^2}{\mu_0 n_0 m_i v^2}$ arises from non-dimensionalization and the dimensional constants should be chosen such that $\frac{B_0^2}{\mu_0 n_0 m_i v^2}$ is $\sim$1. This ensures that the left-hand side of Eqs. \ref{XMHDv1:mass_ND} to \ref{XMHDv1:faraday_ND} is equivalent to that of ideal MHD solvers.

The phenomena of interest in high energy density (HED) plasmas happens at time scales much slower than the characteristic electron plasma and cyclotron frequencies, meaning that $v<<c$ and $\lambda_e<<L_0$, respectively. This forces Eqs. \ref{XMHDv1:ampere_ND} and \ref{XMHDv1:current_ND} to relax to equilibrium as follows,
\begin{align}
\nabla\times\mathbf{B}=\mathbf{J}
\label{XMHDv1:ampere_ND_relax}
\\[1ex]
\mathbf{E}
   + \mathbf{u}\times\mathbf{B}
   - \frac{\lambda_{i}}{L_{0}n}\mathbf{J}\times\mathbf{B}
   = \eta\mathbf{J}.
\label{XMHDv1:current_ND_relax}
\end{align}
Thus, when the electron inertial scale is under-resolved, the solution to Eqs. \ref{XMHDv1:ampere_ND} and \ref{XMHDv1:current_ND} will relax to the non-relativistic Ampere's law and inertia-less Generalized Ohm's law (GOL) in above Eqs. \ref{XMHDv1:ampere_ND_relax} and \ref{XMHDv1:current_ND_relax}. 

\subsection{Strongly collisional plasma} \label{Subsect:Strongly_collision}
One limitation of the above set of XMHD equations is that it does not contain the source term responsible for equilibrating ion and electron temperatures as the plasma becomes strongly collisional. To account for this, a more general equation for the electron entropy equation is considered 
\begin{equation}
    \frac{\partial \tilde{S}_e}{\partial \tilde{t}} + \tilde{\nabla} \cdot (\tilde{\mathbf{u}}_e \tilde{S}_e)=\tilde{Q}_{ei}
\end{equation}
where tilde denotes dimensional variables. The quasi-neutrality assumption then allows the electron-ion interaction to be written as \cite{SANGAM2021110565}
\begin{equation}
    \tilde{Q}_{ei}=\tilde{\nu}_{ei}^g (\tilde{T}_i-\tilde{T}_e)+\tilde{\zeta}_{ei}\tilde{\eta} \tilde{J}^2
\end{equation}
The electron entropy equation can then be written as
\begin{equation} \label{eq:full_electron_entropy}
    \frac{\partial \tilde{S}_e}{\partial \tilde{t}} + \tilde{\nabla} \cdot (\tilde{\mathbf{u}}_e \tilde{S}_e) = \tilde{n}_e^{1 - \gamma} (\gamma-1)\left[\tilde{\nu}_{ei}^g(\tilde{T}_i-\tilde{T}_e) + \tilde{\zeta}_{ei}\tilde{\eta} \tilde{J}^2\right]
\end{equation}
where the electron-ion exchange coefficient rate is $\tilde{\nu}_{ei}^g=\frac{3}{2}\tilde{k}_B \frac{\tilde{n}_e \tilde{n}_i}{\tilde{\tau}_{ie}\tilde{n}_e+\tilde{\tau}_{ei}\tilde{n}_i}$ and take $\tilde{\nu}_{ie}=\frac{\tilde{n}_e \tilde{m}_e}{\tilde{n}_i \tilde{m}_i}\tilde{\nu}_{ei}$ by assuming that both ions and electrons are close to thermal (Maxwellian) distributions where their momentum exhanges are symmetric. As for the fractional Ohmic heating on electron, the following model should be used \cite{SANGAM2021110565},
\begin{equation}
    \tilde{\zeta}_{ei}=\frac{1}{2}\left[\frac{\tilde{\tau}_{ei}\tilde{m}_i}{\tilde{\tau}_{ie}\tilde{m}_e+\tilde{\tau}_{ei}\tilde{m}_i}+\frac{\tilde{\tau}_{ie}}{\tilde{\tau}_{ie}+\tilde{\tau}_{ei}}\right]
\end{equation}
The electron-ion relaxation time can then be obtained using FLASH model \cite{flash_userguide_devel} as follows,
\begin{equation}
    \tilde{\tau}_{ei}=\frac{3\tilde{k}_B^{3/2}}{8\sqrt{2\pi}\tilde{e}^4} \frac{(\tilde{m}_i \tilde{T}_e + \tilde{m}_e \tilde{T}_i)^{3/2}}{(\tilde{m}_e\tilde{m}_i)^{1/2}Z^2\tilde{n}_i \ln \Lambda}(4\pi\tilde{\epsilon}_0)^2
\end{equation}
where Coulomb logarithm is taken as 5 here. Note that in the ideal MHD limit, the macroscopic dynamics depends only on the total pressure (e.g., $p=P_i+P_e$) which enters the momentum (Eq. \ref{XMHDv1:momentum_ND}) and energy (Eq. \ref{XMHDv1:energy_ND}) equations in exactly the same way as in the single-temperature ideal MHD system. This is assuming that Ampere's law and GOL relax to $\mathbf{J}=\nabla \times \mathbf{B}$ and $\mathbf{E}=-\mathbf{u}\times\mathbf{B}$ such that the non-ideal effects go to zero. Subsequently, the electron pressure effects do not have a leading order effect on the macroscopic dynamics (see Sect. \ref{subsec:relax_stiff_src} for more details) and hence, the partition of $p$ into $P_i$ and $P_e$ becomes dynamically irrelevant in this limit.  Therefore, in the ideal MHD limit, only the total pressure controls the dynamics. In other regimes like the Hall MHD limit, the electron-ion energy exchange is naturally small due to low density involved.  However, accurate modeling of $T_i$ and $T_e$ is still crucial, especially in regimes where radiation or other electron-specific physics are significant and thus, are included here.

\section{Numerical formulation}
The final dimensionless XMHD equations to solve numerically are as follows,
\begin{align}
\frac{\partial \rho}{\partial t} + \nabla \cdot (\rho \mathbf{u}) &= 0 \label{XMHDv2:ND_mass} 
\\
\frac{\partial}{\partial t}(\rho \mathbf{u}) + \nabla \cdot \left[\rho \mathbf{u} \mathbf{u} - \mathbf{B} \mathbf{B} + \left(P + \frac{1}{2}B^2 \right)\mathbb{I}\right] &= \left(\mathbf{J} - \nabla \times \mathbf{B}\right) \times \mathbf{B} \label{XMHDv2:ND_momentum} 
\\
\frac{\partial}{\partial t} \left(E_n + \frac{1}{2}B^2 \right) + \nabla \cdot \left[ \mathbf{u} \left(E_n + P + B^2 \right)\right. &- \left.(\mathbf{B} \cdot \mathbf{u})\mathbf{B} \right]   \notag
\\ 
= \left( - \mathbf{B} \cdot [\nabla \times(\mathbf{E}+\mathbf{u}\times \mathbf{B})]\right. &+ \left.\mathbf{u} \cdot \left(\left(\mathbf{J} -  \nabla \times \mathbf{B} 
\right) \times \mathbf{B} \right) + \eta J^2 \right) \label{XMHDv2:ND_energy} 
\\
\frac{\partial \mathbf{B}}{\partial t} - \nabla \cdot (\mathbf{B}\mathbf{u}-\mathbf{u}\mathbf{B}) &= -\nabla \times (\mathbf{E} + \mathbf{u} \times \mathbf{B}) \label{XMHDv2:ND_faraday} 
\\
\frac{\partial \mathbf{E}}{\partial t}
&= \frac{c^{2}}{v^{2}}\bigl(\nabla\times\mathbf{B}-\mathbf{J}\bigr)
\label{XMHDv2:ND_ampere}\\[1ex]
\frac{\partial \mathbf{J}}{\partial t}
+\nabla\cdot\left[
   \mathbf{u}\mathbf{J}
 + \mathbf{J}\mathbf{u}
 - \frac{\lambda_{i}}{L_{0}n_e}\mathbf{J}\mathbf{J}
 - \frac{m_{i}L_{0}}{m_{e}\lambda_{i}} P_e  \mathbb{I}
\right]
&= \frac{L_{0}^{2} n_e}{\lambda_{e}^{2}}
   \left[
     \mathbf{E}
   + \mathbf{u}\times\mathbf{B}
   - \frac{\lambda_{i}}{L_{0}n_e}\mathbf{J}\times\mathbf{B}
   - \eta\mathbf{J}
   \right]
\label{XMHDv2:ND_current} \\
\frac{\partial S_e}{\partial t} + \nabla \cdot (\mathbf{u}_e S_e) = (\gamma - 1) n_e^{1 - \gamma} &\left[\frac{k_B T_0}{m_i v^2}\frac{m_e}{m_i+m_e}\frac{ n_e}{\tau_{ei}} (T_i-T_e) +  \zeta_{ei} \eta J^2 \right] \label{XMHDv2:ND_entropy}
\end{align}
where dimensional constants are chosen such that $\frac{B_0^2}{\mu_0 n_0 m_i v^2}\approx1$. Note that Eq. \ref{XMHDv2:ND_entropy} is obtained by non-dimensionalizing Eq. \ref{eq:full_electron_entropy}.

\begin{figure} [http]
    \centering
    \includegraphics[width=0.8\linewidth]{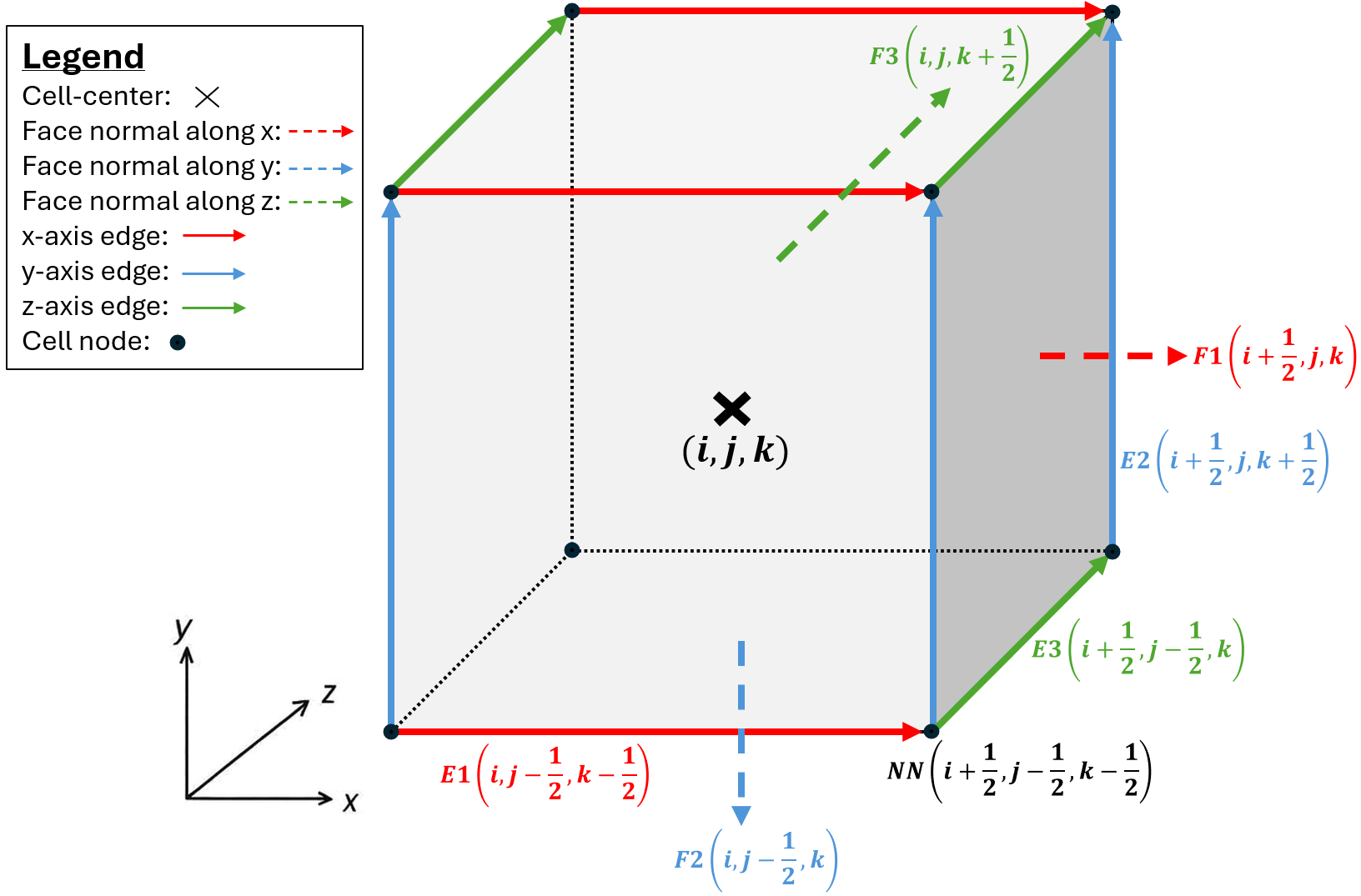}
    \caption{A 3D computational cell on a Cartesian grid is illustrated to indicate the positions of various field variables. The cell center is marked with an ‘x’, while the arrows normal to each cell face represent F1, F2, and F3, corresponding to faces with normal directions along the x, y, and z axes, respectively. The cell edges are shown by arrows labeled E1, E2, and E3, indicating edges oriented along the x, y, and z directions. The cell nodes are depicted as solid black dots and are denoted by NN. $\mathbf{B}$ is placed on the faces (F) while both $\mathbf{E}$ and $\mathbf{J}$ lie on the edges (E). The rest of the conserved variables are placed at the cell-center ($i,j,k$).}
    \label{fig:mesh_TE}
\end{figure}

The XMHD solver is implemented in Artemis, a multifluid radiation hydrodynamics code built on the Parthenon \cite{osti_1903416} high-performance computing (HPC) library, which includes adaptive mesh refinement (AMR) support. This is accomplished by first implementing ideal MHD in Artemis and verifying against common ideal MHD test problems from Athena++ \cite{2020ApJS..249....4S} before extending the MHD solver to include additional physics such as resistivity and Hall effects from extended MHD based on Eqs. \ref{XMHDv2:ND_mass} to \ref{XMHDv2:ND_entropy}. However, several strategies should be considered before discretizing Eqs. \ref{XMHDv2:ND_mass} to \ref{XMHDv2:ND_entropy}. First, in order to ensure we recover ideal MHD behavior appropriately, we choose to evolve Eq.~\ref{XMHDv2:ND_faraday} via a constrained transport \cite{BALSARA1999270} method with face-centered magnetic fields, as shown in Fig.~\ref{fig:mesh_TE}.

Second, spatial interpolation of variables should be high-order and minimized as they introduce additional errors. For instance, $\mathbf{E}$ should be placed at the cell edges so that it can be used to update the magnetic field together with $\mathbf{u}\times \mathbf{B}$ without requiring spatial interpolation in Eq. \ref{XMHDv2:ND_faraday}. Furthermore, taking its curl to update the face-centered magnetic field will preserve $\nabla \cdot \mathbf{B}$ up to machine precision, similar to constrained transport method. As for $\mathbf{J}$, it should be placed at the cell-edges since it will be updated together with $\mathbf{E}$ through implicit source terms and therefore, both must be co-located to be consistent. However, spatial interpolation of the non-ideal MHD ($\mathbf{E}+\mathbf{u}\times\mathbf{B}$) and relativistic source terms ($\mathbf{J}-\nabla \times \mathbf{B}$) from cell edges to cell center in Eqs. \ref{XMHDv2:ND_momentum} and \ref{XMHDv2:ND_energy} are still required.

To compute the intercell numerical inviscid flux, the local Lax-Friedrichs (LLF) Riemann solver \cite{LLF1954} can be applied for all the variables. For improved accuracy, the convective components from the left-hand side of Eqs. \ref{XMHDv2:ND_mass} to \ref{XMHDv2:ND_faraday} can be obtained using the HLLD Riemann solver \cite{HLLD_Miyoshi} instead. Operator splitting is applied to $\mathbf{E}$, $\mathbf{J}$ and electron-ion energy exchange source terms while the rest of the source terms are explicitly updated together with the flux. The algorithm for solving Eqs. \ref{XMHDv2:ND_mass}-\ref{XMHDv2:ND_entropy} is detailed as follows,

\makeatletter
\renewcommand{\fnum@algorithm}{\textbf{Algorithm~\thealgorithm}}
\makeatother

\noindent\rule{\linewidth}{1pt}          
\captionof{algorithm}{Finite Volume scheme for  XMHD equations}         
\noindent\rule{\linewidth}{1pt} 
{\small
\begin{algorithmic}[1]
\State Initialize cell-centered conserved variables $\mathbf{U}^n$ and face-centered magnetic field $\mathbf{B}^n$.
\State Initialized edge-centered $\mathbf{E}$ and $\mathbf{J}$ using the ideal MHD Riemann solver ($-\mathbf{u}\times\mathbf{B}|^n$) and $\nabla\times \mathbf{B}^n$, respectively.

\noindent\hspace*{-\algorithmicindent}\underline{\textbf{Predictor stage:}}
\State Convert conserved variables from $\mathbf{U}^n$ to primitive variables $\mathbf{V}^n$.
\State Compute and store numerical interface fluxes with Riemann solver using $\mathbf{V}^n$, $\mathbf{B}^n$ and $\mathbf{J}^n$.
\State Compute and store ideal MHD's electromotive forces (EMFs) at cell edges using the $\mathbf{B}$-component from the interface fluxes through constrained transport method.
\State Interpolate and store the $\mathbf{J}$-component interface fluxes to the cell nodes.
\State Compute and store explicit source terms from Eqs. \ref{XMHDv2:ND_mass} to \ref{XMHDv2:ND_entropy} using current values including $-\mathbf{u}\times\mathbf{B}|^n$ obtained from Step 5. Note that for Eq. \ref{XMHDv2:ND_entropy}, only the source term due to Ohmic heating is computed and stored.
\State Update edge-centered $\mathbf{J}^n$ to $\mathbf{J}^{**}$ using the divergence of its node fluxes and this is  detailed in Sect. \ref{subsec:Jupdate}.
\State Update face-centered $\mathbf{B}^n$ to $\mathbf{B}^{*}$ by taking the curl of edge-centered EMFs computed from Step 5 and explicit source term from Step 7, similar to constrained transport method.
\State Advance remaining cell-centered variables from $\mathbf{U}^n$ and $S_e^n$ to intermediate (predictor) state $\mathbf{U}^{*}$ and $S_e^{**}$, respectively, using the computed fluxes and explicit source terms.
\State Update edge-centered $\mathbf{E}^n$ using the curl of the face-centered $\mathbf{B}^{*}$ to $\mathbf{E}^{**}$ like in Eq. \ref{XMHDv1:ampere_ND_no_source}. Note that electric field flux update is split from the rest to ensure AP property as described in Sect. \ref{subsec:AP}.
\State Recompute primitive variables $\mathbf{V}^*$ from the updated intermediate conserved variables $\mathbf{U}^*$ earlier since the implicit source term update is split from the flux and explicit source term update. 
\State Recompute and store $-\mathbf{u}^*\times \mathbf{B}^*$ at the edge using ideal MHD Riemann solver and constrained transport method with $\mathbf{V}^*$ and $\mathbf{B}^*$ as the inputs.

\noindent\hspace*{-\algorithmicindent}\underline{\textbf{Implicit source term update:}}
\State Solving for $\mathbf{J}^*$ requires a semi-implicit time advancement to the source terms in Eqs. \ref{XMHDv2:ND_ampere} and \ref{XMHDv2:ND_current} which yields the following spatially local linear algebraic equations for $\mathbf{E}^{*}$ and $\mathbf{J}^{*}$,
\begin{align}
\mathbf{E}^{*} &= \mathbf{E}^{**} - \Delta t \frac{c^2}{v^2}\mathbf{J}^{*} 
\label{XMHD_implicitE} \\[1ex]
\mathbf{J}^{*} &= \mathbf{J}^{**} + \Delta t \frac{L^2_0 \underline{n^*}}{\lambda_e^2}
\Bigl(
\mathbf{E}^{*} +\mathbf{u}^*\times \mathbf{B}^* 
-\frac{\lambda_i}{L_{0}\underline{n^*}}\mathbf{J}^{*}\times \mathbf{B}^*
-\underline{\eta^*}\mathbf{J}^{*}
\Bigr), \label{XMHD_implicitJ}
\end{align}
where $\underline{n^*}=\text{diag}(n^*_x,n^*_y,n^*_z)$, $\underline{\eta^*}=\text{diag}(\eta^*_x,\eta^*_y,\eta^*_z)$ and $\mathbf{u}^*\times \mathbf{B}^*$ is obtained from Step 13. Note that $\mathbf{E}^{**}$ and $\mathbf{J}^{**}$ are the intermediate values before implicit source term update. Substituting $\mathbf{E}^{*}$ from Eq. \ref{XMHD_implicitE} into Eq. \ref{XMHD_implicitJ} gives a 3 by 3 linear system 
    \begin{equation} \label{linear_solve}
        \mathbb{A}\mathbf{J}^{*}=\mathbf{b}^*,
    \end{equation}
    where $k=x,y,z$ corresponds to values at each edge and 
    \begin{align}
        \mathbb{A} &= 
        \begin{bmatrix}
        \alpha_x & \frac{\lambda_i}{L_0 n^*_x}B_z^* & -\frac{\lambda_i}{L_0 n^*_x} B_y^*
        \\
        -\frac{\lambda_i}{L_0 n^*_y}B_z^* & \alpha_y & \frac{\lambda_i}{L_0 n^*_y}B_x^* 
        \\
        \frac{\lambda_i}{L_0 n^*_z}B_y^* & -\frac{\lambda_i}{L_0 n^*_z}B_x^* & \alpha_z
        \end{bmatrix}, 
        \ \ \ \ \ \ 
        \mathbf{b}^*=\frac{\bm{\tau}\bm{\tau} \mathbb{I}}{\Delta t}\mathbf{J}^{**}+\mathbf{E}^{**}+\mathbf{u}^*\times \mathbf{B}^*,
        \notag \\
        \alpha_k &= \frac{\tau^2_k}{\Delta t}+\Delta t \frac{c^2}{v^2}+\eta^*_k, 
        \ \ \ \ \ \
        \tau_k=\frac{\lambda_e}{L_0 \sqrt{n^*_k}}. \notag
    \end{align}
    $\mathbb{A}$ can then be explicitly inverted, allowing $\mathbf{J}^{*}$ to be solved analytically according to Eq. \ref{XMHD_implicitJ}. Here Gaussian elimination with partial pivoting \cite{Trefethen_numerical_linear_algebra} is used to solve Eq. \ref{linear_solve} in case $\mathbb{A}$ becomes poorly-conditioned for improved numerical stability. Additionally, directional splitting of implicit source term update is required due to the Hall term through Algorithm 2 which is explained further in Sect. \ref{Subsect:Direc_split_Hall}.
\State Solve for $\mathbf{E}^*$ using $\mathbf{J}^*$ computed from previous step using Eq. \ref{XMHD_implicitE}.
\State For Eq. \ref{XMHDv2:ND_entropy}, update $S_e^{**}$ to $S_e^*$ through the electron-ion energy exchange source term implicitly using Eq. \ref{eq:Te_eqn2} as detailed in Sect. \ref{subsec:electron_ion_energy_exchange}.

\noindent\hspace*{-\algorithmicindent}\underline{\textbf{Corrector stage:}}
\State Recompute fluxes, source terms, and EMFs at the intermediate state like in the previous steps from 3 to 16.
\State Apply the RK2 (Heun) corrector step to obtain $\mathbf{U}^{n+1}$, $\mathbf{B}^{n+1}$, $\mathbf{E}^{n+1}$ and $\mathbf{J}^{n+1}$.

\noindent\hspace*{-\algorithmicindent}\underline{\textbf{Post-time step update:}}
\State Update simulation time: $t^{n+1} = t^n + \Delta t$
\State Apply boundary conditions.
\State Repeat steps 3 to 20 until simulation end time is reached.
\end{algorithmic}
}
\noindent\rule{\linewidth}{1pt} 

The explicit components of the time evolution can also be performed using a higher-order strong-stability preserving (SSP) integrator like the 3rd-order Shu-Osher Runge-Kutta method \cite{SHU1988439} instead of the SSP-RK2 method used here.

\subsection{Asymptotic-preserving property} \label{subsec:AP}
A scheme is considered Asymptotic-preserving (AP) if it satisfies the following two properties \cite{Jin1999},
\begin{itemize}
    \item For a fixed discretization parameters, it is consistent with the microscopic model for finite scaling parameters and becomes a consistent discretization of the macroscopic model as the scaling parameter approaches zero.
    \item For uniform stability, implicit discretization is required but this can either be implemented explicitly or at least very efficiently (e.g., avoiding the use of iterative nonlinear algebraic solvers). 
\end{itemize}
The 1D XMHD equations can be compactly written as
\begin{equation} \label{eq:AP_ideal}
    \partial_t \mathcal{U}_i+\partial_x \mathcal{F}_i(\mathcal{U}_i)=\mathcal{S}_i(\mathcal{U}_i,\hat{\mathcal{U}}_i) 
\end{equation}
\begin{equation} \label{eq:AP_NI}    \partial_t\hat{\mathcal{U}}_i+\partial_x \hat{\mathcal{F}}_i(\mathcal{U}_i,\hat{\mathcal{U}}_i)=\frac{1}{\epsilon_i}\hat{\mathcal{R}}_i(\mathcal{U}_i,\hat{\mathcal{U}}_i) +\hat{\mathcal{S}}_i(\mathcal{U}_i,\hat{\mathcal{U}}_i)
\end{equation}
where no hat variables denotes ideal MHD components like $(\rho,u_\alpha,p,B_\alpha)$ and $\alpha$ denotes directional components while hat variables denotes non-ideal MHD components like $(E_\alpha,J_\alpha,P_e)$. $\mathcal{S}_i$ are source terms that can be explicitly update while $\hat{\mathcal{R}}_i$ are stiff source terms due to $\epsilon_i$ that cannot be treated similarly to the former. As $\epsilon_i \rightarrow 0$, Eq. \ref{eq:AP_NI} reduces to the inertia-less GOL in Eqs. \ref{XMHDv1:ampere_ND_relax} and \ref{XMHDv1:current_ND_relax}.

However, when $\epsilon_i$ is small but not zero, approximating Eq. \ref{eq:AP_ideal} is challenging due to the presence of stiff relaxation term in Eq. \ref{eq:AP_NI}. Treating the full system explicitly requires very small time step (e.g., $\Delta t\sim\mathcal{O}(\epsilon_i)$) which is computationally costly. Therefore, implicit update is applied to the stiff source term as described earlier by steps 14 to 16 in Algorithm 1. The implicit update through the first-order time-splitting framework is similar to Jin–Xin relaxation model \cite{HU2017103}, except that the convection and relaxation steps are reversed here, since the solution can be initialized on the equilibrium manifold satisfying $\hat{\mathcal{R}}_i(\mathcal{U}_i^{0},\hat{\mathcal{U}}_i^{0})\approx0$. Without including spatial and source discretization, Eqs. \ref{eq:AP_ideal} and \ref{eq:AP_NI} are solved as follows: 
\begin{equation} \label{eq:AP_conv}
\begin{cases}
\dfrac{\mathcal{U}^{n+\frac{1}{2}}_i - \mathcal{U}^n_i}{\Delta t} + \partial_x \mathcal{F}(\mathcal{U}_i^n) = \mathcal{S}_i(\mathcal{U}_i^n,\hat{\mathcal{U}}^n_i), \\[1em]
\dfrac{\hat{\mathcal{U}}^{n+\frac{1}{2}}_i - \hat{\mathcal{U}}^n_i}{\Delta t} +\partial_x\hat{\mathcal{F}}_i(\mathcal{U}_i^n,\hat{\mathcal{U}}_i^n) 
= \hat{\mathcal{S}}_i(\mathcal{U}_i^n,\hat{\mathcal{U}}_i^n),
\end{cases}
\qquad \text{(convection step)}
\end{equation}

\begin{equation} \label{eq:AP_relax}
\begin{cases}
\dfrac{\mathcal{U}^{n+1}_i - \mathcal{U}^{n+\frac{1}{2}}_i}{\Delta t} = 0, \\[1em]
\dfrac{\hat{\mathcal{U}}^{n+1}_i - \hat{\mathcal{U}}^{n+\frac{1}{2}}_i}{\Delta t}  
= \dfrac{1}{\epsilon_i}\hat{\mathcal{R}}_i(\mathcal{U}^{n+1}_i,\hat{\mathcal{U}}_i^{n+1}),
\end{cases}
\qquad \text{(relaxation step)}
\end{equation}
Eq. \ref{eq:AP_relax} may seem non-linear at first glance. However, from the first equation in Eq. \ref{eq:AP_relax}, we have $\mathcal{U}^{n+1}_i = \mathcal{U}^{n+\frac{1}{2}}_i$, so the nonlinear term $\hat{\mathcal{R}}_i(\mathcal{U}^{n+\frac{1}{2}}_i,\hat{\mathcal{U}}_i^{n+1})$ becomes linear and, hence can be solved implicitly without requiring any Newton-type iteration. Therefore, although the relaxation step is implicit, the whole scheme can be implemented without Newton-type iteration, which is an important feature of AP schemes. For the spatial derivative and source term in Eq. \ref{eq:AP_conv}, one can apply the usual finite-difference or finite-volume schemes. Eqs. \ref{eq:AP_conv} and \ref{eq:AP_relax} can then be implemented at every substage of multistage Runge-Kutta method for higher order of temporal accuracy.

The AP property for Algorithm 1 depends on Eqs. \ref{XMHD_implicitE}-\ref{XMHD_implicitJ} and this is investigated here by keeping $\Delta t$ and $\Delta x$ fixed while sending the relaxation parameters ($\epsilon_E=\frac{v^2}{c^2}$, $\epsilon_J=\frac{\lambda_e^2}{L_0^2 n_e}$) to zero. This happens at the ideal MHD limit where $\underline{\eta^*}\rightarrow0$ while $\underline{n^*}$ increases such that the Hall term is negligible. This reduces Eq. \ref{XMHD_implicitJ} to
\begin{equation}
    \mathbf{J}^{*} = \mathbf{J}^{**} + \Delta t \frac{L^2_0 \underline{n^*}}{\lambda_e^2}
\Bigl(
\mathbf{E}^{*} +\mathbf{u}^*\times \mathbf{B}^* 
\Bigr),
\end{equation}
Then sub in Eq. \ref{XMHD_implicitE},
\begin{equation}
    \mathbf{J}^{*} = \mathbf{J}^{**} + \Delta t \frac{L^2_0 \underline{n^*}}{\lambda_e^2}
\left(
\mathbf{E}^{**}-\Delta t \frac{c^2}{v^2}\mathbf{J}^* +\mathbf{u}^*\times \mathbf{B}^* 
\right),
\end{equation}
where $\mathbf{E}^{**}=\mathbf{E}^n+\Delta t\frac{c^2}{v^2}\nabla \times \mathbf{B}^*$ and note that the curl of face-centered $\mathbf{B}$ naturally maps to the edge. Rearranging the above equation such that it is equivalent to Eq. \ref{linear_solve} except without the Hall and resistive term,
\begin{equation}
    \mathbf{J}^*\left(1+\Delta t^2 \frac{c^2}{v^2}\frac{L_0 \underline{n^*}}{\lambda_e^2} \right)=\mathbf{J}^{**}+\Delta t \frac{L_0^2 \underline{n^*}}{\lambda_e^2}\left(\mathbf{E}^n+\Delta t\frac{c^2}{v^2}\nabla \times \mathbf{B}^*+\mathbf{u}^*\times \mathbf{B}^*\right)
\end{equation}
As $\epsilon_E<<\Delta t$ and $\epsilon_J<<\Delta t$, above equation reduces to
\begin{equation}
    \mathbf{J}^*\left(\Delta t^2 \frac{c^2}{v^2}\frac{L_0 \underline{n^*}}{\lambda_e^2} \right)=\mathbf{J}^{**}+\Delta t \frac{L_0^2 \underline{n^*}}{\lambda_e^2}\left(\mathbf{E}^n+\Delta t\frac{c^2}{v^2}\nabla \times \mathbf{B}^*+\mathbf{u}^*\times \mathbf{B}^*\right)
\end{equation}
Then divide both sides by $\Delta t/\epsilon_J$ and as $\epsilon_J<<\Delta t$, the equation simplifies to
\begin{equation}
    \mathbf{J}^*\left(\Delta t \frac{c^2}{v^2} \right)=\mathbf{E}^n+\Delta t\frac{c^2}{v^2}\nabla \times \mathbf{B}^*+\mathbf{u}^*\times \mathbf{B}^*
\end{equation}
Then divide both sides by $\Delta t/\epsilon_E$ and also sub into Eq. \ref{XMHD_implicitE} give
\begin{align} 
    \mathbf{J}^*&=\frac{v^2/c^2}{\Delta t}(\mathbf{E}^n+\mathbf{u}^*\times \mathbf{B}^*)+\nabla \times \mathbf{B}^* \label{eq:J*_int}\\
    \mathbf{E}^*&=\mathbf{E}^n+\frac{\Delta t}{v^2/c^2}\nabla \times \mathbf{B}^*-(\mathbf{E}^n+\mathbf{u}^*\times \mathbf{B}^*)- \frac{\Delta t}{v^2/c^2}\nabla \times\mathbf{B}^* \nonumber\\
    &=-\mathbf{u}^*\times \mathbf{B}^* \label{eq:ideal_E_discrete}
\end{align}
As $\epsilon_E\rightarrow0$, Eq. \ref{eq:J*_int} goes to
\begin{equation} \label{eq:ideal_J_discrete}
    \mathbf{J}^*=\nabla\times \mathbf{B}^*
\end{equation}
$\mathbf{E}^*$ and $\mathbf{J}^*$ from Eqs. \ref{eq:ideal_E_discrete} and \ref{eq:ideal_J_discrete} will then be used to compute the non-ideal source terms (e.g., $\mathbf{E}^*+\mathbf{u}^*\times \mathbf{B}^*$, $\mathbf{J}^*-\nabla\times \mathbf{B}^*$) for Eqs. \ref{XMHDv2:ND_mass} to \ref{XMHDv2:ND_faraday} in the next substage. This will recover the ideal MHD solver since the non-ideal source terms are zero as $\epsilon_E,\epsilon_J \rightarrow 0$. Note that if electric field flux update is not split from the rest (e.g., Step 11 from Algorithm 1), Eq. \ref{eq:ideal_J_discrete} will give $\mathbf{J}^*=\nabla \times \mathbf{B}^n$ instead which may create a non-zero $(\mathbf{J}^*-\nabla\times \mathbf{B}^*)$ source term in the next substage. Additionally, there is also a stiff source term for $S_e$ through the electron-ion energy exchange term but as mentioned earlier, it only affects the partition of electron and ion pressure but not total pressure which is evolved through the energy equation, similar to ideal MHD solver.

\subsection{Slope limiter}
The source term can induce fast dynamics, such as high frequency Whistler wave oscillations caused by Hall effect. Since the Hall effect is larger at low density (as will be shown later in Eq. \ref{eq:Whistler_wave_&_Hall_drift} where the whistler wave scales as $\frac{1}{n_e}$), these rapid variations may not be accounted for when computing the timestep for each update. To prevent the accumulation of numerical artifacts that could eventually contaminate the solution, such oscillations should be damped out. This is done by introducing an additional density-dependent slope limiter together with a total variation diminishing (TVD) slope limiter, which is used to eliminate spurious oscillations from each flux update. The resulting cell-interface primitive variables reconstruction is
\begin{equation}
    \mathbf{U}^L_{i+\frac{1}{2}}=\mathbf{U}_i+\frac{1}{2}\psi(n_i)\phi(\mathbf{r}_i)\Delta \mathbf{U}_i, \ \ \ 
    \mathbf{U}^R_{i+\frac{1}{2}}=\mathbf{U}_{i+1}-\frac{1}{2}\psi(n_{i})\phi(\mathbf{r}_{i})\Delta \mathbf{U}_{i+1},
\end{equation}
where $\phi(\mathbf{r}_i)$ is the TVD slope limiter like Van Leer (VL) \cite{VANLEER1977263} which depends on the ratio of the upstream to downstream slope. $\psi(n_i)$ is an additional density-dependent slope limiter which depends on the local density $n_i$. This slope limiter must be equal to 1 at continuum limit and approach 0 as the density decreases towards the kinetics limit. The rationale for this is to increase the flux diffusivity to damp out small but fast oscillating waves that would otherwise contaminate the solution if not resolved. This effect becomes more significant as the flow density decreases. A simplified model for this is a sigmoid function which restrict the gradient of density away from continuum limit as follows,
\begin{equation} \label{eq: density-dependent-SL}
    \psi(n_i)=\left(1 + \exp\left(\frac{\left(\frac{\lambda_0}{\Delta x}\right)^3-n_i}{L_0}\right)\right)^{-1},
\end{equation}
where $\left(\frac{\lambda_0}{\Delta x} \right)^3$ is a free parameter and from the shock tube problem, 0.005 is a suitable value. This can be seen akin to a local Knudsen number of around 0.17 which lies in the transition regime away from the continuum limit. Therefore, as the flow density decreases and approaches the molecular limit, the above limiter approaches zero, giving rise to a 1st-order diffusive flux to stabilize the solution by damping out small amplitude but high frequency oscillations. Note that this density-dependent slope limiter is only applicable to problems that contain both high- and low-density regions. For problems in which all densities lie within the low-density regime (e.g., Sect. \ref{subsect:Huba}), the limiter reduces to 1 after normalization by the reference low density $n_0$. Future work could be focused on a more rigorous density-dependent slope limiter without any free parameter.

\subsection{Primitive form for flux Jacobian}
We can compute the signal speeds of the XMHD model by computing the eigenvalues of the flux Jacobian.  Consider the Jacobian of the conserved system without the source term from Eqs. \ref{XMHDv1:ampere_ND} - \ref{XMHDv1:entropy_ND},
\begin{equation}
    \frac{\partial \mathbf{U}}{\partial t} + \mathbb{A}(\mathbf{U})\frac{\partial \mathbf{U}}{\partial x}=\mathbf{0}, \ \ \ \mathbb{A}(\mathbf{U})=\frac{\partial \mathbf{F}}{\partial \mathbf{U}}
\end{equation}
which can be written in primitive form 
\begin{equation}
    \frac{\partial \mathbf{V}}{\partial t} + \mathbb{A}_p(\mathbf{V})\frac{\partial \mathbf{V}}{\partial x}=\mathbf{0},
    \end{equation}
where the primitive variables for 1D XMHD are $\mathbf{V}=[\rho,u,v,w,B_y,B_z,p,E_x,E_y,E_z,J_x,J_y,J_z,P_e]^T$ and the flux Jacobian is
\[
\mathbb{A}_p = \left[
\begin{array}{*{15}{c}}  % 14 columns
u & \rho & 0 & 0 & 0 & 0 & 0 & 0 & 0  & 0 & 0 & 0 & 0 & 0 \\
0 & u & 0 & 0 & \frac{B_y}{\rho} & \frac{B_z}{\rho} & \frac{1}{\rho} & 0 & 0  & 0 & 0 & 0 & 0 & 0 \\
0 & 0 & u & 0 & \frac{B_x}{\rho} & 0 & 0 & 0 & 0 & 0 & 0 & 0 & 0 & 0 \\
0 & 0 & 0 & u & 0 & \frac{B_x}{\rho} & 0 & 0 & 0 & 0 & 0 & 0 & 0 & 0 \\
0 & B_y & -B_x & 0 & u & 0 & 0 & 0 & 0  & 0 & 0 & 0 & 0 & 0 \\
0 & B_z & 0 & -B_x & 0 & u & 0 & 0 & 0  & 0 & 0 & 0 & 0 & 0 \\
0 & \gamma p & 0 & 0 & -B_x & 0 & u & 0 & 0  & 0 & 0 & 0 & 0 & 0 \\
0 & 0 & 0 & 0 & 0 & 0 & 0 & 0 & 0  & 0 & 0 & 0 & 0 & 0 \\
0 & 0 & 0 & 0 & 0 & \frac{c^2}{v^2} & 0 & 0 & 0 & 0  & 0 & 0 & 0 & 0 \\
0 & 0 & 0 & 0 & -\frac{c^2}{v^2} & 0 & 0 & 0 & 0 & 0  & 0 & 0 & 0 & 0 \\
0 & 2J_x & 0 & 0 & 0 & 0 & 0 & 0 & 0 & 0  & 2u-2\frac{\lambda_i J_x}{L_0 \rho} & 0 & 0 & \frac{-m_i L_0}{m_e \lambda_i} \\
0 & J_y & J_x & 0 & 0 & 0 & 0 & 0 & 0 & 0  & v-\frac{\lambda_i J_y}{L_0 \rho} & u-\frac{\lambda_i J_x}{L_0 \rho} & 0 & 0 \\
0 & J_z & 0 & J_x & 0 & 0 & 0 & 0 & 0 & 0  & w-\frac{\lambda_i J_z}{L_0\rho} & 0 & u-\frac{\lambda_i J_x}{L_0 \rho} & 0 \\
C_1 & C_2 & 0 & 0 & 0 & 0 & 0 & 0 & 0 & 0  & -\frac{J_0}{en_0 v}\frac{P_e}{Z\rho} & 0 & 0 & u-\frac{J_0}{en_0 v}\frac{J_x}{Z\rho} \\
\end{array}
\right]
\]
where 
\begin{equation}
    C_1 = (\gamma-1)P_e \rho u + (1-\gamma)P_e\rho \left(u-\frac{J_0}{en_0 v}\frac{J_x}{Z\rho}\right) + P_e \frac{J_0}{en_0 v} \frac{J_x}{Z\rho^2}
\end{equation}
\begin{equation}
    C_2 = (\gamma-1)P_e \rho^2 + P_e.
\end{equation}
The Jacobian can then be written as four sub-blocks,
\begin{equation}
    \mathbb{A}_p=\begin{bmatrix}
        \underline{A} & \underline{0} \\
        \underline{B} & \underline{C}
    \end{bmatrix}.
\end{equation}
This allows one to obtain the eigenvalues of the Jacobian much more easily since it is a block lower-triangular matrix.
\begin{equation} \label{eq:split_Jacobian}
    det(\lambda \mathbb{I}-\mathbb{A}_p)=det(\lambda \mathbb{I}-\underline{A})det(\lambda \mathbb{I}-\underline{C}).
\end{equation}
The  eigenvalues corresponding to  ideal MHD are  preserved since they are decoupled from the additional equations from XMHD. The characteristic speeds for the flux Jacobian are
\begin{align}
\lambda_1 &= u - c_f, & \lambda_2 &= u - c_a, & \lambda_3 &= u - c_s, &
\lambda_4 &= u, \notag
\\ 
\lambda_5 &= u + c_s, & \lambda_6 &= u + c_a, & \lambda_7 &= u + c_f, & \lambda_{11} &= u - \frac{\lambda_i J_x}{L_0 \rho}, \label{eq:eigenvalues}
\\
\lambda_{12} &= u - \frac{\lambda_i J_x}{L_0 \rho}, & \lambda_{13} &= u + c_m, & \lambda_{14} &= u + c_p,
\notag 
\end{align}
where $c_f$ and $c_s$ are the fast and slow magnetosonic speeds, respectively:  
\begin{align}
c_{f}^2 &= \frac{1}{2} \Big( a^{*2} + \sqrt{a^{*4} - 4 a^2 b_x^2} \Big), \\
c_{s}^2 &= \frac{1}{2} \Big( a^{*2} - \sqrt{a^{*4} - 4 a^2 b_x^2} \Big),
\label{eq:cfcs}
\end{align}
with
\begin{align}
a^2 = \frac{\gamma p}{\rho}, \ \  \mathbf{b} = \frac{1}{\sqrt{\rho}}(B_x, B_y, B_z),  \ \  a^{*2} = a^2 + |\mathbf{b}|^2,
\label{eq:a_defs}
\end{align}
and $c_a$ is the Alfvén speed,
\begin{align}
c_a = \frac{B_x}{\sqrt{\rho}}.
\label{eq:ca}
\end{align}
There are three new wavespeeds ($\lambda_{11}=\lambda_{12}$,$\lambda_{13}$,$\lambda_{14}$) where
\begin{equation}
    c_m=\frac{3}{2}u-\frac{\lambda_i J_x}{L_0 \rho}-\frac{J_0}{2en_0 v}\frac{J_x}{Z\rho}-\sqrt{\left(\frac{J_0 J_x}{2en_0 v Z \rho} -\frac{\lambda_i J_x}{L_0 \rho}\right)^2 +\frac{J_0 P_e m_i L_0}{e n_0 v Z \rho m_e \lambda_i}}
\end{equation}
\begin{equation}
    c_p=\frac{3}{2}u-\frac{\lambda_i J_x}{L_0 \rho}-\frac{J_0}{2en_0 v}\frac{J_x}{Z\rho} +\sqrt{\left(\frac{J_0 J_x}{2en_0 v Z \rho} -\frac{\lambda_i J_x}{L_0 \rho}\right)^2 +\frac{J_0 P_e m_i L_0}{e n_0 v Z \rho m_e \lambda_i}}.
\end{equation}
Note that $v$ here is the constant used to non-dimensionalize velocity. Additionally, the characteristic speed does not depend on the speed of light which means that the timestep is not overly restricted. However, the source term may introduce signal speeds much faster than the ones above, especially at near vacuum where the signal speed will approach the speed of light. A safer approach will be to use the electron sound speed which is also used in this code as was generally found to be larger than the above three new wavespeeds in our experiments. The wavespeeds based on the electron sound speeds are 
\begin{equation} \label{eq:elec_sound_speed}
    \lambda_{es}=u \pm \sqrt{\frac{m_i}{m_e}}a.
\end{equation}

Note that the explicit source terms depend on $\mathbf{E}$ and $\mathbf{J}$ which are implicitly updated by the diffusive backward Euler to relax the stiff source term to the equilibrium value where the characteristic speeds of the flux Jacobian become more relevant. Moreover, the relaxation accounts for the speed of light according to Eqs. \ref{eq:taylor_expEJ} which shows that larger $\Delta t$ results in faster relaxation since more higher-order terms are damped out. However, these are not sufficient and thus, density-dependent slope limiters are required for the shock tube problem to further damp out small and fast oscillations. More rigorous analysis and testing will be needed in future.

\subsection{Relaxation of stiff source term} \label{subsec:relax_stiff_src}
The extended MHD can be seen as deviation from the equilibrium values of $\mathbf{E}_0=-\mathbf{u}\times\mathbf{B}$ and $\mathbf{J}_0=\nabla \times\mathbf{B}$ derived from ideal MHD. To show relaxation and neglecting Hall term (e.g., sufficiently high $n_e$ for it to be negligible) here to simplify analysis, the implicit source term update can be written as the following discretized linear differential equations
\begin{align}
    \mathbf{J}^{n+1} &=\mathbf{J}^*+\frac{\Delta t}{\epsilon_J}(\mathbf{E}^{n+1}-\mathbf{E}_0-\eta \mathbf{J}^{n+1}) \\
    &=\mathbf{J}^n + \Delta t \nabla \cdot \mathbb{J} +\frac{\Delta t}{\epsilon_J}(\mathbf{E}^{n+1}-\mathbf{E}_0-\eta \mathbf{J}^{n+1}) \\
    \mathbf{E}^{n+1} &= \mathbf{E}^* + \frac{\Delta t}{\epsilon_E} (\mathbf{J}_0 - \mathbf{J}^{n+1}) 
\end{align}
where $\eta$ comes from Ohmic resistivity ($\eta_{Ohmic}$), $\mathbf{E}^*=\mathbf{E}^n$ and $\mathbb{J}$ is the convective flux from Eq. \ref{XMHDv1:current_ND_no_source}. If $\Delta t<<\epsilon_E$ and $\Delta t << \epsilon_J$, $\mathbf{J}^*$ and $\mathbf{E}^*$ will be advanced in time by backward Euler equation. If $\Delta t>>\epsilon_E$ and $\Delta t >> \epsilon_J$, the above equations will reduce to 
\begin{align}
    \mathbf{J}^{n+1} =\mathbf{J}_0 \ \ ; \ \ 
    \mathbf{E}^{n+1} = \mathbf{E}_0 + \eta \mathbf{J}_0.
\end{align}
where $\mathbf{J}^{n+1}$ and $ \mathbf{E}^{n+1}$ are independent of electron inertia and pressure effects. For ideal MHD, $\eta$ is very small and this will render the non-ideal and relativistic source terms to zero, recovering the ideal MHD solver.

Solving for $\mathbf{J}^{n+1}$ and $\mathbf{E}^{n+1}$ using the above equations and expanding the solution to second order in $\frac{\epsilon_J}{\Delta t}$ and $\frac{\epsilon_E}{\Delta t}$ yields corrections to the equilibrium states $\mathbf{E}_0$ and $\mathbf{J}_0$. These deviations are of order $\frac{\epsilon_J}{\Delta t}$ or $\frac{\epsilon_E}{\Delta t}$ and arise from non-ideal effects. 
\begin{align} \label{eq:taylor_expEJ}
    \mathbf{J}^{n+1} &\approx \left(1-\frac{\eta \epsilon_E}{\Delta t} +\frac{\epsilon_E (\eta^2\epsilon_E-\epsilon_J)}{\Delta t^2}\right)\mathbf{J}_0 + \frac{\epsilon_E}{\Delta t}\left(\frac{\eta \epsilon_E}{\Delta t}-1\right)(\mathbf{E}_0-\mathbf{E}^*)+\frac{\epsilon_J \epsilon_E}{\Delta t^2}\mathbf{J}^* \\
    \mathbf{E}^{n+1} & \approx \mathbf{E}_0+\eta\mathbf{J}_0+\frac{\epsilon_J}{\Delta t} \left(1-\frac{\eta \epsilon_E}{\Delta t} \right)(\mathbf{J}_0-\mathbf{J}^*) +  \frac{\epsilon_E}{\Delta t}\left(\frac{\eta^2 \epsilon_E}{\Delta t}-\eta- \frac{\epsilon_J}{\Delta t}\right)(\mathbf{E}_0-\mathbf{E}^*) \\
    & + \mathbf{J}_0 \left(-\frac{\eta^2\epsilon_E}{\Delta t}+\frac{\eta^3 \epsilon_E^2}{\Delta t^2}-\frac{\epsilon_e\epsilon_J \eta}{\Delta t^2} +\frac{\epsilon_J^2}{\eta \Delta t^2}\right) \nonumber
\end{align}
The validity of the quasi-neutral assumption can be roughly determined by $\nabla\cdot \mathbf{J}^{n+1}$ from Eq. \ref{eq:taylor_expEJ} which gives
\begin{equation} \label{eq:divJ^n+1}
    \nabla \cdot \mathbf{J}^{n+1}\approx \frac{\epsilon_E}{\Delta t}\left(\frac{\eta \epsilon_E}{\Delta t}-1\right)\nabla \cdot(\mathbf{E}_0-\mathbf{E}^*)+\frac{\epsilon_J \epsilon_E}{\Delta t^2}\nabla \cdot\mathbf{J}^*
\end{equation}
In the ideal MHD limit, $\nabla\cdot \mathbf{J}^{n+1}$ will go to zero as $\epsilon_E<<\Delta t$ and $\epsilon_J<<\Delta t$. This is consistent with the ideal MHD limit where $\mathbf{J}=\nabla \times\mathbf{B}$ enforces quasi-neutrality since $\nabla \cdot (\nabla \times \mathbf{B})=0$. At initial timestep of $n=0$, when electric field and current density are initialized with the ideal MHD condition such that $\mathbf{E}^*=\mathbf{E}^0=\mathbf{E}_0$ and $\mathbf{J}^0=\nabla \times \mathbf{B}$, Eq. \ref{eq:divJ^n+1} then reduces to
\begin{equation} \label{eq:divJ^1}
    \nabla \cdot \mathbf{J}^{1}\approx \frac{\epsilon_J \epsilon_E}{\Delta t}\nabla \cdot(\nabla \cdot \mathbb{J})
\end{equation}
In the Resistive MHD limit, the magnitude of $\nabla\cdot \mathbf{J}^{n+1}$ is proportional to $\frac{\epsilon_E\epsilon_J}{\Delta t}$ initially and this contribution may be small if the source terms are stiff. Additionally, if electron inertia is under-resolved, $\nabla \cdot (\nabla\cdot \mathbb{J})$ is expected to be small and may be neglected at leading order. However, at low density, the magnitude of $\nabla\cdot \mathbf{J}^{n+1}$ may increase further due to the Hall term. This leads to a larger effective $\eta = \eta_{Ohmic} + \eta_{Hall}$, where $\eta_{Hall} \sim \frac{\lambda_i}{L_0 n_e} |\mathbf{B}|$ in Eq. \ref{eq:divJ^n+1}, indicating higher sensitivity to non-neutral effects.

\subsection{Temporal order of accuracy}
The temporal update of the current algorithm uses operator splitting of explicit and implicit components at each substage. The explicit second-order Runge–Kutta (RK2) method achieves second-order accuracy for the unsplit explicit flux and source term updates but only first-order accuracy for the implicit source term update. As a result, the overall temporal accuracy remains first order. A formal proof of the temporal accuracy is shown below and a slightly more general scheme (e.g., introducing $\alpha$ and $\gamma$) is used to avoid an over-determined system of equations.
\begin{align} 
    U_1 &= U^n - \alpha \frac{\Delta t}{\epsilon} R(U_1) \label{eq: LA1} \\ 
    U_2 &= U_1 - \tilde{\alpha}\Delta t DF(U_1) \label{eq: LA3} \\
    U_3 &= U_2 - \beta \frac{\Delta t}{\epsilon}R(U_3) -\gamma \frac{\Delta t}{\epsilon} R(U_1) \label{eq: LA4}\\
    U_4 &= \xi U^n + \eta U_3 \label{eq: LA7}\\
    U_5 &= U_4 - \tilde{\beta}\Delta t DF(U_4) \label{eq: LA6}\\
    U^{n+1} &= U_5 - \mu \frac{\Delta t}{\epsilon}R(U^{n+1}), \label{eq: LA8}
\end{align}
where $R(U)$ and $DF(U)$ are the implicit and explicit update, respectively. The explicit source and flux are left unsplit to simplify the problem of finding a solution to this linear analysis. Note that although $\mathbf{E}$-flux update is split from the other variables flux and explicit source term update, the $\mathbf{E}$-flux can be seen as inputs (e.g., components of $U_3$, $U^{n+1}$) to the implicit source term update in Eq. \ref{eq: LA4} and \ref{eq: LA8}. The advantage of operator splitting of implicit and explicit parts at each substage is that it ensures that the stiff source term relaxes to $R(U)\approx0$ which in the ideal MHD limit will preserve the specific form of the Ohm's law, $\mathbf{E}\approx -\mathbf{u}\times \mathbf{B}$. This will then be used to update the flux while the explicit non-ideal source term will go to zero. The above steps should ideally be combined such that they are 2nd-order accurate in the linear regime.

Applying Eqs. \ref{eq: LA1} - \ref{eq: LA8} to a linear system gives \cite{Broadwell}
\begin{equation} \label{eq: linear_sys}
    \partial_t U+AU+BU=0,
\end{equation}
where A and B are constant matrices but A is treated implicitly. This gives the following exact solution at time $t=\Delta t$,
\begin{equation}
    U(\Delta t) = e^{-(A+B)\Delta t} U(0).
\end{equation}
For 2nd-order of accuracy, the following condition must be imposed
\begin{equation}
    (\mathcal{C}(\Delta t)-e^{-(A+B)\Delta t})U(0)=\mathcal{O}(\Delta t^3),
\end{equation}
where $\mathcal{C}$ is the numerical scheme from Eqs. \ref{eq: LA1} - \ref{eq: LA8}. This gives the following simultaneous equations that must be solved,
\begin{align}
    \xi+\eta &= 1, \ \ \tilde{\alpha}\eta + \tilde{\beta}\eta+\tilde{\beta}\xi=1, \ \ \eta(\alpha+\gamma+\beta)+\mu(\xi + \eta)=1  \label{eq:1st_order}\\
    2\eta \tilde{\alpha} \tilde{\beta} &= 1, \ \  2\mu(\tilde{\alpha}\eta+\tilde{\beta}\eta+\tilde{\beta}\xi) + 2\tilde{\alpha}\beta\eta=1, \ \ 2\eta(\tilde{\alpha}\alpha + \tilde{\beta}\alpha + \tilde{\beta}\gamma+\tilde{\beta}\beta)=1, 
    \label{eq:2nd_order_a}\\
    \eta (\alpha^2 &+ \alpha \gamma +\beta \alpha +\beta \gamma+\beta^2) + \eta \mu (\alpha +\gamma+\beta) + \mu^2 (\xi +\eta)=\frac{1}{2}. \label{eq:2nd_order_b}
\end{align}
This gives an under-determined system with 7 equations with 8 unknowns to be solved. Satisfying Eq. \ref{eq:1st_order} gives 1st-order accuracy while satisfying Eqs. \ref{eq:1st_order}-\ref{eq:2nd_order_b} gives 2nd-order accuracy. The explicit RK2 method with operator splitting of implicit substage update has the following coefficients  $\left(\xi,\eta,\alpha,\mu,\beta,\gamma,\tilde{\alpha},\tilde{\beta} \right) =$ $\left(\frac{1}{2},\frac{1}{2},0,\frac{1}{2},1,0,1,\frac{1}{2}\right)$ which only satisfies Eq. \ref{eq:1st_order}. Therefore, when non-ideal effects become more significant, the solution approaches 1st-order of accuracy temporally.

\subsection{Directional splitting of implicit source term update} \label{Subsect:Direc_split_Hall}
The Hall term contains the cross product ($\mathbf{J}\times\mathbf{B}$) which must be treated carefully since the x, y and z components lie on different edges orthogonal to each other. One simply cannot just take the cross-product of these spatially-separated edge components which will lead to some spatial biasing that may lead to bad results especially when the non-ideal effects dominate. Instead, for n-dimensional spatial problem, the implicit update is done n times which is described as follows for 3D problem,

\noindent\rule{\linewidth}{1pt}          
\captionof{algorithm}{Directional splitting of implicit source term update}         
\noindent\rule{\linewidth}{1pt} 
{\small
\begin{algorithmic}[1] 
\State Interpolate density and electron pressure (required to compute Spitzer resistivity) from cell-center to the edge along x. 
\State Interpolate face-centered $\mathbf{B}$ to the edge.
\State Interpolate $\mathbf{J}_\perp$ and $\mathbf{E}_\perp$ to the edge along x.
\State Interpolate $\mathbf{u}\times\mathbf{B}|_{\perp}$ obtained from the ideal MHD Riemann solver through constrained transport method to the edge along x.
\State Implicitly update $J_x$, $E_x$ using Eq. \ref{linear_solve}. This will also produce new $\mathbf{J}_\perp$ and $\mathbf{E}_\perp$ which lie along x-edge that are discarded.
\State Steps 1 to 5 are then repeated for edge along y and z using old $\mathbf{E}$ and $\mathbf{J}$ values before their implicit update.
\end{algorithmic}
}
\noindent\rule{\linewidth}{1pt}
Interpolation across different topological elements (TE) will be done through simple aver-
aging except for interpolating from cell center to node which will be done by choosing the
cell center with the largest density value. Likewise, other cell-center variables like $P_e$ can be interpolated to the cell-edge by choosing the cell-edge's four neighbouring cell-centers with the largest density value. 

\begin{figure}[http]
    \centering
    \includegraphics[width=0.8\linewidth]{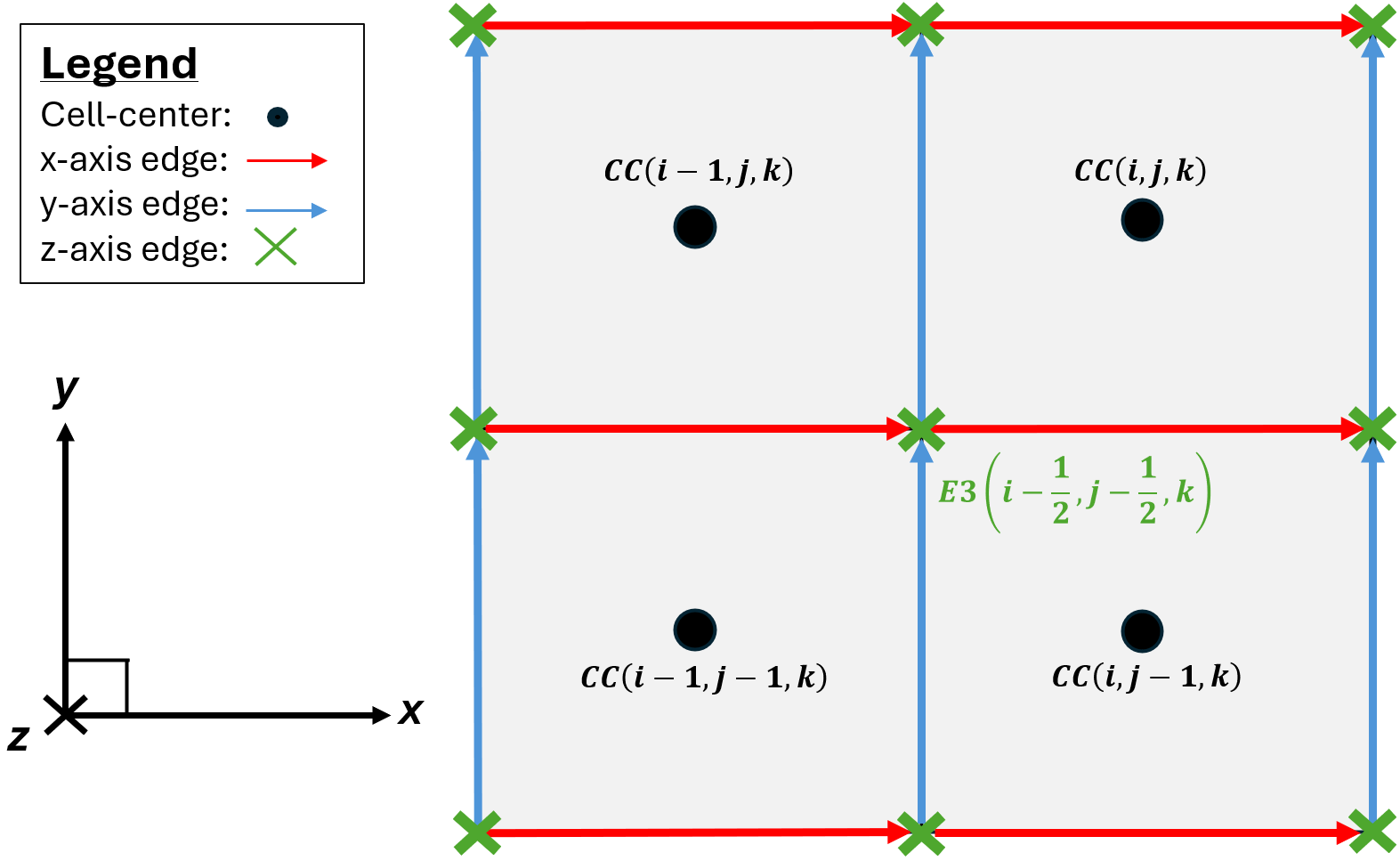}    
    \caption{2D diagram illustrating the 4 cell centers with a shared edge along z at the center. \label{fig:mesh_interp_CC_2_Edge}}
\end{figure}

The rationale for this can be seen as "upwinding" using Fig. \ref{fig:mesh_interp_CC_2_Edge} for visualization. For instance, at cell edge $E3(i-\frac{1}{2},j-\frac{1}{2},k)$, the fluid can have 4 different Hall velocity which is given by $V_{H,z}=-J_z/n_ee$ where $n_e$ can be obtained from any of the closest 4 neighbouring cell-centers with shared $J_z$. The smaller the $n_e$, the higher the magnitude of $V_{H,z}$ and thus, it is more likely to be 'advected' away. Therefore, the smallest Hall velocity which corresponds to the quadrant with the largest density, as well as other cell-center variables like $P_e$ (required for computing Spitzer resistivity), will be used for the non-ideal MHD terms at that cell edge through the GOL. Since the Hall term is proportional to the reciprocal of the density, using the maximum density instead of averaged density will decrease the Hall term and therefore, helps improve numerical stability.

Directional splitting of the implicit update means that the implicit source term update is no longer fully implicit which may affect the stability of the solver. Moreover, some values are discarded and the implicit update does not guarantee that $\nabla\cdot \mathbf{J}$ remains small at the cell node for the quasi-neutrality assumption to hold. Future work will focus on directionally unsplit implicit source-term update, which may be important for preserving, or at least limiting the growth of $\nabla \cdot\mathbf{J}$ at those locations.

\subsection{Interpolation of variables across different topological elements}
Interpolation is required as the variables are all not co-located. For instance,
\begin{itemize}
    \item $\mathbf{B}$ field lies on the cell faces, for example F1, F2, F3 from Fig. \ref{fig:mesh_TE}.
    \item $\mathbf{E}$, $\mathbf{J}$ and $-\mathbf{u}\times\mathbf{B}$ (obtained from ideal MHD Riemann solver through constrained transport method) fields lie on the cell edges, for example E1, E2, E3 from Fig. \ref{fig:mesh_TE}.
    \item The flux for $\mathbf{J}$ lies on cell nodes, for example NN from Fig. \ref{fig:mesh_TE}. This is obtained by interpolating the cell's interface flux obtained from the LLF Riemann solver to the cell nodes. 
\end{itemize}
The rest of the variables lies at the cell center and their corresponding fluxes lie at the cell face. Here, simple arithmetic averaging is generally sufficient which is also applied for the constrained transport method implemented here to ensure divergence-free magnetic field.

\subsection{Explicit update of edge-centered variables} \label{subsec:Jupdate}
The curl of face-centered magnetic field naturally maps to the edge which can then be directly used to update $\mathbf{E}$ as follows,
\begin{align}
\frac{\partial \mathbf{E}}{\partial t}
- \frac{c^{2}}{v^{2}}\bigl(\nabla\times\mathbf{B}\bigr) = 0
\label{XMHDv1:ampere_ND_no_source}
\end{align}

\begin{figure} [http]
    \centering
    \includegraphics[width=0.8\linewidth]{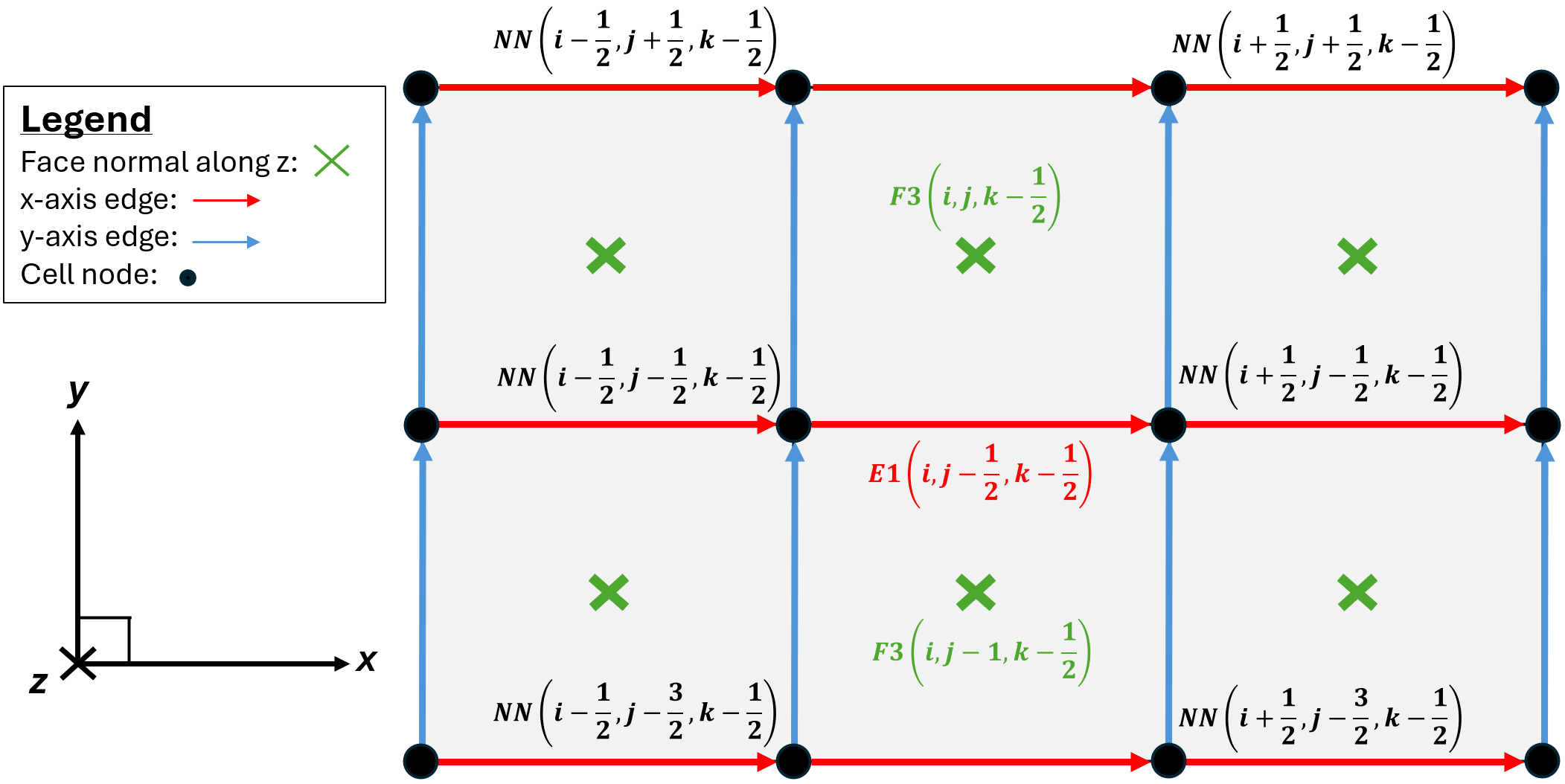}
    \caption{2D diagram illustrating the faces $F3$ of six computational cells, with their normals directed into the paper along the z-axis. In this configuration, the faces $F3(i,j,k-\frac{1}{2})$ and $F3(i,j-1,k-\frac{1}{2})$ correspond to cell centers at $(i,j,k)$ and $(i,j-1,k)$, respectively. Each computational cell is bounded by edges that define its geometry: the horizontal arrows represent the edges $E1$, while the vertical arrows denote the edges $E2$. The edges $E3$, which align with the z-axis, are positioned at the cell nodes and point into the paper. The intersections of these edges are depicted as solid black dots to indicate cell nodes $NN$.}
    \label{fig:mesh_Jflux}
\end{figure}

However, the update of edge-centered $\mathbf{J}$ is less straightforward as the divergence of the flux in Eq. \ref{XMHDv1:current_ND_no_source} indicates that the flux must be placed at the cell nodes.
\begin{align}
\frac{\partial \mathbf{J}}{\partial t}
+\nabla\cdot\Bigl(
   \mathbf{u}\mathbf{J}
 + \mathbf{J}\mathbf{u}
 - \frac{\lambda_{i}}{L_{0}n}\mathbf{J}\mathbf{J}
 - \frac{m_{i}L_{0}}{m_{e}\lambda_{i}} P_e  \mathbb{I}
\Bigr)
= 0,
\label{XMHDv1:current_ND_no_source}
\end{align}
The flux for $\mathbf{J}$ is first obtained from the LLF Riemann solver at the cell faces and then interpolated to the cell nodes. The divergence of the flux at the cell nodes is subsequently used to update the edge-centered $\mathbf{J}$ values. For instance, for $J_x$ located at $E1(i,j-\frac{1}{2},k-\frac{1}{2})$ as shown in Fig.~\ref{fig:mesh_Jflux}, the change in $J_x$ due to the divergence of the parallel component of the convective flux can be computed directly by taking the finite difference of that component between the nodes $NN(i-\frac{1}{2},j-\frac{1}{2},k-\frac{1}{2})$ and $NN(i+\frac{1}{2},j-\frac{1}{2},k-\frac{1}{2})$. However, the perpendicular components cannot be evaluated using the same two nodes, as they do not vary spatially along the $y$ and $z$ directions. Therefore, for example, for the $y$-contribution, the finite central difference must be applied twice: the first difference is taken between $NN(i-\frac{1}{2},j+\frac{1}{2},k-\frac{1}{2})$ and $NN(i-\frac{1}{2},j-\frac{3}{2},k-\frac{1}{2})$, and the second between $NN(i+\frac{1}{2},j+\frac{1}{2},k-\frac{1}{2})$ and $NN(i+\frac{1}{2},j-\frac{3}{2},k-\frac{1}{2})$, as shown in Fig.~\ref{fig:mesh_Jflux}. The average of these two finite central difference values is then used to update $J_x$. The same procedure applies for the $z$-contribution. Note that the stencil used for computing the perpendicular components of the convective flux is rather large while ignoring the center node which may affect numerical stability and this will be investigated further in future.

\subsection{Electron-Ion energy exchange implicit source term update} \label{subsec:electron_ion_energy_exchange}
Eq. \ref{eq:full_electron_entropy} is non-dimensionalized and expanded to get
\begin{equation}\label{eq:Se_imp_vs_exp}
    \frac{\partial S_e}{\partial t} + \nabla \cdot (\mathbf{u}_e S_e) = (\gamma - 1) n_e^{1 - \gamma} \left[\underbrace{\frac{k_B T_0}{m_i v^2}\frac{m_e}{m_i+m_e}\frac{ n_e}{\tau_{ei}} (T_i-T_e)}_{\text{implicit source}} + \underbrace{\frac{B_0^2}{\mu_0 n_0 m_i v^2} \zeta_{ei} \eta J^2}_{\text{explicit source}} \right] 
\end{equation}
where Ohmic heating is denoted as "explicit source" which will be updated explicitly. At the ideal MHD limit, the electron-ion energy exchange source term will become stiff as $\tilde{\tau}_{ei}$ becomes very small and hence, operator split implicit update will be applied to $S_e$ and this is denoted as "implicit source" in above Eq. \ref{eq:Se_imp_vs_exp}. The discrete form of this implicit update is written as
\begin{equation} \label{eq:Te_eqn}
    T_e^{*}=T_e^{**}+\Delta t f \frac{1}{\tau_{ei}}(T_i^{*}-T_e^{*})
\end{equation}
where $T^{**}$ and $T^*$ denote the intermediate states after the explicit flux and source update, and after the subsequent implicit source term update, respectively. The dimensionless collision time scale is $\tau_{ei}=\tilde{\tau}_{ei}/t_0$ and
\begin{equation*}
    f=(\gamma-1)\frac{k_B T_0}{m_i v^2}\frac{m_e}{m_i+m_e}
\end{equation*}
which arises from assuming that the rate of momentum transfer between ions and electrons are symmetric. Similarly, for ion temperature,
\begin{equation} \label{eq:Ti_eqn}
    T_i^{*}=T_i^{**}+\Delta t f \frac{c_{v,e}}{c_{v,i}} \frac{1}{\tau_{ei}}(T_e^{*}-T_i^{*})
\end{equation}
Note that since we are assuming ideal gas, $c_{v,e}/c_{v,i}$ can be taken as the ratio of the number of electrons to ions (e.g., 3 for Al$^{3+}$). However, this is not a good assumption at the low density limit where $T_i$ and $T_e$ are not necessary equal but for simplicity, $c_{v,e}/c_{v,i}$ will be taken as equal to $Z$. Rearranging Eq. \ref{eq:Ti_eqn} then gives
\begin{equation} \label{eq:Ti_eqn2}
    T_i^{*}=\frac{1}{1+\Delta t f \frac{c_{v,e}}{c_{v,i}} \frac{1}{\tau_{ei}}}T_i^{**}+\frac{\Delta t f \frac{c_{v,e}}{c_{v,i}} \frac{1}{\tau_{ei}}}{1+\Delta t f \frac{c_{v,e}}{c_{v,i}} \frac{1}{\tau_{ei}}}T_e^{*}
\end{equation}
Then sub Eq. \ref{eq:Ti_eqn2} into Eq. \ref{eq:Te_eqn} and re-arrange to obtain updated electron temperature,
\begin{equation} \label{eq:Te_eqn2}
    T_e^{*}=\left(T_e^{**}+\frac{\Delta t f \frac{1}{\tau_{ei}}}{1+\Delta t f \frac{c_{v,e}}{c_{v,i}} \frac{1}{\tau_{ei}}}T_i^{**} \right) / \left(1+ \Delta t f \frac{1}{\tau_{ei}} - \frac{\Delta t^2 f^2 \frac{c_{v,e}}{c_{v,i}} \frac{1}{\tau_{ei}^2}}{1+\Delta t f \frac{c_{v,e}}{c_{v,i}} \frac{1}{\tau_{ei}}}\right)
\end{equation}
Note that $\tau_{ei}$ depends on both $T_e$ and $T_i$ but for simplicity, we assumed that $\tau_{ei}\equiv\tau_{ei}(T_e^{**}, T_i^{**})$ which does its job in ensuring that both electron and ion temperature equilibrate to the equilibrium temperature under suitable timescales. Another thing to note is that Eqs. \ref{eq:Ti_eqn2} and \ref{eq:Te_eqn2} do not guarantee that they contributes to the same total pressure as $p^*$ and thus, $P_e$ may exceed $p$ which will cause negative $T_i$. To ensure consistency with total energy, $T_i^*$ is not obtained from Eq. \ref{eq:Ti_eqn2} but rather from $p$ and $T_e^*$. Hence, $T_{i,floor}=10^{-10}$ and $T_e^*=\min(T_e^*,(p^*-n^* T_{i,floor})/(Zn^*))$ should be applied to ensure that $T_i$ does not become negative.

\section{Numerical tests}
This section focuses on testing the ability of the XMHD solver to address linear problems with known analytical solutions across different physical limits—ideal, resistive, and Hall MHD.
For nonlinear problems involving discontinuities, the Brio–Wu shock tube test \cite{BRIO1988400} is used as a stringent benchmark to assess the solver’s capability to accurately capture nonlinear flow features and discontinuous structures. However, note that for conventional ideal MHD test, the MHD equations are typically written such that $\mu_0=1$ \cite{2008ApJS..178..137S} but this is not the case here which must be accounted for. Here, LLF Riemann solver will be employed for the ideal MHD components unless stated otherwise. This is because LLF Riemann solver is numerically more stable than HLLD Riemann solver as it is more diffusive than the latter.

For the tests done here, Eqs. \ref{XMHDv1:mass} to \ref{XMHDv1:entropy} will be non-dimensionalized to $\mathbf{U}=U_0 \mathbf{\tilde{U}}$ where $\tilde{\mathbf{U}}$ is the set of dimensionless variables while $U_0$ is the set of dimensional normalization factors with the values in Table \ref{table:ref_values} unless stated otherwise:

\begin{table}[htbp]
\centering
\caption{Reference parameters used for non-dimensionalization.}
\begin{tabular}{llc}
\hline
\textbf{Parameter} & \textbf{Description} & \textbf{Value} \\
\hline
$n_0$ & Reference number density & $6 \times 10^{28}~\mathrm{m}^{-3}$ \\
$t_0$ & Reference time & $100~\mathrm{ns}$ \\
$L_0$ & Reference length & $1~\mathrm{mm}$ \\
$v_0$ & Reference velocity & $10^4~\mathrm{m/s}$ \\
$B_0$ & Reference magnetic field & $580~\mathrm{T}$ \\
$E_0$ & Reference electric field & $5.8 \times 10^6~\mathrm{V/m}$ \\
$J_0$ & Reference current density & $4.6 \times 10^{11}~\mathrm{A/m}^2$ \\
$T_0$ & Reference temperature & $28~\mathrm{eV}$ \\
$\sigma_0$ & Reference conductivity & $\frac{1}{\mu_0 L_0 v_0}~\Omega^{-1} \cdot \mathrm{m}^{-1}$ \\
\hline
\end{tabular}
\label{table:ref_values}
\end{table}

Unless stated otherwise, the material chosen here is Aluminium (Al) which has an ion mass number of 27 and an ionisation number ($Z$) of 3. As for the timestepping, it is done using the CFL condition ($C$),
\begin{equation}
    \Delta t=C\frac{\Delta x}{\max(\lambda_i)}
\end{equation}
where $\lambda_i$ comes from Eqs. \ref{eq:eigenvalues}
 and \ref{eq:elec_sound_speed}. 2nd-order Runge-Kutta time integration is employed for the explicit time-update of the conserved variables. 2nd-order piecewise-linear method (PLM) is used in conjunction with the density-dependent slope limiter from Eq. \ref{eq: density-dependent-SL} and will then be referred to as modified PLM for brevity. 
 
\subsection{Brio-Wu shock tube problem}
A common test case for ideal MHD solver is the magnetized shock tube with the following initial conditions,
\[ \left( \rho, u_x, u_y, u_z, P, P_e, B_x, B_y, B_z \right) =
\begin{cases}
\left( 1.000\,\rho_0,\, 0,\, 0,\, 0,\, 1.0\,\rho_0,\, 0.50\,\rho_0,\, 0.75\,\sqrt{\rho_0},\, +\sqrt{\rho_0},\, 0 \right), & \text{if } x \leq 0 \\
\left( 0.125\,\rho_0,\, 0,\, 0,\, 0,\, 0.1\,\rho_0,\, 0.05\,\rho_0,\, 0.75\,\sqrt{\rho_0},\, -\sqrt{\rho_0},\, 0 \right), & \text{if } x > 0
\end{cases} \]

\begin{figure}[http]
    \centering
    \includegraphics[width=\linewidth]{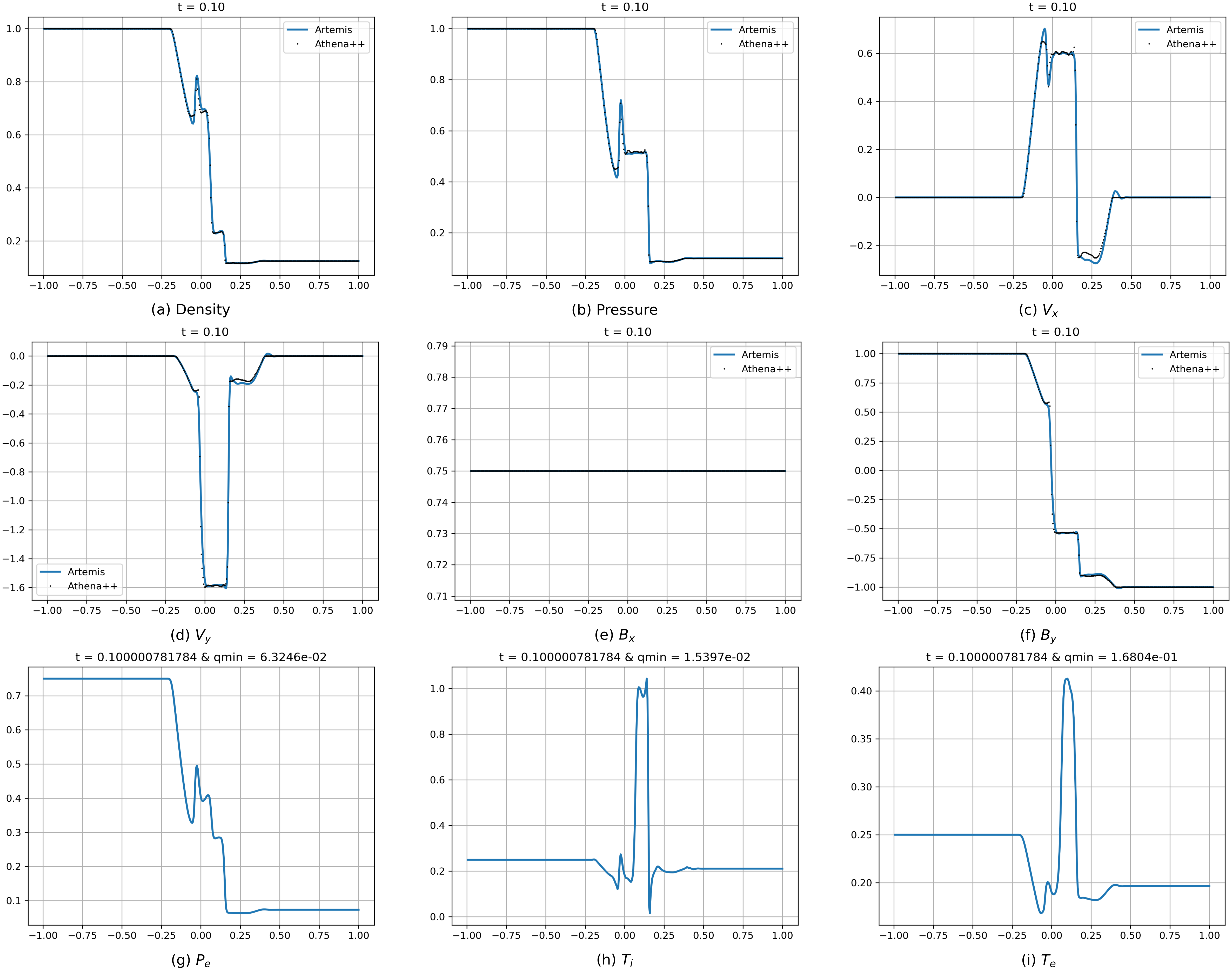}    
    \caption{For the standard ideal MHD test case on shock tube, the XMHD (Artemis) is quite similar to ideal MHD (Athena++) but there is some discrepancy especially at lower-density region where non-ideal effects become more crucial. \label{fig:XMHD_BWtube_rho1p0}}
\end{figure}

\begin{figure}[http]
    \centering
    \includegraphics[width=\linewidth]{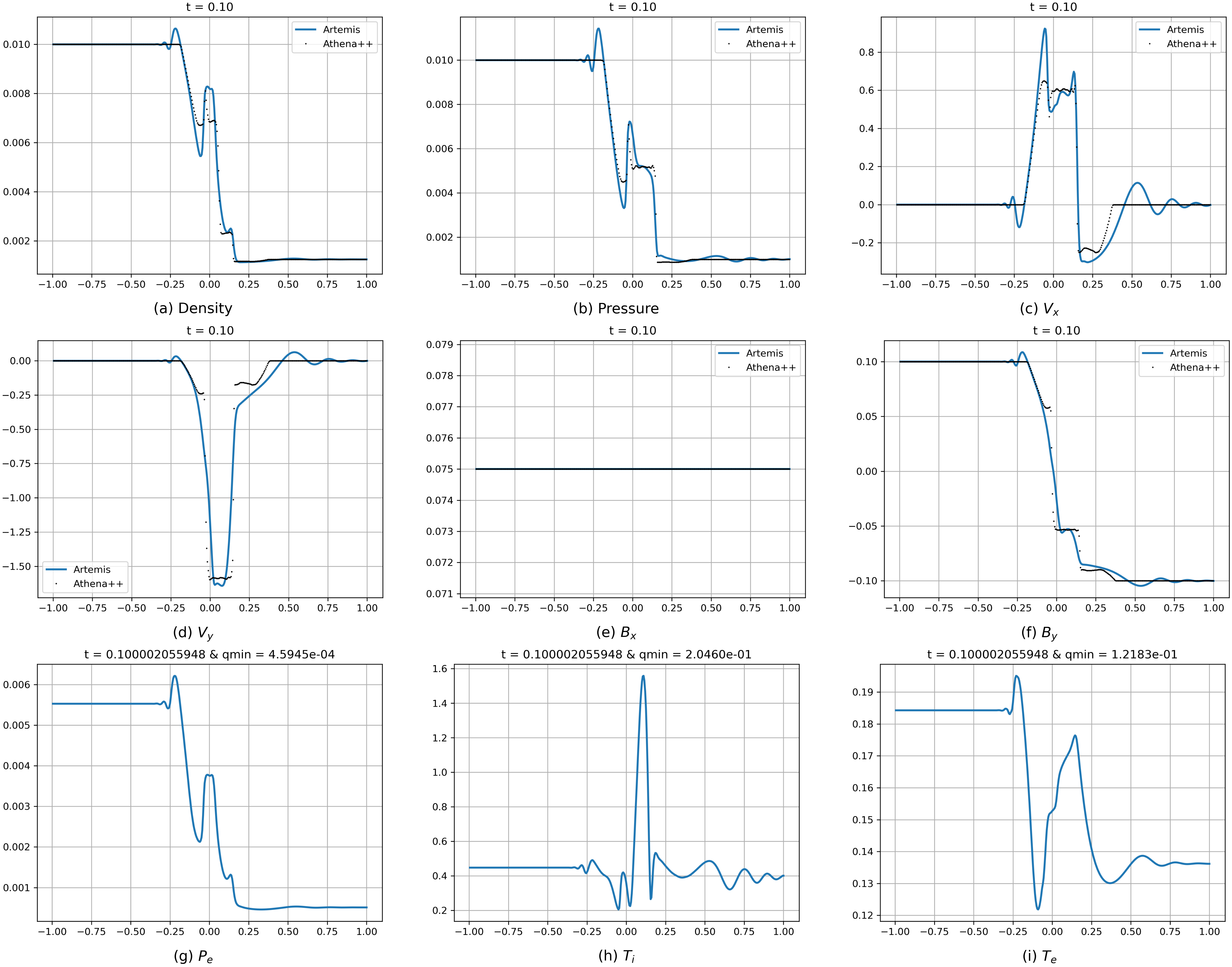}    
    \caption{At low density, XMHD (Artemis) begins to deviate significantly from ideal MHD (Athena++) due to more significant non-ideal MHD effects. \label{fig:XMHD_BWtube_rho1pm2}}
\end{figure}

For all the ideal MHD test cases, the resistivity is set to $\eta=10^{-8}$ \cite{Ferraro2010} unless stated otherwise. Note that ion temperature is initially higher than electron temperature: $T_i=ZT_e$. XMHD solver will then be tested on $\rho_0=10^4$, $\rho_0=1.0$ and $\rho_0=0.01$. Outflow boundaries are used here. The results of XMHD on Artemis are then compared to the ideal MHD solver on Athena++ to compare the discrepancy at low density regions due to non-ideal effects as shown in Fig. \ref{fig:XMHD_BWtube_rho1p0} and \ref{fig:XMHD_BWtube_rho1pm2}.

\begin{figure}[http]
    \centering
    \includegraphics[width=\linewidth]{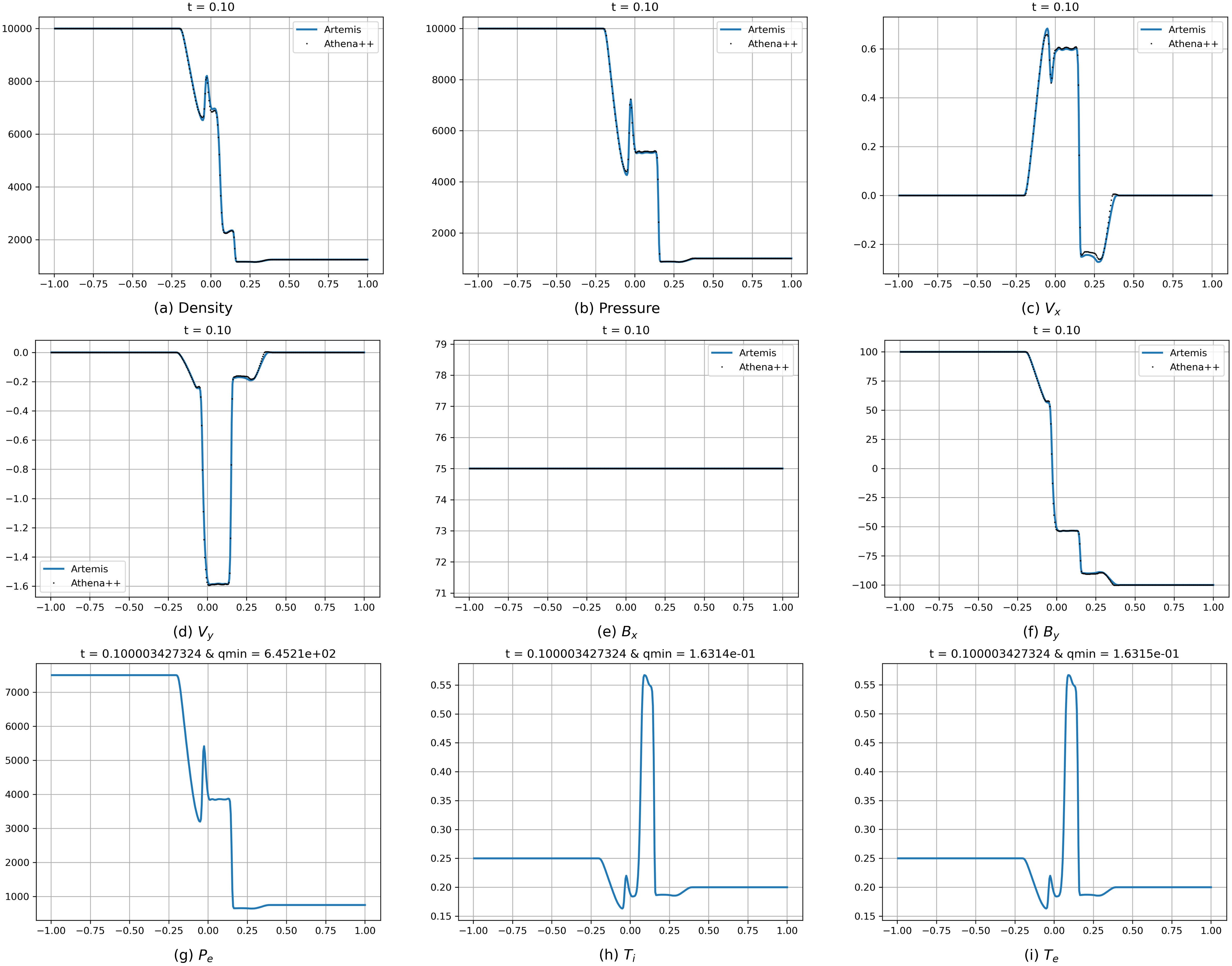}    
    \caption{At high density of $\rho_0=10^4$, both ions and electrons temperatures are expected to equilibrate to equilibrium temperature. \label{fig:XMHD_BWtube_rho1e4}}
\end{figure}

The post-shock electron temperature is also expected to be lower than the ion temperature, since electrons experience primarily adiabatic compression heating without irreversible shock dissipation. This can be seen in Fig. \ref{fig:XMHD_BWtube_rho1pm2} where $T_e$ is much lower than $T_i$. However, as $\rho_0$ increases, $T_e$ increases while $T_i$ decreases until equilibrium temperature is reached due to the more significant electron-ion energy exchange source term. From Fig. \ref{fig:XMHD_BWtube_rho1p0}, $T_i$ and $T_e$ have equilibrated to common temperature at smooth region while $T_i$ is still larger than $T_e$ at the developed discontinuous region. Also, $T_i$ is generally larger than $T_e$ except across the slow shock when there is a sharp spike to the ion temperature floor. This is due to the misalignment of the slow shock between $p$ and $P_e$ due to the use of split Riemann solver (refer to Eq. \ref{eq:split_Jacobian}), hence causing numerical diffusivity between the two variables to be "asynchronous". As $P_e$ is more diffusive than $p$ across the slow shock, it can exceed $p$, resulting in negative $T_i$ which then leads to the imposition of $T_i$ floor. At $\rho_0=10^4$ which is shown in Fig. \ref{fig:XMHD_BWtube_rho1e4}, both $T_i$ and $T_e$ have equilibrated to the same temperature everywhere, thus matching the single-fluid assumption of ideal MHD.

\subsection{Ideal MHD linear wave test} \label{SubSect::ideal_MHD_linear_wave}
In the ideal MHD limit, the XMHD is ideally expected to have 2nd-order convergence of L1 errors for the propagation of ideal MHD linear waves since RK2 and modified PLM are used here. The initial conserved variables state vector is \cite{GARDINER20084123}
\begin{equation}
    q^0=\overline{q}+\epsilon R_p \cos\left(\frac{2\pi x_1}{\lambda}\right)
\end{equation}
where $\overline{q}$ is the mean background state, $\epsilon=10^{-6}$ is wave amplitude and $R_p$ is the right eigenvector in conserved variables for wave mode $p$. Here, density and pressure are set to 1 and $\frac{1}{\gamma}$, respectively. Velocity components are set to 0 except for $v_x$ which is set to 1 for entropy mode test because Artemis run the test by wave period. After propagating each of the waves by a period, their L1 error w.r.t. their initial conditions for each component k of the conserved variables are computed as follows,
\begin{equation}
    \delta q_k=\frac{1}{2N^2}\sum_i \sum_j |q_{i,j,k}^n-q_{i,j,k}^0|
\end{equation}
by summing up over all the grid cells. The root-mean-square (RMS) error is then obtained by taking the norm of the above error vector
\begin{equation}
    |\delta q|=\sqrt{\sum_k (\delta q_k)^2}
\end{equation}
for each of the wave modes.

For 1D problem, a computational domain of $[0,1]$ with grid resolutions of $[8,16,32,64,128,256,512,1024]$ are tested. For multi-dimensional test problem of linear wave test, the linear wave is propagated such that it lies oblique to the grid and the coordinate transformation based on spherical coordinate is employed as follows
\begin{align}
    x &= x_1 \cos \alpha \cos \beta - x_2 \sin \beta - x_3 \sin \alpha \cos \beta \\
    y &= x_1 \cos \alpha \sin \beta + x_2 \cos \beta - x_3 \sin \alpha \sin \beta \\
    z &= x_1 \sin \alpha + x_3 \cos \alpha
\end{align}
For 2D problem, it is done on a computational domain of $[0,2.236068]\times [0,1.118034]$ with grid resolutions along x ranging from $[16,32,64,128,256,512]$.

\begin{figure}[http]
    \centering
    \centering
    \begin{subfigure}[b]{\linewidth}
    \includegraphics[width=\linewidth]{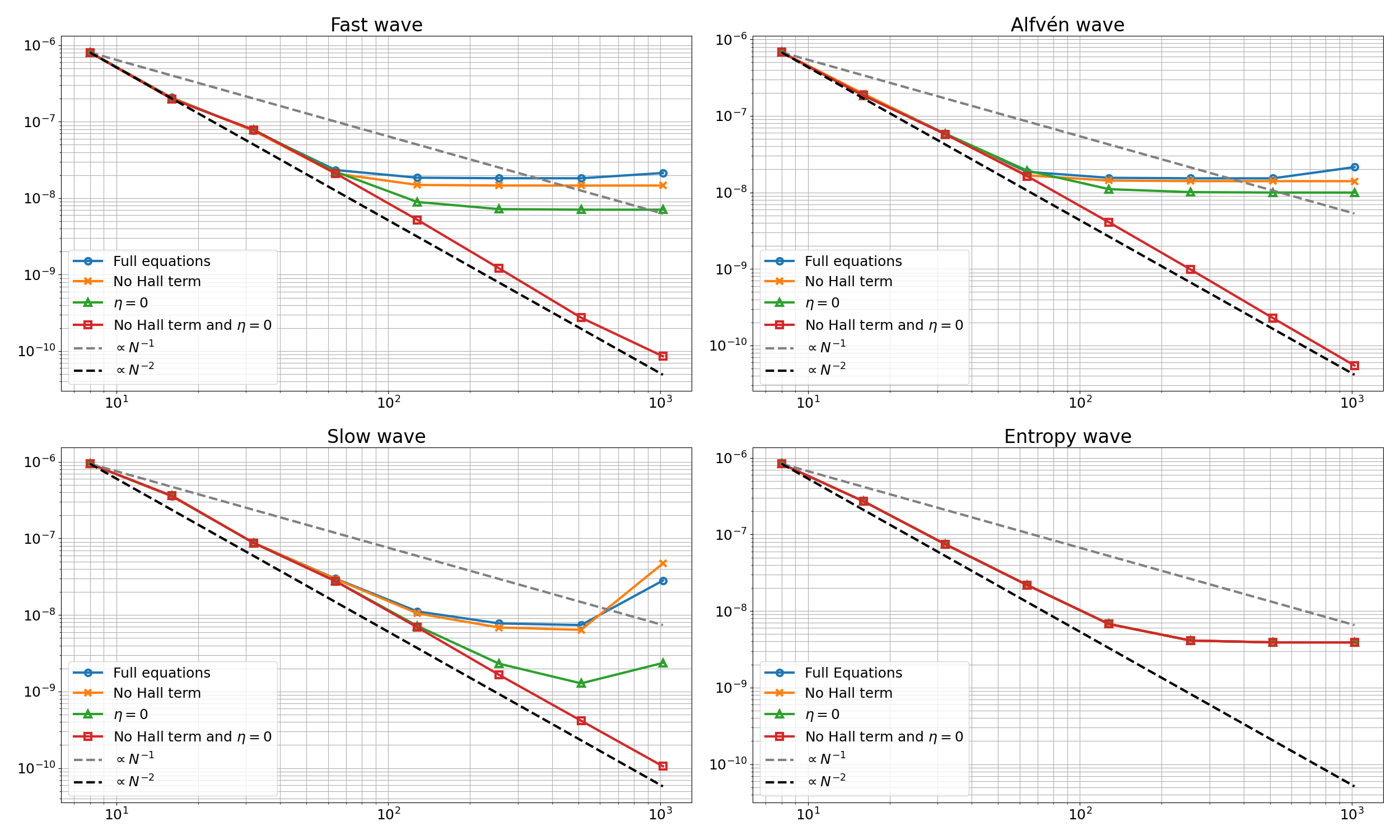}    
    \caption{To test for 2nd-order convergence in the ideal MHD limit, the code is tested by (i) using the full equations, (ii) setting Hall term to zero and, (iii) setting resitivity to zero and (iv) setting both Hall and resistive terms to zero. Only entropy wave is independent of the Hall or resistive terms as it is not specific to ideal MHD. \label{fig:XMHD_1D_linear_wave}}
    \end{subfigure}

    \begin{subfigure}[b]{0.6\linewidth}
    \includegraphics[width=\linewidth]{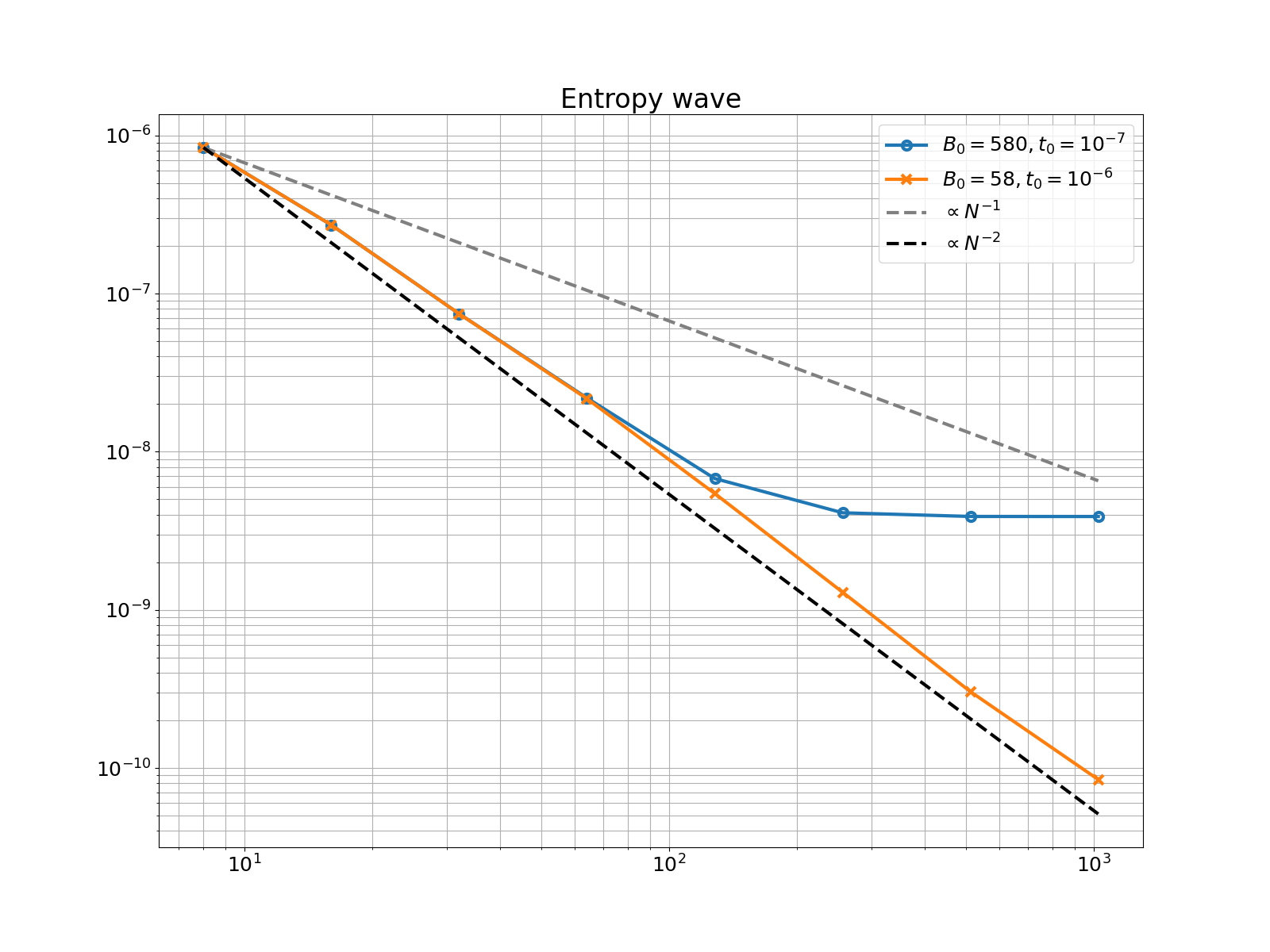}    
    \caption{Since entropy wave did not fully converge to 2nd-order especially at high resolution, a different set of dimensional constants are used to ensure that the XMHD solver is able to relax sufficiently to the ideal MHD limit for 2nd-order convergence. \label{fig:XMHD_1D_entropy_wave}}
    \end{subfigure}
    \caption{For 1D problem at grid resolution of $[16, 32, 64, 128, 256, 512, 1024]$.}
\end{figure}

As expected, the results in Fig. \ref{fig:XMHD_1D_linear_wave} show that without Hall and resistive terms, the coupling of ideal MHD and other components through the implicit source term update is correct in ensuring that $\mathbf{E}\approx-\mathbf{u}\times \mathbf{B}$ due to the 2nd-order convergence from employing modified PLM and RK2 time integration. However, the entropy wave ceases to exhibit second-order convergence at very fine resolutions. This may be a result of GOL not relaxing to the ideal MHD's Ohm law as the grid resolution increases. For instance, when different $B_0$ and $t_0$ are chosen such that $\epsilon_E$ is smaller while $\frac{B_0^2}{\mu_0 n_0 m_i v^2}$ remains approximately 1 (see  Table~\ref{tab:relax_param_L0}), second-order convergence is observed in Fig.~\ref{fig:XMHD_1D_entropy_wave}.

\begin{table}[h!]
\centering
\begin{tabular}{|c|c|c|}
\hline
             & $B_0=580$, $t_0=10^{-7}$ & $B_0=58$, $t_0=10^{-6}$ \\ \hline
$\epsilon_E$ & $1.11 \times 10^{-9}$ & $1.11 \times 10^{-11}$  \\ \hline
$\epsilon_J$ & $4.71 \times 10^{-10}$ & $4.71 \times 10^{-10}$ \\ \hline 
$\frac{B_0^2}{\mu_0 n_0 m_i v^2}$ & $9.95 \times 10^{-1}$ & $9.95 \times 10^{-1}$ \\ \hline
\end{tabular}
\caption{Different relaxation parameters where the second column is the original dimensional constants used for this problem. }
\label{tab:relax_param_L0}
\end{table}

\begin{figure}[http]
    \centering
    \includegraphics[width=\linewidth]{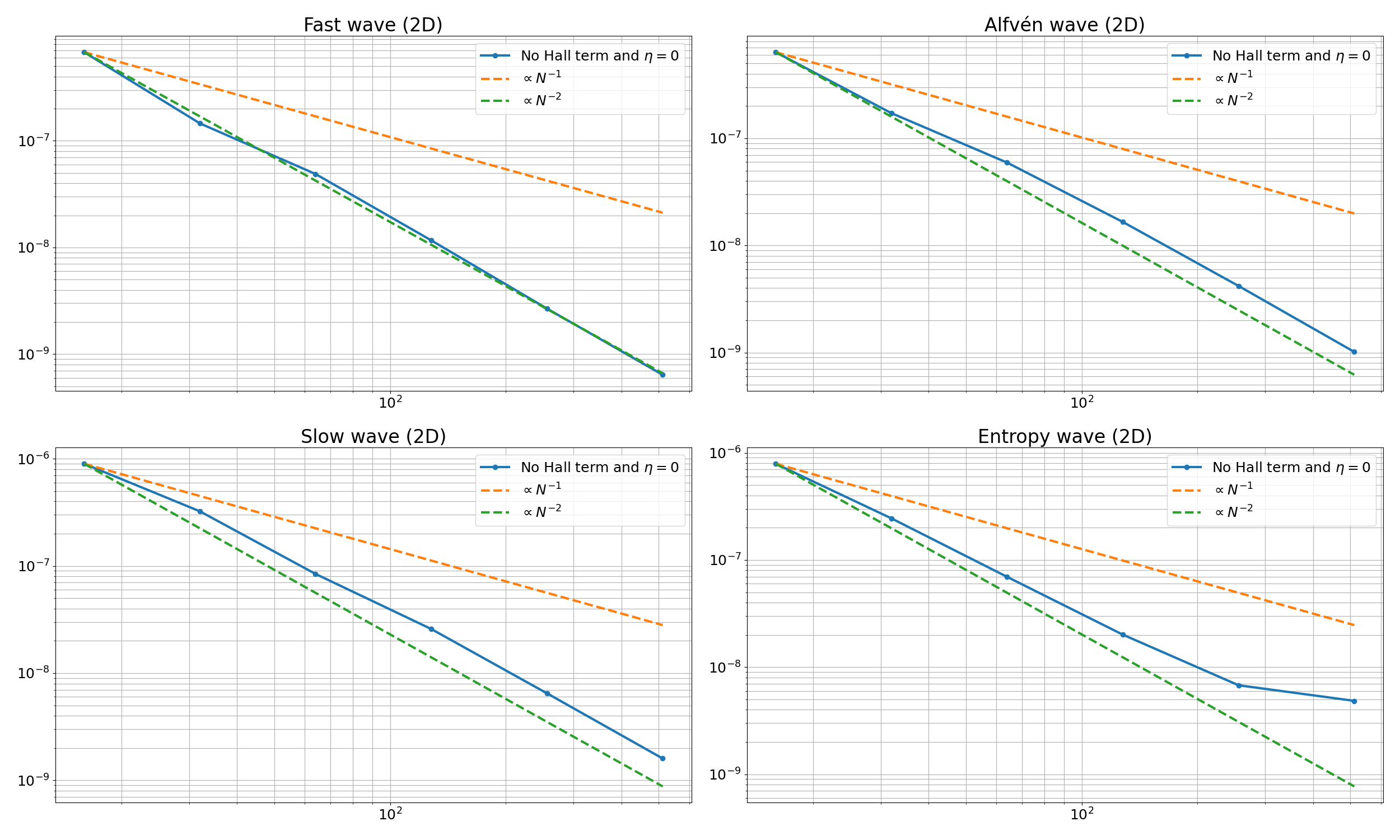}    
    \caption{For 2D problem, the code is tested by setting both Hall and resistive terms to zero. 2nd-order convergence is expected since modified PLM and RK2 time integration are used. \label{fig:XMHD_2D_linear_wave}}
\end{figure}

Note that from Fig. \ref{fig:XMHD_1D_linear_wave}, entropy wave retains the same convergence order with or without the Hall and/or resistive term as it is not specific to ideal MHD. When Hall or resistive terms are involved, not all the waves have 2nd-order convergence especially for slow wave which gets worse at higher resolution when resistivity is involved. The deviation of 2nd-order convergence may also be a result of the Hall term which makes the phase speeds of waves wavenumber-dependent \cite{Iwasaki_2025} and increasing resolution increases the Hall effect on the solution. Since Hall and resistive terms impede 2nd-order convergence for ideal MHD linear wave test, they will be turned off for multidimensional linear wave test like in Fig. \ref{fig:XMHD_2D_linear_wave} where 2nd-order convergence is expected.

\subsection{Field Diffusion} \label{SubSect:FieldDiffuse}
To test the resistive component of the XMHD, it is used to diffuse magnetic field with a n-dimensional Gaussian distribution as it has an analytical solution as follows \cite{Mullen2021}
\begin{equation} \label{eq:FD_Gaussian}
    \mathbf{G}(x,y,z,t)=\frac{A}{\sqrt{4\pi \eta^* (t_0 +t)}^n}\exp \left[-\frac{x^2+y^2+z^2}{4\eta^* (t_0+t)}\right]
\end{equation}
where n is the number of dimensions of the problem and $\eta^*=\eta \sigma_0$ is the dimensionless resistivity used in the GOL equation. Here $t_0$ is taken as 0.5 and the problem is evolve to $t_f=t_0+2$. For 1D tests, the L1 error of the result is computed as
\begin{equation}
    L_{1,\xi}=\sum^N_{i=1}|\xi_i - \xi_{exact}|\Delta x
\end{equation}
where $\xi_i$ and $\xi_{exact}$ are the numerical and exact solution at grid cell i, respectively.

\begin{figure}[http]
    \centering
    \includegraphics[width=\linewidth]{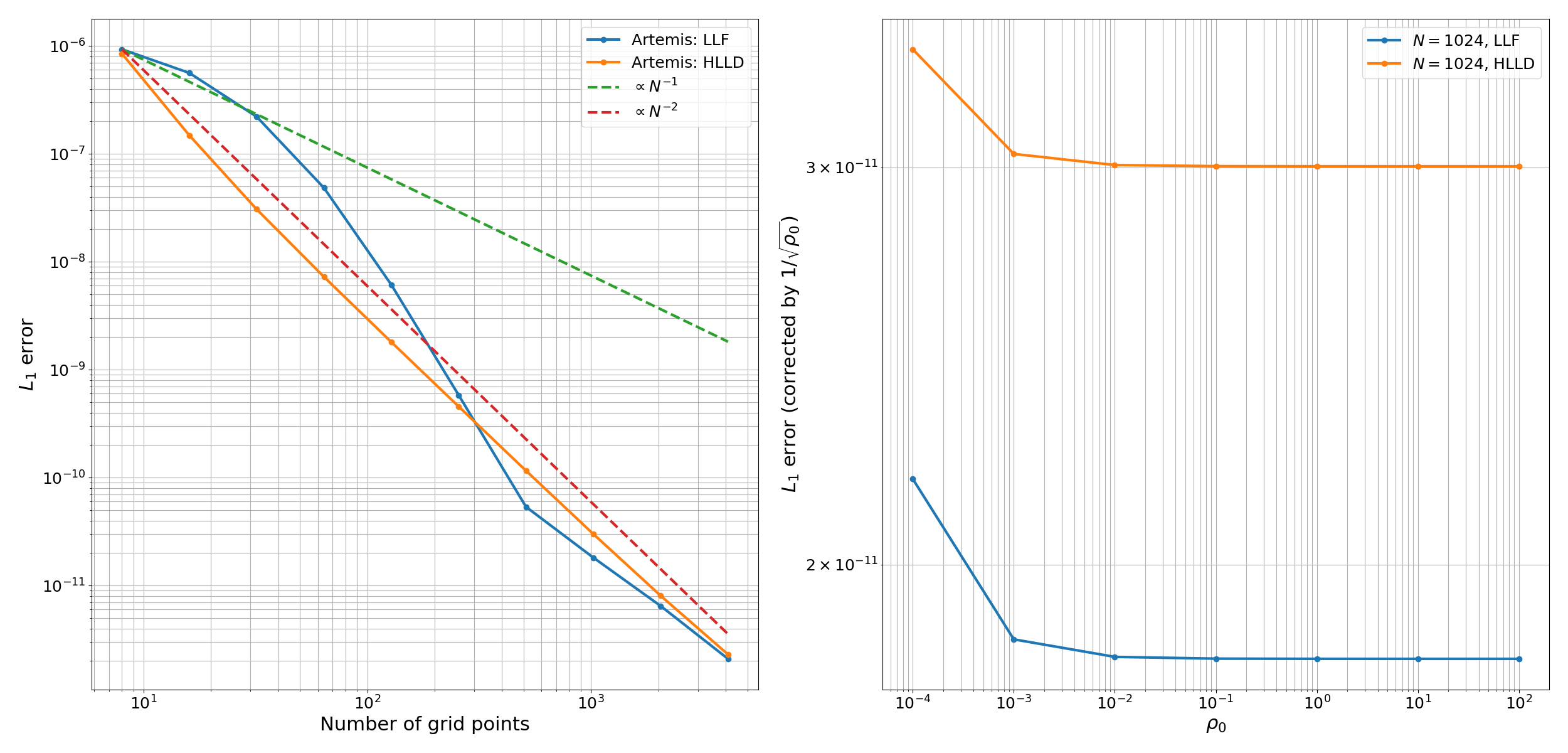}    
    \caption{To test for order of convergence in 1D field diffusion problem, several grid resolution and their L1 error norms at $t=2.5$ are plotted here. \label{fig:XMHD_1D_field_diffusion}}
\end{figure}

The problem is set up on a computational domain of $[-6,6]$ with grid resolution of $N=[8,16,32,64,128,256,512,1024,2048,4096]$. The density and pressure are set to 1 while the velocities are set to 0 across the whole domain. $B_x$ and $B_z$ are set to 0 while $B_y$ is initialized using the Gaussian distribution from Eq. \ref{eq:FD_Gaussian}. The dimensional resistivity is set to $\eta=10^{-6}$ while the amplitude from Eq. \ref{eq:FD_Gaussian} is set to $A=10^{-6}$. From the left diagram in Fig. \ref{fig:XMHD_1D_field_diffusion}, the algorithm has 2nd-order of convergence with increasing grid resolution due to the 2nd-order accurate PLM reconstruction method used. $\rho_0$ is also varied to illustrate the deviation from resistive MHD as Hall term and electron inertia effects become more important at lower density, causing both LLF and HLLD solvers to have higher L1 errors at lower densities. 

\subsection{Hall MHD limit}
To test the algorithm in the limit where the Hall term dominates resistive and ideal effects, it is crucial to consider the two wave modes (e.g., Whistler and Hall drift waves) associated with the Hall term. For instance, the GOL can be simplified by assuming $L_0>>\frac{c}{\omega_{pe}}$, $L_0>>\rho_e$ and $\eta\rightarrow0$ where $\omega_{pe}$ and $\rho_e$ are the electron plasma frequency and electron Larmor radius, respectively. This allows one to neglect the electron inertia, electron pressure and resistive term, reducing to the simplified Ohm's law
\begin{equation}
    \mathbf{E}=-\mathbf{u}\times \mathbf{B}+\frac{1}{n_e e}\mathbf{J}\times \mathbf{B}
\end{equation}
where the final term is the Hall term which decouples ion and electron motion on ion inertial length scales $L<c/\omega_{pi}$ where $\omega_{pi}=\frac{4\pi n_i e^2}{m_i}$ is the Hydrogen ion plasma frequency which is used for this problem. This will reduce Eq. \ref{XMHDv1:faraday_ND} to 
\begin{equation}
    \frac{\partial \mathbf{B}}{\partial t} - \nabla \cdot (\mathbf{B}\mathbf{u}-\mathbf{u}\mathbf{B}) = -\nabla \times \left(\frac{1}{n_e e}\mathbf{J}\times\mathbf{B} \right)
\end{equation}
Assuming stationary ($\mathbf{u}=0$) and neutralizing background, the above equation simplifies to
\begin{align}
     \frac{\partial \mathbf{B}}{\partial t} &= -\nabla \times \left(\frac{1}{n_e e}\mathbf{J}\times\mathbf{B} \right) \\
     &= -\frac{1}{n_e e}\nabla \times (\mathbf{J}\times \mathbf{B}) + \frac{1}{n_e^2 e} \nabla n \times (\mathbf{J}\times \mathbf{B}) \label{eq:Whistler_wave_&_Hall_drift}
\end{align}
where the first and second terms correspond to whistler and Hall drift waves, respectively. Thus, Hall drift wave is associated with inhomogeneous plasmas since it depends on $\nabla n$.

\subsubsection{1D Whistler wave problem} 
\label{subsec:whistler_wave}
Since we are considering only whistler wave here, we assume that the plasma's density is homogeneous and have an ambient magnetic field along x ($\mathbf{B}=B_0\hat{e}_x$). The magnetic field is then perturbed along y and z ($\propto\exp(ik_xx-i\omega t)$). The linear dispersion relation can be solved from
\begin{equation}
    \frac{\partial \delta \mathbf{B}}{\partial t}=-\frac{1}{n_ee}\nabla \times (\delta \mathbf{J}\times \mathbf{B})
\end{equation}
Then linearize the above equation to get
\begin{equation}
    \omega \delta \mathbf{B}=\frac{1}{n_e e}(k_x B \delta J_y \hat{e}_y+k_x B \delta J_z \hat{e}_z)
\end{equation}
Then sub in Ampere's law (e.g., $\delta \mathbf{J}=(1/\mu_0)\nabla \times \delta \mathbf{B}$) to get the following coupled equations
\begin{align} 
    \omega \delta B_y&=-i\frac{k_x^2 B}{\mu_0 n_e e}\delta B_z \\
    \omega \delta B_z&=i\frac{k_x^2 B}{\mu_0 n_e e}\delta B_y
\end{align}
Solving the above two coupled equations then give the whistler wave dispersion relation 
\begin{equation}
    \omega=\frac{k_x^2 B}{\mu_0n_e e}=\frac{k_x^2 V_A}{\omega_{pi}}
\end{equation}

\begin{figure}
    \centering
    \includegraphics[width=0.8\linewidth]{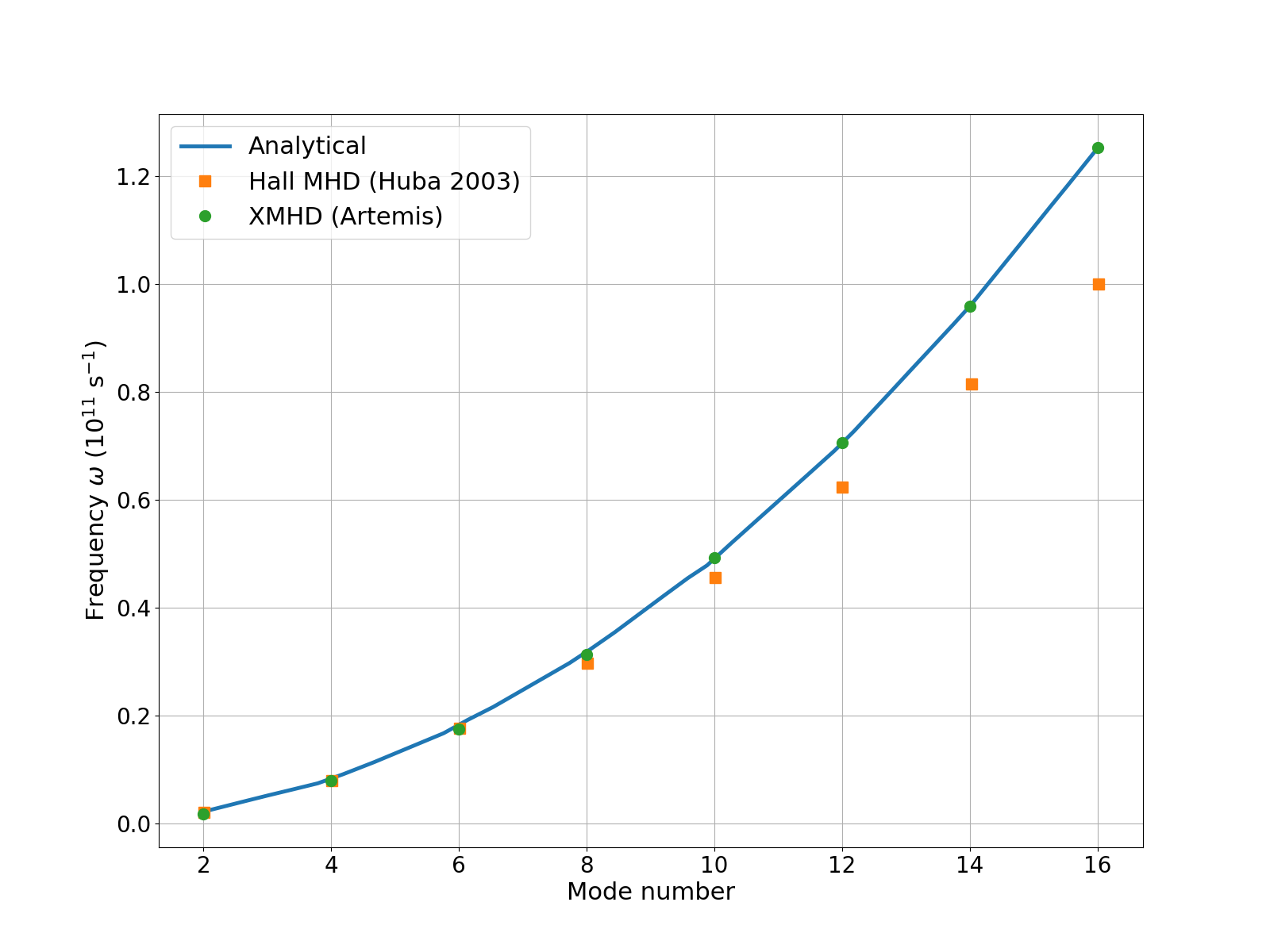}
    \caption{Plot of whistler wave dispersion after running the simulation to $t=0.1$. }
    \label{fig:whistler_wave}
\end{figure}

The problem is then setup with ambient density of $n_0=10^{18} m^{-3}$ and ambient magnetic field of $\mathbf{B}=B_0\hat{e}_x$ with $B_0=1000$G ($\beta=10^{-4}$) along the z-axis. The 1D domain has a length of $L_x=20$cm and 2400 grid points are used along the x-direction. The plasma used is Hydrogen and it is perturbed with $\delta B_y = \delta B \sin(2\pi mx/L_x)$ and $\delta B_z =\delta B \cos(2\pi m x/L_x)$ where $m$ is the mode number and $\delta B=10$G. The simulation is run to $t=0.1$ before determining $\omega$ to compare against the analytical solution as shown in Fig. \ref{fig:whistler_wave}. The XMHD solver shows good agreement with the analytical solution, even at high mode numbers, whereas the Hall MHD solution \cite{Huba2003} employs a much lower resolution of 120 grid points, which may contribute to the observed differences.

\subsubsection{2D Huba's Hall drift problem} \label{subsect:Huba}
Consider the inhomogeneous plasma density along x with an ambient magnetic field in the z-direction. The ambient magnetic field ($\mathbf{B}=B_0 \hat{e}_z$) is then perturbed with $\delta B_z \propto exp(ik_y y - i \omega t)$ giving rise to the following linear dispersion equation
\begin{align}
    \frac{\partial \delta \mathbf{B}}{\partial t} &=\frac{1}{n_e^2 e}\nabla n \times(\delta \mathbf{J}\times \mathbf{B}) 
\end{align}
Linearizing above gives 
\begin{align}
    i\omega\delta B_z &= \frac{1}{n_e^2 e} \frac{\partial n}{\partial x} \delta J_xB_z
\end{align}
Since $\delta J_x=\frac{ik_y}{\mu_0}\delta B_z$ and $n=n_e$ for Hydrogen plasma, the dispersion relation is then
\begin{align}
    \omega &= \frac{k_y B_z}{n\mu_0}\frac{\partial n}{\partial x}=k_y V_A \frac{c}{L_n \omega_{pi}}
\end{align}
where $L_n=(\frac{\partial \ln n}{\partial x})^{-1}$ is density gradient length scale, $V_A=\frac{B_z}{\sqrt{\mu_0 m_i n}}$ is Alfven wave speed and $\omega_{pi}=\sqrt{\frac{e^2 n}{\epsilon_0 m_i}}$ is the Hydrogen ion plasma frequency. Since Hall term is crucial for this problem, $L_n<\frac{c}{\omega_{pi}}$ to generate Hall drift wave which propagates in the $\mathbf{B}\times \nabla n$ direction.

The problem is then set up on a computational domain of $[-0.3,0]\times[0.3,0.2]$ where the number of grid points are $(200,120)$. Hydrogen plasma is used and has a number density profile of \cite{Huba2003}
\begin{equation}
    n(x) = \frac{n_0}{2}\left[(1+A)+(1-A)\tanh\left(\frac{x-x_0}{\Delta x}\right) \right]
\end{equation}
where $x_0=10$cm, $\Delta x=3$cm and Atwood number is set as $A=20$ with a low number density of $n_0=10^{18} \text{m}^{-3}$ where Hall effect is significant. This means that the dimensional constants for the PDEs are also adjusted such that $n_0=10^{18}$, $L_0=1$ and $B_0$ set to 0.4568 such that the dimensionless coefficient ($\frac{B_0^2}{\mu_0 n_0 m_i v^2}$) is approximately 1. The density profile is then maintained by setting $u_x=0$ every timestep. The plasma is also assumed to be isothermal. Magnetic field is set along the z direction where $B_0=0.1 \text{T}$ with perturbation of $\delta B_z = \delta B \cos \left(\frac{2\pi m y}{L_y} \right)$ in the y-direction where $\delta B=0.001 \text{T}$ and m is the mode number. The plasma beta ($\beta$) is then set to 0.0001 at peak density region where $x=0$. This allows the Hall drift wave to propagate in the y-direction.

\begin{figure}[htbp]
    \centering
    \begin{subfigure}[b]{0.8\linewidth}
        \centering
        \includegraphics[width=\linewidth]{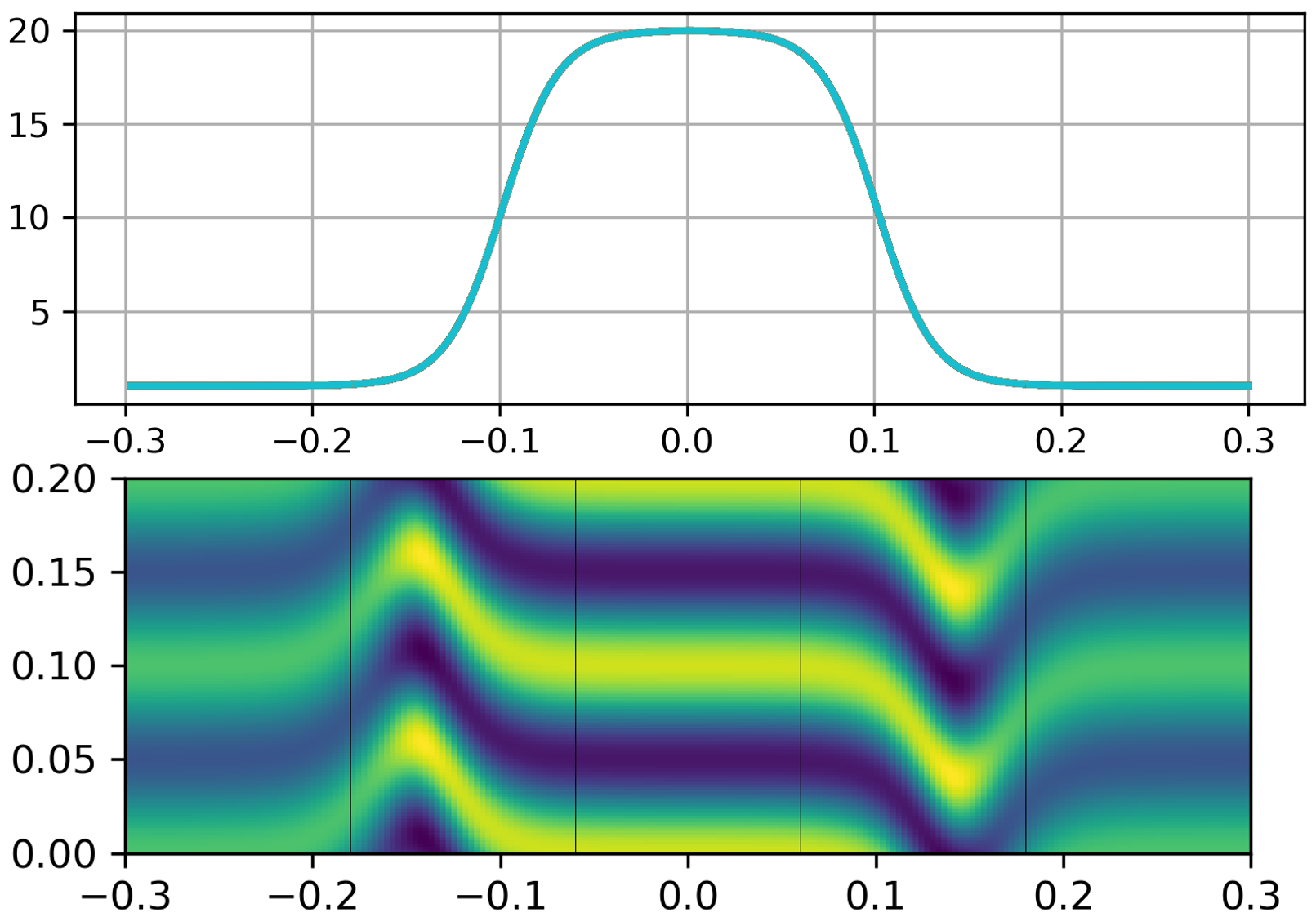}
        \caption{Huba’s Hall drift problem at $t = 8\times10^{-9}$ s where
        the top and bottom panels show density and magnetic field
        $B_z(x,y)$, respectively. Note that both are normalized by $n_0$ and $B_0$ here.}
        \label{fig:XMHD_Huba_t8em2}
    \end{subfigure}
    \begin{subfigure}[b]{0.8\linewidth}
        \centering
        \includegraphics[width=\linewidth]{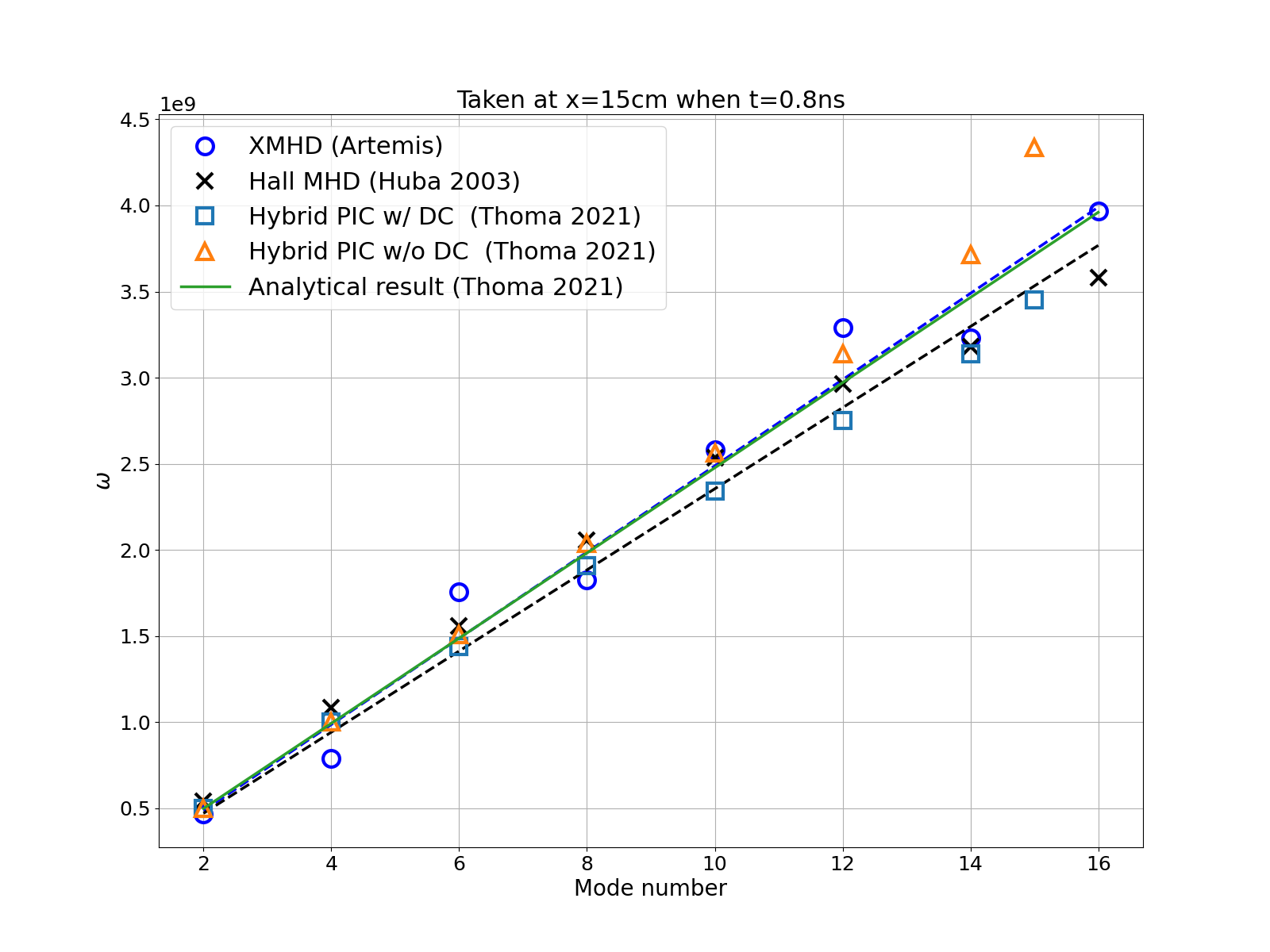}
        \caption{Dashed line based on the linear fit of XMHD (Artemis) and Hall MHD \cite{Huba2003} are added to show linear dispersion relation obtained from the Hall drift problem. Results from Hybrid PIC \cite{THOMA2021} with or without displacement current (DC) are also included for comparison.}
        \label{fig:XMHD_Huba_dispersion}
    \end{subfigure} 
    \caption{Results for Huba’s Hall drift problem obtained from XMHD implemented on Artemis and compared to Hall MHD and Hybrid PIC results.}
    \label{fig:XMHD_Huba_combined}
\end{figure}

For this problem, the maximum phase velocity of the Hall drift wave is expected to occur at $x\simeq 15$cm. From Fig. \ref{fig:XMHD_Huba_t8em2}, the magnetic field ($B_z$) contours are distorted in the y-direction in the density gradient region and the maximum distortion occurs at around 15cm along x-axis. Some discrepancy is expected since electron inertia effects and Ohmic resistivity are not set to zero ($\eta=10^{-8}$) here. For instance, from Fig. \ref{fig:XMHD_Huba_t8em2}, there appear to be some $B_z$ diffusion at $|x|>0.2$. 

\subsection{2D MHD blast wave problem}
\begin{figure}[http]
    \centering
    \includegraphics[width=\linewidth]{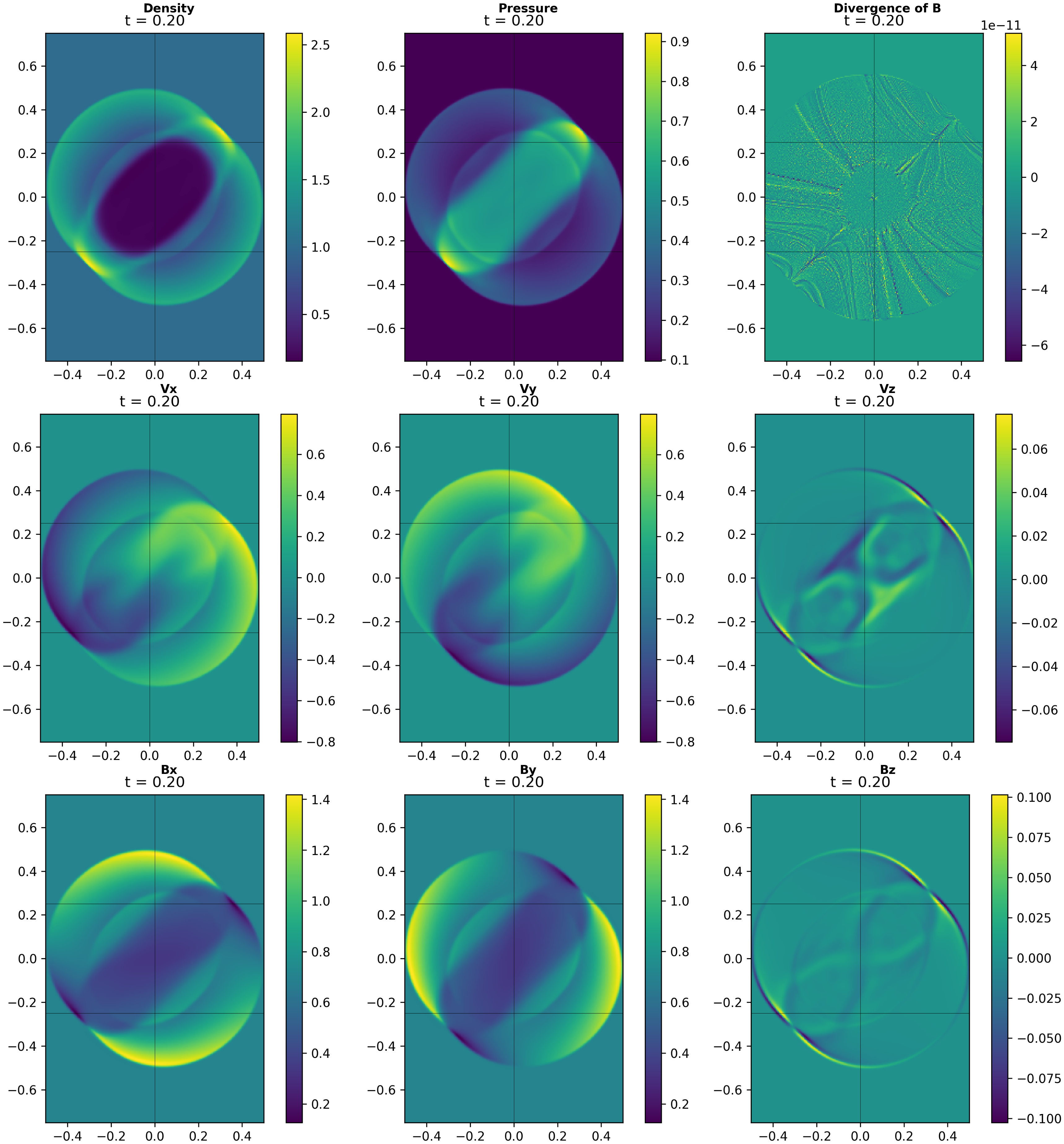}    
    \caption{Default 2D ideal MHD blast wave problem with slanted magnetic field along x-y direction. \label{fig:XMHD_2Dblast_rho1e0}}
\end{figure}

\begin{figure}[http]
    \centering
    \includegraphics[width=\linewidth]{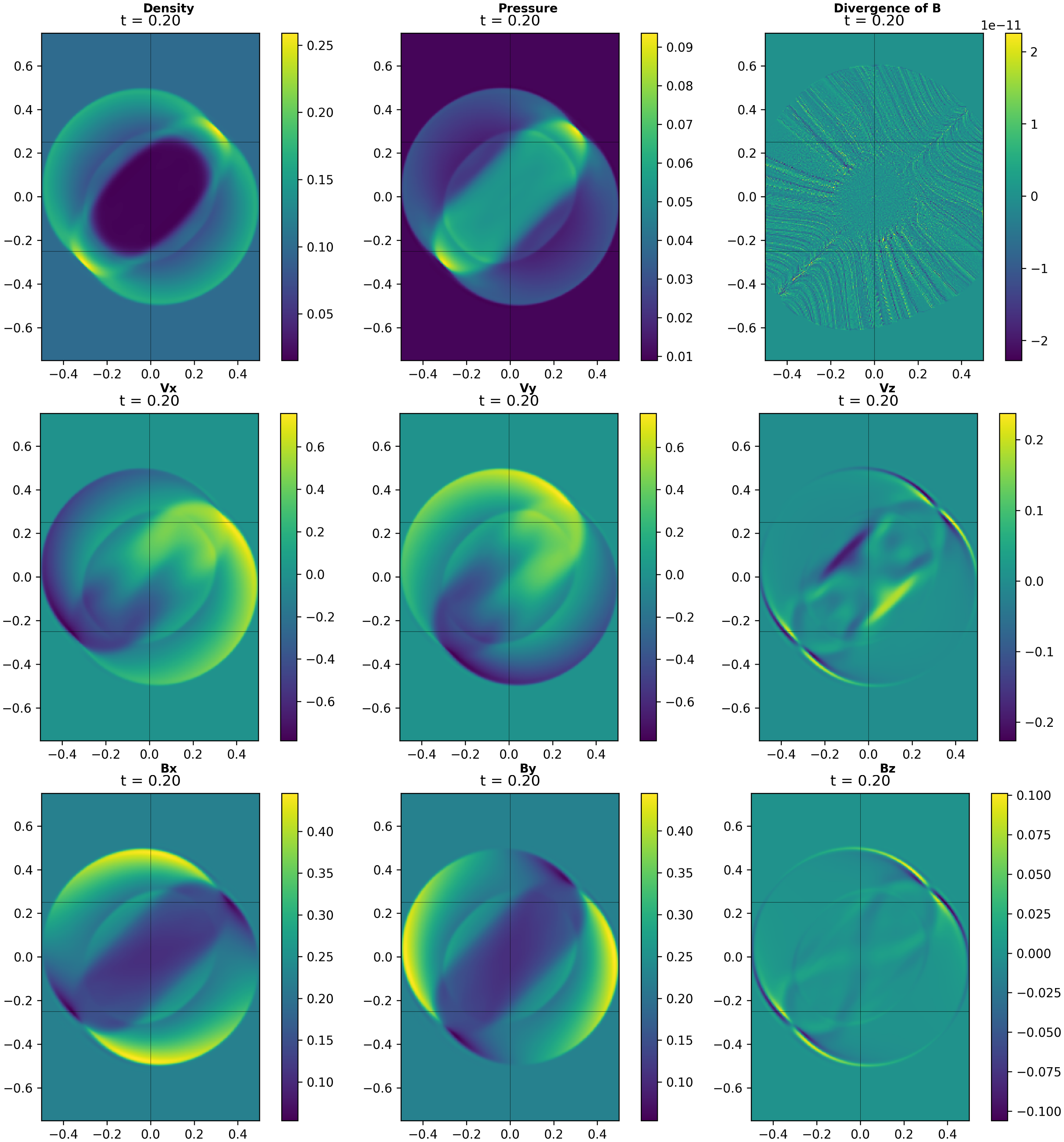}    
    \caption{2D MHD blast wave problem at lower density of $\rho_0=0.1$ where non-ideal effects become more significant. \label{fig:XMHD_2Dblast_rho1em1}}
\end{figure}

This is done on a computational domain of $[-0.5,0.5] \times [-0.75, 0.75]$ with grid resolution of 200 by 300. The background pressure is set to $0.1\rho_0$ while the center spherical region of radius 0.1 has a pressure of $10\rho_0$. The density of the gas is $\rho_0$ while the velocities are 0 everywhere. Uniform magnetic field along y and z components are applied and they have values of $\sqrt{0.5\rho_0}$ and $\sqrt{0.5\rho_0}$, respectively. The heat capacity ratio is set to $\frac{5}{3}$ here. For the ideal MHD case, $\rho_0$ is 1 and to test for the increment of non-ideal effects, $\rho_0$ is decreased to 0.1. This can be seen from Figs. \ref{fig:XMHD_2Dblast_rho1e0} and \ref{fig:XMHD_2Dblast_rho1em1} where the increment of non-ideal effects causes the ratio of the out-of-plane z-components for both velocity and magnetic field to the in-plane components to increase.

\subsection{2D MHD rotor problem}
\begin{figure}[http]
    \centering
    \includegraphics[width=\linewidth]{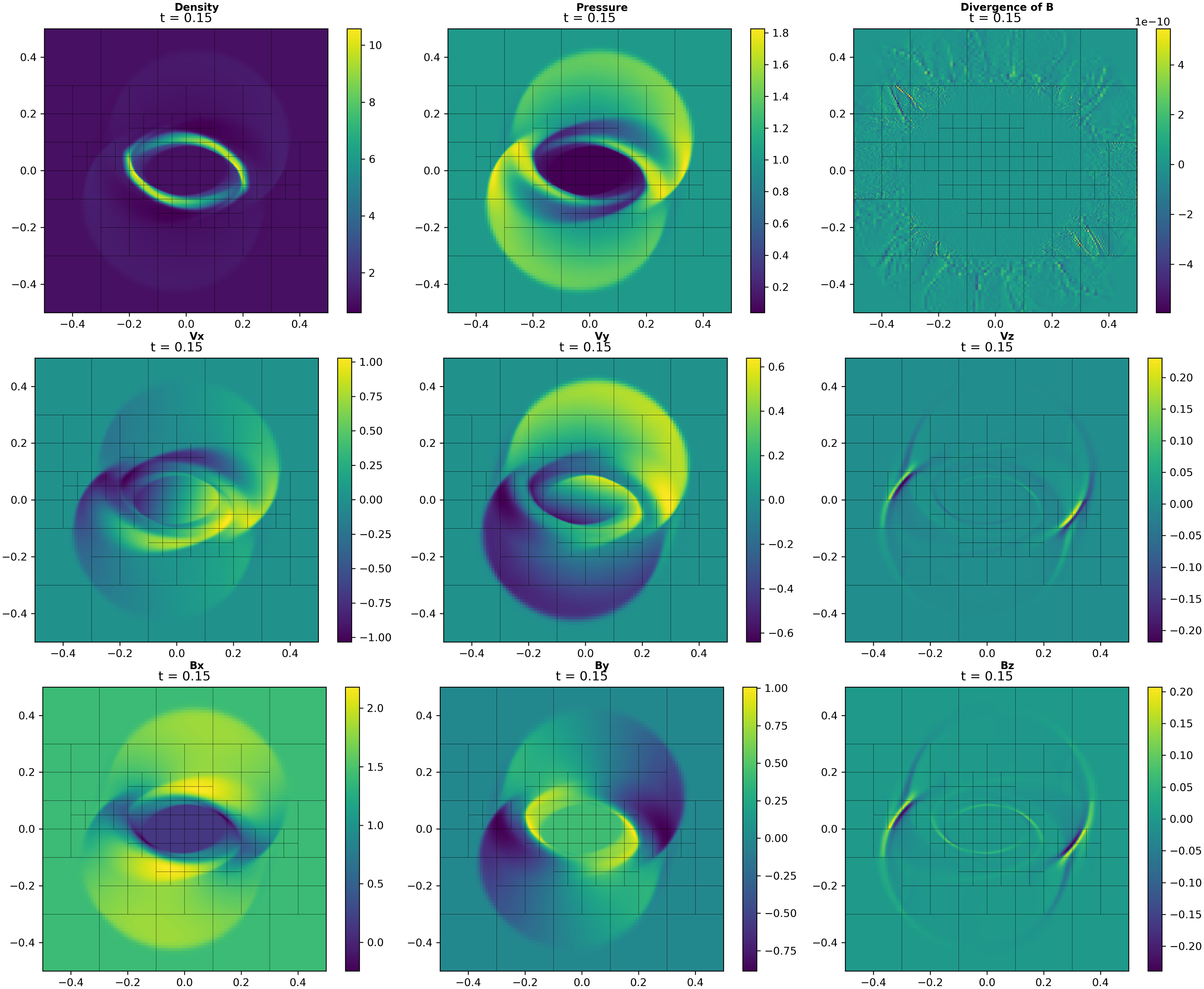}    
    \caption{Default 2D ideal MHD rotor problem with $\rho_0=1$. \label{fig:XMHD_2Drotor_rho1e0}}
\end{figure}

\begin{figure}[http]
    \centering
    \includegraphics[width=\linewidth]{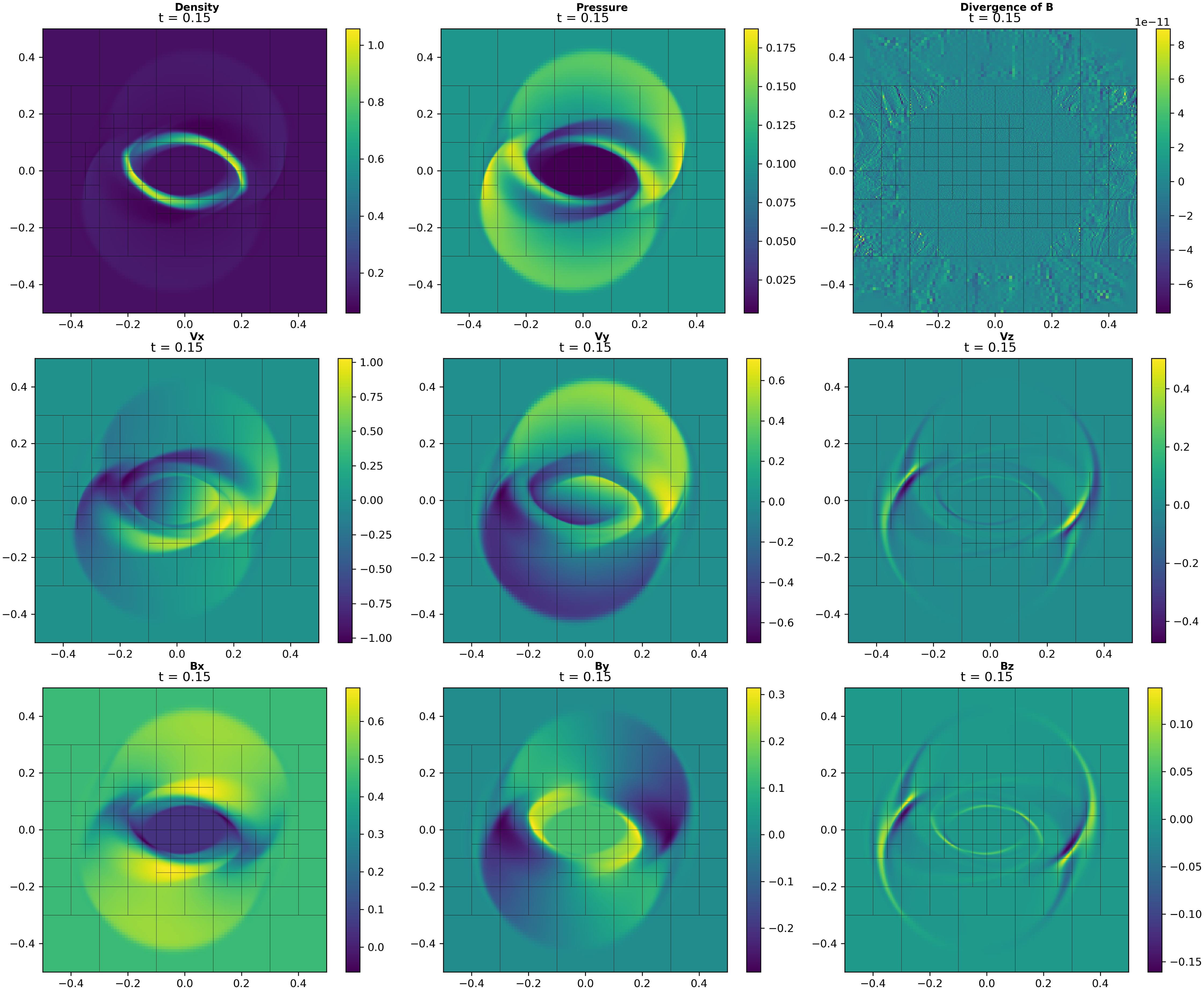}    
    \caption{Default 2D ideal MHD rotor problem with $\rho_0=0.1$ when non-ideal effects become more significant. \label{fig:XMHD_2Drotor_rho1em1}}
\end{figure}

Rotor problem is also included to study the onset and propagation of strong torsional Alfven waves in both ideal and non-ideal regimes. Parthenon’s built-in adaptive mesh refinement (AMR), featuring a uniquely designed second-order prolongation and restriction scheme that preserves $\nabla \cdot \mathbf{B}$ to machine precision \cite{Toth2002}, is also tested with the XMHD solver. This is done using a computational domain of $[-0.5,0.5]\times[-0.5,0.5]$ with base grid resolution of 100 by 100 and 3 levels of adaptive grid refinement. The heat capacity ratio is set to $1.4$ here. For the ideal MHD case, $\rho_0$ is 1 and to test for the increment of non-ideal effects, $\rho_0$ is decreased to 0.1. The problem is initialized with

\begin{align}
\rho(x,y) &= 
\begin{cases}
10, & r \le r_0,\\[6pt]
1 + 9\,f(r), & r_0 < r < r_1,\\[6pt]
1, & r \ge r_1
\end{cases}
\qquad
p(x,y) = 1
\\
u(x,y) &= 
\begin{cases}
-\,u_0\,y, & r \le r_0,\\[6pt]
-\,f(r)\,u_0\,y, & r_0 < r < r_1,\\[6pt]
0, & r \ge r_1
\end{cases}
\qquad
v(x,y) = 
\begin{cases}
u_0\,x, & r \le r_0,\\[6pt]
f(r)\,u_0\,x, & r_0 < r < r_1,\\[6pt]
0, & r \ge r_1
\end{cases}
\\
B_x(x,y) &= \dfrac{5}{\sqrt{4\pi}}
\qquad
B_y(x,y) = 0
\end{align}

where
\[
u_0 = 20, \qquad
r_0 = 0.1,\qquad r_1 = 0.115,\qquad r=\sqrt{x^2+y^2},\qquad w=B_z=0,
\]
and the linear taper function which may be used to prevent strong transients start-up during computation \cite{BALSARA1999270},
\[
f(r)=\frac{r_1-r}{r_1-r_0}.
\]
Since the pressure is uniform initially unlike density, density gradient is used as the criteria for AMR here. Similar to the blast wave problem, decreasing density causes the out-of-plane z-components for both velocity and magnetic field to the in-plane components to increase as shown in Figs. \ref{fig:XMHD_2Drotor_rho1e0} and \ref{fig:XMHD_2Drotor_rho1em1}.

\section{Conclusion}
The goal of this work is to develop a semi-implicit finite-volume scheme for solving the XMHD equations while asymptotically preserving in the ideal MHD limit. Moreover, it produces promising results at the resistive and Hall MHD limits. This potentially enables the current solver to handle a wider range of density regimes, which is particularly important for high-energy-density (HED) plasma systems. Such systems often exhibit large density variations that must be accurately captured to improve the understanding and control of the plasma’s energy and stability.

This is accomplished by first re-writing the governing equations such that it preserves most of the ideal MHD framework, allowing existing ideal MHD solvers or algorithms to be re-used. For instance, the advantage of preserving locally divergence-free magnetic field by placing magnetic-field at face-center through constrained transport method is employed in this XMHD solver. Naturally, the $\mathbf{E}$ field is placed at cell edges, together with $\mathbf{J}$ field for consistency during implicit source term updates. However, the implicit source term update requires directional splitting to avoid spatial biasing since it requires all of the edge-centered $\mathbf{J}$ and $\mathbf{E}$ components which are not co-located. The Jacobian matrix for the governing equations without the source terms is also obtained to determine the diffusive stabilization for the non-ideal components of the LLF Riemann solver evaluated at the cell faces. But this is not sufficient stability-wise, especially at low density where non-ideal effects become significant and therefore, a density-dependent slope limiter is used to increase flux diffusivity away from the continuum limit. Finally, since all the variables are not co-located, interpolation of values across different topological elements of the computational cell is required and simple arithmetic averaging seems sufficient for the test problems here. 

The XMHD solver is then validated using the Brio-Wu shock tube test, producing results consistent with the ideal MHD solver in Athena++. At high densities, ions and electrons temperature immediately equilibrate to a common temperature that is consistent with the single fluid assumption of ideal MHD. At low densities, the solution departs noticeably from the ideal MHD results but remains stable and free of oscillations. The solver is further tested in the ideal MHD limit by propagating linear MHD waves, which exhibit the expected second-order convergence consistent with the PLM reconstruction and RK2 time integration. Second-order convergence is also confirmed in the resistive MHD limit through the diffusion of a Gaussian magnetic-field profile. In the low density regime corresponding to the Hall MHD limit, the XMHD solver reproduces Hall drift waves in regions with density gradients and captures the linear dispersion relation in agreement with analytical predictions. Lastly, the solver is fully compatible with Parthenon’s AMR framework, preserving the divergence-free condition of the magnetic field via specialized prolongation and restriction operators during refinement.

There are other aspects of the current algorithm that require further understanding and improvement. For example, the solver includes two stiff terms (i.e., $\epsilon_E$ and $\epsilon_J$) that depend on the dimensional constants used to non-dimensionalize the XMHD equations. These constants must be chosen such that the term $\frac{B_0^2}{\mu_0 n_0 m_i v^2} \approx 1$, as this allows one to retain the existing ideal MHD solver without explicitly including the term. The ratio of the timestep $dt$ to $\epsilon_E$ and $\epsilon_J$ may influence how effectively the system relaxes toward the appropriate physical limit, and future work will involve better understanding this. Furthermore, there are several potential avenues for enhancing the current algorithm such as
\begin{itemize}
    \item Positive-preserving scheme for ion temperature by increasing and decreasing the numerical diffusivity of total and electron pressure, respectively, across shock.
    \item IMEX or linear analysis can be used to increase the order of accuracy for the implicit source term update, which is currently at most 1st-order.
    \item Upwind constrained transport method can be used instead. Additionally, the concept of upwind constrained transport method can be applied to the non-ideal source term used to update the magnetic field, although the $\mathbf{E}$ field must be accounted for during the upwinding.
    \item Adding ambipolar diffusion to the right-hand side of GOL and testing it against field diffusion with an initial Gaussian profile because it has an analytical solution.
\end{itemize}

\section*{Acknowledgement}
The authors are grateful to Jonah Miller, Patrick Mullen, and Luke Roberts for valuable advise on this work.  This work has been assigned document
release number LA-UR-25-31337. Support was provided by the U.S. Department of Energy through the Los Alamos
National Laboratory. Los Alamos National Laboratory is operated by Triad National Security, LLC, for the National
Nuclear Security Administration of U.S. Department of Energy (Contract No. 89233218CNA000001).  

\appendix
\section{Re-writing XMHD equations}
\label{App1}
The derivation of Eqs. \ref{XMHDv1:mass} - \ref{XMHDv1:entropy} from Eqs. \ref{DG_eq:mass} - \ref{DG_eq:entropy} requires the following vector and vector calculus identities,
\begin{align}
    \nabla \cdot (\mathbf{A} \mathbf{B}) &= \mathbf{B}(\nabla \cdot \mathbf{A}) +(\mathbf{A}\cdot \nabla)\mathbf{B} \label{VCI_1} 
    \\
    \nabla \times (\mathbf{A} \times \mathbf{B}) &= \nabla \cdot (\mathbf{B}\mathbf{A}-\mathbf{A}\mathbf{B}) \label{VCI_2}
    \\
    \nabla(\mathbf{A} \cdot \mathbf{B}) &= (\mathbf{A} \cdot \nabla) \mathbf{B} + \mathbf{A} \times (\nabla \times \mathbf{B}) + (\mathbf{B} \cdot \nabla) \mathbf{A} + \mathbf{B} \times (\nabla \times \mathbf{A}) \label{VCI_3} 
    \\
    \nabla \cdot (\mathbf{A} \times \mathbf{B}) &= (\nabla \times \mathbf{A})\cdot \mathbf{B} - (\nabla \times \mathbf{B}) \cdot \mathbf{A} \label{VCI_4}
    \\
    \mathbf{A} \cdot (\mathbf{B} \times \mathbf{C}) &=
    (\mathbf{A} \times \mathbf{B}) \cdot \mathbf{C} =  -\mathbf{A} \cdot (\mathbf{C} \times \mathbf{B}) \label{VI_1} 
    \\
    \mathbf{A}\times(\mathbf{B}\times \mathbf{C}) &=(\mathbf{A}\cdot \mathbf{C})\mathbf{B}-(\mathbf{A}\cdot \mathbf{B})\mathbf{C} \label{VI_2} 
\end{align}
where $\mathbf{A}\mathbf{B} \equiv \mathbf{A}\otimes\mathbf{B}$.

From Eq. \ref{DG_eq:momentum},
\begin{align}
    \frac{\partial}{\partial t}(\rho \mathbf{u}) + \nabla \cdot (\rho \mathbf{u} \mathbf{u} + P\mathbb{I}) &= \mathbf{J} \times \mathbf{B} \notag 
    \\
    &= \frac{1}{\mu_0} \left( \nabla \times \mathbf{B} \right) \times \mathbf{B} + \left(\mathbf{J} - \frac{1}{\mu_0}\nabla \times \mathbf{B} \right) \times \mathbf{B} \notag
    \\
    &= \frac{1}{\mu_0} \left[ (\mathbf{B} \cdot \nabla) \mathbf{B} - \frac{1}{2}\nabla (\mathbf{B}\cdot \mathbf{B}) \right]  + \left(\mathbf{J} - \frac{1}{\mu_0}\nabla \times \mathbf{B} \right) \times \mathbf{B} \notag
    \\
     &= \frac{1}{\mu_0} \left[ \nabla \cdot  (\mathbf{B}\mathbf{B}) - \frac{1}{2}\nabla \cdot (B^2 \mathbb{I}) \right] + \left(\mathbf{J} - \frac{1}{\mu_0}\nabla \times \mathbf{B} \right) \times \mathbf{B} \notag
\end{align}
where bringing over the divergence term on the right-hand side to the left side of the equation recovers Eq. \ref{XMHDv1:momentum}.

As for Eq. \ref{DG_eq:faraday},
\begin{align}
    \frac{\partial \mathbf{B}}{\partial t} &= - \nabla \times \mathbf{E} \notag
    \\
    &= \nabla \times (\mathbf{u} \times \mathbf{B}) - \nabla \times (\mathbf{E} + \mathbf{u} \times \mathbf{B}) \notag
    \\
    &= \nabla \cdot (\mathbf{B}\mathbf{u}-\mathbf{u}\mathbf{B}) - \nabla \times (\mathbf{E} + \mathbf{u} \times \mathbf{B}) \notag
\end{align}
Similarly, the divergence on the right-hand side of the equation is carried to the left to give Eq. \ref{XMHDv1:faraday}.

Finally for energy equation from Eq. \ref{DG_eq:energy},
\begin{align}
    \frac{\partial E_n}{\partial t} + \nabla \cdot \left[ \mathbf{u}(E_n + P) \right] &= \mathbf{u} \cdot (\mathbf{J} \times \mathbf{B}) + \eta J^2 \notag 
    \\
    &= \mathbf{u} \cdot \left[\frac{1}{\mu_0} (\nabla \times \mathbf{B}) \times \mathbf{B}\right] + \mathbf{u} \cdot \left[ \left(\mathbf{J} - \frac{1}{\mu_0}\nabla \times \mathbf{B} \right) \times \mathbf{B} \right] + \eta J^2 \label{XMHD_A1energy}
\end{align}
To further expand the right-hand side of Eq. \ref{XMHD_A1energy}, take the dot product of Eq. \ref{XMHDv1:faraday} with $\frac{1}{\mu_0}\mathbf{B}$ as follows,
\begin{align}
    \frac{\partial}{\partial t}\left(\frac{1}{2\mu_0}B^2 \right) &= \frac{1}{\mu_0} \mathbf{B}\cdot [\nabla \times (\mathbf{u}\times \mathbf{B})]-\frac{1}{\mu_0} \mathbf{B} \cdot [\nabla \times(\mathbf{E}+\mathbf{u}\times \mathbf{B})] \notag
    \\
    &= \frac{1}{\mu_0} \left[ \nabla \cdot [(\mathbf{u}\times \mathbf{B})\times \mathbf{B}] + (\mathbf{u}\times\mathbf{B})\cdot (\nabla \times \mathbf{B}) \right] -\frac{1}{\mu_0} \mathbf{B} \cdot [\nabla \times(\mathbf{E}+\mathbf{u}\times \mathbf{B})] \notag 
    \\
     &= \frac{1}{\mu_0} \left[ \nabla \cdot [(\mathbf{u}\times \mathbf{B})\times \mathbf{B}] + \mathbf{u}\cdot(\mathbf{B} \times(\nabla \times \mathbf{B}) \right] -\frac{1}{\mu_0} \mathbf{B} \cdot [\nabla \times(\mathbf{E}+\mathbf{u}\times \mathbf{B})]
\end{align}
This gives 
\begin{equation}
    \mathbf{u} \cdot \left[\frac{1}{\mu_0} (\nabla \times \mathbf{B}) \times \mathbf{B}\right] = -\frac{\partial}{\partial t}\left(\frac{1}{2\mu_0}B^2 \right) + \frac{1}{\mu_0} \nabla \cdot [(\mathbf{u}\times \mathbf{B})\times \mathbf{B}] -\frac{1}{\mu_0} \mathbf{B} \cdot [\nabla \times(\mathbf{E}+\mathbf{u}\times \mathbf{B})]
\end{equation}
which can be substituted into the right-hand side of Eq. \ref{XMHD_A1energy},
\begin{align}
    \frac{\partial E_n}{\partial t} + \nabla \cdot \left[ \mathbf{u}(E_n + P) \right] &= -\frac{\partial}{\partial t}\left(\frac{1}{2\mu_0}B^2 \right) + \frac{1}{\mu_0} \nabla \cdot [(\mathbf{u}\times \mathbf{B})\times \mathbf{B}] -\frac{1}{\mu_0} \mathbf{B} \cdot [\nabla \times(\mathbf{E}+\mathbf{u}\times \mathbf{B})]  \notag 
    \\
    &+ \mathbf{u} \cdot \left[ \left(\mathbf{J} - \frac{1}{\mu_0}\nabla \times \mathbf{B} \right) \times \mathbf{B} \right] + \eta J^2 \label{XMHD_A2energy}
\end{align}
Bringing over the first two right-hand terms of above Eq. \ref{XMHD_A2energy} to the left will recover Eq. \ref{XMHDv1:energy}.

\section{Deriving dimensionless XMHD equations}
\label{App2}
To non-dimensionalized XMHD equations from Eqs. \ref{XMHDv1:mass} - \ref{XMHDv1:entropy} to Eqs. \ref{XMHDv1:ampere_ND} -\ref{XMHDv1:entropy_ND}, the following conversions must be used.
\begin{align}
    P_0&=n_0 m_i v^2, \ \ E_0=vB_0, \ \ J_0=\frac{B_0}{\mu_0L_0}, \ \ v_A=\frac{B_0}{\sqrt{n_0 m_i \mu_0}}, \ \ v=\frac{L_0}{t_0} \notag
    \\
    x&=L_0 x^*, \ \ t=t_0 t^*, \ \ 
    n_e=Zn_0\rho^*, \ \ P=P_0P^*, \ \ P_e=P_0P^*_e,\\
    \mathbf{u}&=v\mathbf{u}^*,  \ \  \mathbf{B}=B_0 \mathbf{B}^*, \ \ \mathbf{E}=E_0 \mathbf{E}^*, \ \ \mathbf{J}=J_0 \mathbf{J}^* \notag
\end{align}
where $\epsilon_0\approx 8.854\times 10^{-12}$, $\mu_0 = 4\pi \times 10^{-7}$ and $(\cdot)^\ast$ denotes dimensionless variables. Note that $L_0$ and $t_0$ are chosen such that $v=v_A$ which allows the dimensional coefficients to be largely canceled out.




\bibliographystyle{ieeetr}
\bibliography{refs}

@article{ZHAO2014400,
title = {A positivity-preserving semi-implicit discontinuous Galerkin scheme for solving extended magnetohydrodynamics equations},
journal = {Journal of Computational Physics},
volume = {278},
pages = {400-415},
year = {2014},
issn = {0021-9991},
doi = {https://doi.org/10.1016/j.jcp.2014.08.044},
url = {https://www.sciencedirect.com/science/article/pii/S0021999114006202},
author = {Xuan Zhao and Yang Yang and Charles E. Seyler}
}

@article{10.1063/1.3543799,
    author = {Seyler, C. E. and Martin, M. R.},
    title = {Relaxation model for extended magnetohydrodynamics: Comparison to magnetohydrodynamics for dense Z-pinches},
    journal = {Physics of Plasmas},
    volume = {18},
    number = {1},
    pages = {012703},
    year = {2011},
    month = {01},
    issn = {1070-664X},
    doi = {10.1063/1.3543799},
    url = {https://doi.org/10.1063/1.3543799},
    eprint = {https://pubs.aip.org/aip/pop/article-pdf/doi/10.1063/1.3543799/14687083/012703\_1\_online.pdf},
}

@article{SHU1988439,
title = {Efficient implementation of essentially non-oscillatory shock-capturing schemes},
journal = {Journal of Computational Physics},
volume = {77},
number = {2},
pages = {439-471},
year = {1988},
issn = {0021-9991},
doi = {https://doi.org/10.1016/0021-9991(88)90177-5},
url = {https://www.sciencedirect.com/science/article/pii/0021999188901775},
author = {Chi-Wang Shu and Stanley Osher}
}

@ARTICLE{2008ApJS..178..137S,
       author = {{Stone}, James M. and {Gardiner}, Thomas A. and {Teuben}, Peter and {Hawley}, John F. and {Simon}, Jacob B.},
        title = "{Athena: A New Code for Astrophysical MHD}",
      journal = {The Astrophysical Journal Supplement Series},
         year = 2008,
        month = sep,
       volume = {178},
       number = {1},
        pages = {137-177},
          doi = {10.1086/588755},
archivePrefix = {arXiv},
       eprint = {0804.0402},
 primaryClass = {astro-ph},
       adsurl = {https://ui.adsabs.harvard.edu/abs/2008ApJS..178..137S}
}

@ARTICLE{2020ApJS..249....4S,
       author = {{Stone}, James M. and {Tomida}, Kengo and {White}, Christopher J. and {Felker}, Kyle G.},
        title = "{The Athena++ Adaptive Mesh Refinement Framework: Design and Magnetohydrodynamic Solvers}",
      journal = {The Astrophysical Journal Supplement Series},
         year = 2020,
        month = jul,
       volume = {249},
       number = {1},
          eid = {4},
        pages = {4},
          doi = {10.3847/1538-4365/ab929b},
archivePrefix = {arXiv},
       eprint = {2005.06651},
 primaryClass = {astro-ph.IM},
       adsurl = {https://ui.adsabs.harvard.edu/abs/2020ApJS..249....4S}
}

@article{BALSARA1999270,
title = {A Staggered Mesh Algorithm Using High Order Godunov Fluxes to Ensure Solenoidal Magnetic Fields in Magnetohydrodynamic Simulations},
journal = {Journal of Computational Physics},
volume = {149},
number = {2},
pages = {270-292},
year = {1999},
issn = {0021-9991},
doi = {https://doi.org/10.1006/jcph.1998.6153},
url = {https://www.sciencedirect.com/science/article/pii/S0021999198961538},
author = {Dinshaw S Balsara and Daniel S Spicer}
}

@article{Ferraro2010,
    author = {Ferraro, N. M. and Jardin, S. C. and Snyder, P. B.},
    title = {Ideal and resistive edge stability calculations with M3D-C1},
    journal = {Physics of Plasmas},
    volume = {17},
    number = {10},
    pages = {102508},
    year = {2010},
    month = {10},
    issn = {1070-664X},
    doi = {10.1063/1.3492727},
    url = {https://doi.org/10.1063/1.3492727},
    eprint = {https://pubs.aip.org/aip/pop/article-pdf/doi/10.1063/1.3492727/15712128/102508\_1\_online.pdf},
}

@article{GARDINER20084123,
title = {An unsplit Godunov method for ideal MHD via constrained transport in three dimensions},
journal = {Journal of Computational Physics},
volume = {227},
number = {8},
pages = {4123-4141},
year = {2008},
issn = {0021-9991},
doi = {https://doi.org/10.1016/j.jcp.2007.12.017},
url = {https://www.sciencedirect.com/science/article/pii/S0021999107005669},
author = {Thomas A. Gardiner and James M. Stone}
}

@phdthesis{Mullen2021,
  title        = {Magnetized models for the formation of the Moon},
  author       = {Mullen, Patrick Dean},
  school       = {University of Illinois at Urbana-Champaign},
  year         = {2021},
  month        = {August},
  address      = {Urbana, IL},
  advisor      = {Gammie, Charles F},
  committee    = {Ricker, Paul M and Fields, Brian D and Looney, Leslie W},
  url          = {http://hdl.handle.net/2142/113014},
  note         = {Ph.D. Dissertation},
}

@article{Broadwell,
author = {Caflisch, Russel E. and Jin, Shi and Russo, Giovanni},
title = {Uniformly Accurate Schemes for Hyperbolic Systems with Relaxation},
journal = {SIAM Journal on Numerical Analysis},
volume = {34},
number = {1},
pages = {246-281},
year = {1997},
doi = {10.1137/S0036142994268090},

URL = { 
    
        https://doi.org/10.1137/S0036142994268090
    
    

},
eprint = { 
    
        https://doi.org/10.1137/S0036142994268090
    
    

}
}

@article{Iwasaki_2025,
doi = {10.3847/1538-4357/adc2f7},
url = {https://dx.doi.org/10.3847/1538-4357/adc2f7},
year = {2025},
month = {apr},
publisher = {The American Astronomical Society},
volume = {984},
number = {1},
pages = {50},
author = {Iwasaki, Kazunari and Tomida, Kengo},
title = {Comparative Analysis of Hall Effect Implementations in Hall Magnetohydrodynamics},
journal = {The Astrophysical Journal}
}

@Inbook{Huba2003,
author="Huba, Joseph D.",
editor="B{\"u}chner, J{\"o}rg
and Scholer, Manfred
and Dum, Christian T.",
title="Hall Magnetohydrodynamics - A Tutorial",
bookTitle="Space Plasma Simulation",
year="2003",
publisher="Springer Berlin Heidelberg",
address="Berlin, Heidelberg",
pages="166--192",
isbn="978-3-540-36530-3",
doi="10.1007/3-540-36530-3_9",
url="https://doi.org/10.1007/3-540-36530-3_9"
}

@book{Trefethen_numerical_linear_algebra,
author = {Trefethen, Lloyd N. and Bau, David},
title = {Numerical Linear Algebra, Twenty-fifth Anniversary Edition},
publisher = {Society for Industrial and Applied Mathematics},
year = {2022},
doi = {10.1137/1.9781611977165},
address = {Philadelphia, PA},
edition   = {},
URL = {https://epubs.siam.org/doi/abs/10.1137/1.9781611977165},
eprint = {https://epubs.siam.org/doi/pdf/10.1137/1.9781611977165}
}

@ARTICLE{HLLD_Miyoshi,
       author = {{Miyoshi}, Takahiro and {Kusano}, Kanya},
        title = "{A multi-state HLL approximate Riemann solver for ideal magnetohydrodynamics}",
      journal = {Journal of Computational Physics},
         year = 2005,
        month = sep,
       volume = {208},
       number = {1},
        pages = {315-344},
          doi = {10.1016/j.jcp.2005.02.017},
       adsurl = {https://ui.adsabs.harvard.edu/abs/2005JCoPh.208..315M}
}

@article{THOMA2021,
title = {Implicit highly-coupled single-ion Hall-MHD formulation for hybrid particle-in-cell codes},
journal = {Computer Physics Communications},
volume = {261},
pages = {107823},
year = {2021},
issn = {0010-4655},
doi = {https://doi.org/10.1016/j.cpc.2021.107823},
url = {https://www.sciencedirect.com/science/article/pii/S0010465521000011},
author = {C. Thoma and D.R. Welch and D.V. Rose}
}

@article{osti_1903416,
title = {Parthenon—a performance portable block-structured adaptive mesh refinement framework},
author = {Grete, Philipp and Dolence, Joshua C. and Miller, Jonah M. and Brown, Joshua and Ryan, Ben and Gaspar, Andrew and Glines, Forrest and Swaminarayan, Sriram and Lippuner, Jonas and Solomon, Clell J. and Shipman, Galen and Junghans, Christoph and Holladay, Daniel and Stone, James M. and Roberts, Luke F.},
abstractNote = {On the path to exascale the landscape of computer device architectures and corresponding programming models has become much more diverse. While various low-level performance portable programming models are available, support at the application level lacks behind. To address this issue, we present the performance portable block-structured adaptive mesh refinement (AMR) framework Parthenon, derived from the well-tested and widely used Athena++ astrophysical magnetohydrodynamics code, but generalized to serve as the foundation for a variety of downstream multi-physics codes. Parthenon adopts the Kokkos programming model, and provides various levels of abstractions from multidimensional variables, to packages defining and separating components, to launching of parallel compute kernels. Parthenon allocates all data in device memory to reduce data movement, supports the logical packing of variables and mesh blocks to reduce kernel launch overhead, and employs one-sided, asynchronous MPI calls to reduce communication overhead in multi-node simulations. Using a hydrodynamics miniapp, we demonstrate weak and strong scaling on various architectures including AMD and NVIDIA GPUs, Intel and AMD x86 CPUs, IBM Power9 CPUs, as well as Fujitsu A64FX CPUs. At the largest scale on Frontier (the first TOP500 exascale machine), the miniapp reaches a total of 1.7 × 10 13 zone-cycles/s on 9216 nodes (73,728 logical GPUs) at [Formula: see text] weak scaling parallel efficiency (starting from a single node). In combination with being an open, collaborative project, this makes Parthenon an ideal framework to target exascale simulations in which the downstream developers can focus on their specific application rather than on the complexity of handling massively-parallel, device-accelerated AMR.},
doi = {10.1177/10943420221143775},
journal = {International Journal of High Performance Computing Applications},
number = 5,
volume = 37,
place = {United States},
year = {2022},
month = {12}
}

@article{LLF1954,
author = {Lax, Peter D.},
title = {Weak solutions of nonlinear hyperbolic equations and their numerical computation},
journal = {Communications on Pure and Applied Mathematics},
volume = {7},
number = {1},
pages = {159-193},
doi = {https://doi.org/10.1002/cpa.3160070112},
url = {https://onlinelibrary.wiley.com/doi/abs/10.1002/cpa.3160070112},
eprint = {https://onlinelibrary.wiley.com/doi/pdf/10.1002/cpa.3160070112},
year = {1954}
}

@book{Hazeltine2004,
  author    = {Hazeltine, Richard D.},
  title     = {The Framework of Plasma Physics},
  edition   = {1st},
  year      = {2004},
  publisher = {CRC Press},
  address   = {Boca Raton},
  doi       = {10.1201/9780429502804},
  isbn      = {9780429502804},
  pages     = {344},
  note      = {eBook published March 7, 2018},
  url       = {https://doi.org/10.1201/9780429502804}
}

@book{krall1973principles,
  title={Principles of Plasma Physics},
  author={Krall, N.A. and Trivelpiece, A.W.},
  isbn={9780070353466},
  lccn={lc72005695},
  series={International series in pure and applied physics},
  url={https://books.google.com/books?id=QmxztAEACAAJ},
  year={1973},
  publisher={McGraw-Hill}
}

@BOOK{1956pfig.book.....S,
       author = {{Spitzer}, L.},
        title = "{Physics of Fully Ionized Gases}",
         year = 1956,
       adsurl = {https://ui.adsabs.harvard.edu/abs/1956pfig.book.....S}
}

@article{CHACON2003573,
title = {A 2D high-$\beta$ Hall MHD implicit nonlinear solver},
journal = {Journal of Computational Physics},
volume = {188},
number = {2},
pages = {573-592},
year = {2003},
issn = {0021-9991},
doi = {https://doi.org/10.1016/S0021-9991(03)00193-1},
url = {https://www.sciencedirect.com/science/article/pii/S0021999103001931},
author = {L. Chacón and D.A. Knoll}
}

@article{HAKIM2006418,
title = {A high resolution wave propagation scheme for ideal Two-Fluid plasma equations},
journal = {Journal of Computational Physics},
volume = {219},
number = {1},
pages = {418-442},
year = {2006},
issn = {0021-9991},
doi = {https://doi.org/10.1016/j.jcp.2006.03.036},
url = {https://www.sciencedirect.com/science/article/pii/S0021999106001707},
author = {A. Hakim and J. Loverich and U. Shumlak}
}

@article{Hakim2008,
  author    = {Hakim, Ammar H.},
  title     = {Extended MHD Modelling with the Ten-Moment Equations},
  journal   = {Journal of Fusion Energy},
  year      = {2008},
  volume    = {27},
  number    = {1},
  pages     = {36--43},
  doi       = {10.1007/s10894-007-9116-z},
  url       = {https://doi.org/10.1007/s10894-007-9116-z},
  issn      = {1572-9591}
}

@inproceedings{Huba2005,
  author    = {Huba, J. D.},
  title     = {Numerical Methods: Ideal and Hall MHD},
  booktitle = {Proceedings of ISSS},
  volume    = {7},
  year      = {2005},
  month     = {March},
  pages     = {26--31}
}

@article{LOVERICH2005251,
title = {A discontinuous Galerkin method for the full two-fluid plasma model},
journal = {Computer Physics Communications},
volume = {169},
number = {1},
pages = {251-255},
year = {2005},
note = {Proceedings of the Europhysics Conference on Computational Physics 2004},
issn = {0010-4655},
doi = {https://doi.org/10.1016/j.cpc.2005.03.058},
url = {https://www.sciencedirect.com/science/article/pii/S0010465505001554},
author = {J. Loverich and U. Shumlak}
}

@article{SHUMLAK2003620,
title = {Approximate Riemann solver for the two-fluid plasma model},
journal = {Journal of Computational Physics},
volume = {187},
number = {2},
pages = {620-638},
year = {2003},
issn = {0021-9991},
doi = {https://doi.org/10.1016/S0021-9991(03)00151-7},
url = {https://www.sciencedirect.com/science/article/pii/S0021999103001517},
author = {U. Shumlak and J. Loverich}
}

@article{BRIO1988400,
title = {An upwind differencing scheme for the equations of ideal magnetohydrodynamics},
journal = {Journal of Computational Physics},
volume = {75},
number = {2},
pages = {400-422},
year = {1988},
issn = {0021-9991},
doi = {https://doi.org/10.1016/0021-9991(88)90120-9},
url = {https://www.sciencedirect.com/science/article/pii/0021999188901209},
author = {M Brio and C.C Wu}
}

@article{walsh2020,
    author = {Walsh, C. A. and Chittenden, J. P. and Hill, D. W. and Ridgers, C.},
    title = {Extended-magnetohydrodynamics in under-dense plasmas},
    journal = {Physics of Plasmas},
    volume = {27},
    number = {2},
    pages = {022103},
    year = {2020},
    month = {02},
    issn = {1070-664X},
    doi = {10.1063/1.5124144},
    url = {https://doi.org/10.1063/1.5124144},
    eprint = {https://pubs.aip.org/aip/pop/article-pdf/doi/10.1063/1.5124144/15769877/022103_1_online.pdf},
}

@article{young2021,
    author = {Young, J. R. and Adams, M. B. and Hasson, H. and West-Abdallah, I. and Evans, M. and Gourdain, P.-A.},
    title = {Using extended MHD to explore lasers as a trigger for x-pinches},
    journal = {Physics of Plasmas},
    volume = {28},
    number = {10},
    pages = {102703},
    year = {2021},
    month = {10},
    issn = {1070-664X},
    doi = {10.1063/5.0060581},
    url = {https://doi.org/10.1063/5.0060581},
    eprint = {https://pubs.aip.org/aip/pop/article-pdf/doi/10.1063/5.0060581/15881059/102703_1_online.pdf},
}

@article{Woolstrum2020,
    author = {Woolstrum, J. M. and Yager-Elorriaga, D. A. and Campbell, P. C. and Jordan, N. M. and Seyler, C. E. and McBride, R. D.},
    title = {Extended magnetohydrodynamics simulations of thin-foil Z-pinch implosions with comparison to experiments},
    journal = {Physics of Plasmas},
    volume = {27},
    number = {9},
    pages = {092705},
    year = {2020},
    month = {09},
    issn = {1070-664X},
    doi = {10.1063/5.0012170},
    url = {https://doi.org/10.1063/5.0012170},
    eprint = {https://pubs.aip.org/aip/pop/article-pdf/doi/10.1063/5.0012170/15813920/092705_1_online.pdf},
}

@article{Kleiner2025,
  author    = {A. Kleiner and N. M. Ferraro and R. Sweeney and B. C. Lyons and M. Reinke},
  title     = {Extended-{MHD} simulations of disruption mitigation via massive gas injection in {SPARC}},
  journal   = {Nuclear Fusion},
  volume    = {65},
  number    = {2},
  pages     = {026015},
  year      = {2025},
  publisher = {IOP Publishing Ltd on behalf of the IAEA},
  doi       = {10.1088/1741-4326/ad9ec4},
  url       = {https://doi.org/10.1088/1741-4326/ad9ec4},
  note      = {Published 30 December 2024. {\textcopyright} 2024 The Author(s). Open access.}
}

@ARTICLE{Hamlin2019,
  author={Hamlin, Nathaniel D. and Seyler, Charles E.},
  journal={IEEE Transactions on Plasma Science}, 
  title={Power Flow in Pulsed-Power Systems: The Influence of Hall Physics and Modeling of the Plasma–Vacuum Interface}, 
  year={2019},
  volume={47},
  number={5},
  pages={2064-2073},
  doi={10.1109/TPS.2019.2903843}}

@ARTICLE{Toth2002,
       author = {{T{\'o}th}, G. and {Roe}, P.~L.},
        title = "{Divergence- and Curl-Preserving Prolongation and Restriction Formulas}",
      journal = {Journal of Computational Physics},
         year = 2002,
        month = aug,
       volume = {180},
       number = {2},
        pages = {736-750},
          doi = {10.1006/jcph.2002.7120},
       adsurl = {https://ui.adsabs.harvard.edu/abs/2002JCoPh.180..736T}
}

@article{VANLEER1977263,
title = {Towards the ultimate conservative difference scheme III. Upstream-centered finite-difference schemes for ideal compressible flow},
journal = {Journal of Computational Physics},
volume = {23},
number = {3},
pages = {263-275},
year = {1977},
issn = {0021-9991},
doi = {https://doi.org/10.1016/0021-9991(77)90094-8},
url = {https://www.sciencedirect.com/science/article/pii/0021999177900948},
author = {Bram {Van Leer}}
}

@article{Huba1999,
  author    = {J. D. Huba and J. G. Lyon},
  title     = {A new 3D MHD algorithm: the distribution function method},
  journal   = {Journal of Plasma Physics},
  year      = {1999},
  volume    = {61},
  number    = {3},
  pages     = {391--405},
  publisher = {Cambridge University Press},
  address   = {United Kingdom},
  note      = {Received 20 May 1998; revised 17 November 1998}
}

@article{SANGAM2021110565,
title = {Derivation and numerical approximation of two-temperature Euler plasma model},
journal = {Journal of Computational Physics},
volume = {444},
pages = {110565},
year = {2021},
issn = {0021-9991},
doi = {https://doi.org/10.1016/j.jcp.2021.110565},
url = {https://www.sciencedirect.com/science/article/pii/S0021999121004605},
author = {Afeintou Sangam and Élise Estibals and Hervé Guillard}
}

@manual{flash_userguide_devel,
  title        = {FLASH User’s Guide (Development Version)},
  author       = {{FLASH Center for Computational Science, University of Rochester}},
  organization = {University of Rochester},
  year         = {2024},
  note         = {Development release, updated daily},
  url          = {https://flash.rochester.edu/site/flashcode/user_support/flash_ug_devel.pdf}
}

@article{Jin1999,
author = {Jin, Shi},
title = {Efficient Asymptotic-Preserving (AP) Schemes For Some Multiscale Kinetic Equations},
journal = {SIAM Journal on Scientific Computing},
volume = {21},
number = {2},
pages = {441-454},
year = {1999},
doi = {10.1137/S1064827598334599},

URL = { 
    
        https://doi.org/10.1137/S1064827598334599
    
    

},
eprint = { 
    
        https://doi.org/10.1137/S1064827598334599
    
    

}
}

@incollection{HU2017103,
title = {Chapter 5 - Asymptotic-Preserving Schemes for Multiscale Hyperbolic and Kinetic Equations},
editor = {Rémi Abgrall and Chi-Wang Shu},
series = {Handbook of Numerical Analysis},
publisher = {Elsevier},
volume = {18},
pages = {103-129},
year = {2017},
booktitle = {Handbook of Numerical Methods for Hyperbolic Problems},
issn = {1570-8659},
doi = {https://doi.org/10.1016/bs.hna.2016.09.001},
url = {https://www.sciencedirect.com/science/article/pii/S1570865916300102},
author = {J. Hu and S. Jin and Q. Li}
}

\end{document}